\documentclass[12pt]{article}%
\usepackage{amsfonts}
\usepackage{amsmath}
\usepackage{amssymb}
\usepackage{sectsty}
\usepackage[round,authoryear]{natbib}
\usepackage[left=0.6in, right=0.6in, top=1in, bottom=1in]{geometry}
\usepackage[sc,center,compact,noindentafter,medium]{titlesec}
\usepackage[onehalfspacing,nodisplayskipstretch]{setspace}
\usepackage{graphicx}
\usepackage{paralist}
\usepackage[colorlinks,bookmarksnumbered]{hyperref}%
\providecommand{\U}[1]{\protect\rule{.1in}{.1in}}
\newtheorem{theorem}{\normalfont\scshape Theorem}[section]
\newtheorem{corollary}{\normalfont\scshape Corollary}[section]
\newtheorem{lemma}{\normalfont\scshape Lemma}[section]
\newtheorem{condition}{\normalfont\scshape Assumption}

\newtheorem{remark}{\normalfont\scshape Remark}[section]

\expandafter
\def \expandafter \normalsize \expandafter{\normalsize \setlength \abovedisplayskip{10pt plus 2pt minus 7pt}}
\expandafter
\def \expandafter \normalsize \expandafter{\normalsize \setlength \abovedisplayshortskip{0pt plus 2pt}}
\expandafter
\def \expandafter \normalsize \expandafter{\normalsize \setlength \belowdisplayskip{10pt plus 2pt minus 7pt}}
\expandafter
\def \expandafter \normalsize \expandafter{\normalsize \setlength \belowdisplayshortskip{5pt plus 2pt minus 3pt}}

\numberwithin{equation}{section}
\pltopsep=\medskipamount
\begin{document}

\title{\textsc{Bootstrap Inference in the Presence of Bias}}
\author{Giuseppe Cavaliere\thanks{Corresponding author. Address: Dipartimento di
Scienze Economiche, Piazza Scaravilli 2, Bologna, Italy; email:
\texttt{giuseppe.cavaliere@unibo.it}.}\\University of Bologna, Italy\\\& Exeter Business School, UK
\and S\'{\i}lvia Gon\c{c}alves\\McGill University, Canada\linebreak
\and Morten \O rregaard Nielsen\\Aarhus University, Denmark
\and Edoardo Zanelli\\University of Bologna, Italy}
\maketitle

\begin{abstract}
We consider bootstrap inference for estimators which are (asymptotically)
biased. We show that, even when the bias term cannot be consistently
estimated, valid inference can be obtained by proper implementations of the
bootstrap. Specifically, we show that the prepivoting approach of Beran (1987,
1988), originally proposed to deliver higher-order refinements, restores
bootstrap validity by transforming the original bootstrap p-value into an
asymptotically uniform random variable. We propose two different
implementations of prepivoting (plug-in and double bootstrap), and provide
general high-level conditions that imply validity of bootstrap inference. To
illustrate the practical relevance and implementation of our results, we
discuss five examples: (i)~inference on a target parameter based on model
averaging; (ii)~ridge-type regularized estimators; (iii)~nonparametric
regression; (iv)~a location model for infinite variance\ data; and (v)~dynamic
panel data models.

\end{abstract}

\medskip\noindent\textsc{Keywords}:\ Asymptotic bias, bootstrap, incidental
parameter bias, model averaging, nonparametric regression, prepivoting.%

\def\spacingset#1{\renewcommand{\baselinestretch}{#1}\small\normalsize}
\spacingset{1}
\thispagestyle{empty}
\newpage\spacingset{1.1}%

\section{Introduction}

\textsc{Suppose that} $\theta$ is a scalar parameter of interest and let
$\hat{\theta}_{n}$ denote an estimator for which
\begin{equation}
T_{n}:=g(n)(\hat{\theta}_{n}-\theta)\overset{d}{\rightarrow}B+\xi_{1},
\label{eq 1}%
\end{equation}
where $g(n)\rightarrow\infty$ is the rate of convergence of $\hat{\theta}_{n}%
$, $\xi_{1}$ is a continuous random variable centered at zero, and $B$ is an
asymptotic bias (our theory in fact allows for a more general formulation of
the bias). A typical example is $g(n)=n^{1/2}$ and $\xi_{1}\sim N(0,\sigma
^{2})$. Unless $B$ can be consistently estimated, which is often difficult or
impossible, classic (first-order) asymptotic inference on $\theta$ based on
quantiles of $\xi_{1}$ in (\ref{eq 1}) is not feasible. Furthermore, the
bootstrap, which is well known to deliver asymptotic refinements over
first-order asymptotic approximations as well as bias corrections (Hall, 1992;
Horowitz, 2001; Cattaneo and Jansson, 2018, 2022; Cattaneo, Jansson, and Ma,
2019), cannot in general be applied to solve the asymptotic bias problem when
a consistent estimator of $B$ does not exist. Examples are given below.

Our goal is to justify bootstrap inference based on $T_{n}$ in the context of
asymptotically biased estimators and where a consistent estimator of $B$ does
not exist. Consider the bootstrap statistic $T_{n}^{\ast}:=g(n)(\hat{\theta
}_{n}^{\ast}-\hat{\theta}_{n})$, where $\hat{\theta}_{n}^{\ast}$ is a
bootstrap version of $\hat{\theta}_{n}$, such that%
\begin{equation}
T_{n}^{\ast}-\hat{B}_{n}\overset{d^{\ast}}{\rightarrow}_{p}\xi_{1},
\label{eq 2}%
\end{equation}
where $\hat{B}_{n}$ is the implicit bootstrap bias, and `$\overset{d^{\ast
}}{\rightarrow}_{p}$' denotes weak convergence in probability (defined below).
When $\hat{B}_{n}-B=o_{p}(1)$, the bootstrap is asymptotically valid in the
usual sense that the bootstrap distribution of $T_{n}^{\ast}$ is consistent
for the asymptotic distribution of $T_{n}$, i.e., $\sup_{x\in\mathbb{R}%
}|P^{\ast}(T_{n}^{\ast}\leq x)-P(T_{n}\leq x)|=o_{p}(1)$.

We consider situations where $\hat{B}_{n}-B$ is not asymptotically negligible
so the bootstrap fails to replicate the asymptotic bias. For example, this
happens when the asymptotic bias term in the bootstrap world includes a random
(additive) component, i.e.\
\begin{equation}
\hat{B}_{n}-B\overset{d}{\rightarrow}\xi_{2}\text{ (jointly with
(\ref{eq 1})),} \label{eq 3}%
\end{equation}
where $\xi_{2}$ is a random variable centered at zero. In this case, the
bootstrap distribution is random in the limit and hence cannot mimic the
asymptotic distribution given in (\ref{eq 1}). Moreover, the distribution of
the bootstrap p-value, $\hat{p}_{n}:=P^{\ast}(T_{n}^{\ast}\leq T_{n})$, is not
asymptotically uniform, and the bootstrap cannot in general deliver hypothesis
tests (or confidence intervals) with the desired null rejection probability
(or coverage probability).

In this paper, we show that in this non-standard case valid inference can
successfully be restored by proper implementation of the bootstrap. This is
done by focusing on properties of the bootstrap p-value rather than on the
bootstrap as a means of estimating limiting distributions, which is infeasible
due to the asymptotic bias. In particular, we show that such implementations
lead to bootstrap inferences that are valid in the sense that they provide
asymptotically uniformly distributed p-values.

Our inference strategy is based on the fact that, for some bootstrap schemes,
the large-sample distribution of the bootstrap p-value, say $H(u)$,
$u\in\lbrack0,1]$, although not uniform, does not depend on~$B$. That is, we
can search for bootstrap algorithms which generate bootstrap p-values that, in
large samples, are not affected by unknown bias terms. When this is possible,
we can make use of the prepivoting approach of Beran (1987, 1988), which ---
as we will show in this paper --- allows to restore bootstrap validity.
Specifically, our proposed modified p-value is defined as%
\[
\tilde{p}_{n}:=\hat{H}_{n}(\hat{p}_{n}),
\]
where $\hat{H}_{n}(u)$ is any consistent estimator of $H(u)$, uniformly over
$u\in\lbrack0,1]$. The (asymptotic) probability integral transform $\hat
{p}_{n}\mapsto H(\hat{p}_{n})$, continuity of $H(u)$, and consistency of
$\hat{H}_{n}(u)$ then guarantee that $\tilde{p}_{n}$ is asymptotically
uniformly distributed. Interestingly, Beran (1987, 1988) proposed this
approach to obtain asymptotic refinements for the bootstrap, but did not
consider asymptotically biased estimators as we do here.

We propose two approaches to estimating $H$. First, if $H=H_{\gamma}$, where
$\gamma$ is a finite-dimensional parameter vector, and a consistent estimator
$\hat{\gamma}_{n}$ of $\gamma$ is available, then a `plug-in' approach setting
$\hat{H}_{n}=H_{\hat{\gamma}_{n}}$ can deliver asymptotically uniform
p-values. Second, if estimation of $\gamma$ is difficult (e.g., when $\gamma$
does not have a closed form expression), we can use a `double bootstrap'
scheme (Efron, 1983; Hall, 1986), where estimation of $H$ is achieved by
resampling from the bootstrap data originated in the first level.

For both methods, we provide general high-level conditions that imply validity
of the proposed approach. Our conditions are not specific to a given bootstrap
method; rather, they can in principle be applied to any bootstrap scheme
satisfying the proposed sufficient conditions for asymptotic validity.

Our approach is related to recent work by Shao and Politis (2013) and
Cavaliere and Georgiev (2020). In particular, a common feature is that the
distribution function of the bootstrap statistic, conditional on the original
data, is random in the limit. Cavaliere and Georgiev (2020)\ emphasize that
randomness of the limiting bootstrap measure does not prevent the bootstrap
from delivering an asymptotically uniform p-value (bootstrap `unconditional'
validity), and provide results to assess such asymptotic uniformity. Our
context is different, since the presence of an asymptotic bias term renders
the distribution of the bootstrap p-value non-uniform, even asymptotically. In
this respect, our work is related to Shao and Politis (2013), who show that
$t$-statistics based on subsampling or block bootstrap methods with bandwidth
proportional to sample size may deliver non-uniformly distributed p-values
that, however, can be estimated.

To illustrate the practical relevance of our results and to show how to
implement them in applied problems, we consider three examples involving
estimators that feature an asymptotic bias term. In the first two examples
(model averaging and ridge regression), $B$ is not consistently estimable due
to the presence of local-to-zero parameters and the standard bootstrap fails.
In the third example (nonparametric regression), the bootstrap fails because
$B$ depends on the second-order derivative of the conditional mean function,
whose estimation requires the use of a different (suboptimal) bandwidth. In
these examples, $\xi_{1}$ is normal, but $g(n)$ and $B$ are example-specific.
Two additional examples are presented in the supplement. The fourth is a
simple location model without the assumption of finite variance, where
$\xi_{1}$ is not normal and estimators converge at an unknown rate. The fifth
example considers inference for dynamic panel data models, where $B$ is the
incidental parameter bias.

The remainder of the paper is organized as follows. In
Section~\ref{sec:examples} we introduce our three leading examples.
Section~\ref{Sec:general} contains our general results, which we apply to the
three examples in Section~\ref{sec:examples2}. Section~\ref{Sec conc}
concludes. The supplemental material contains two appendices.
Appendix~\ref{Sec asympt Gaussian} specializes the general theory to the case
of asymptotically Gaussian statistics, and Appendix~\ref{Sec Applications}
contains details and proofs for the three leading examples, as well as two
additional examples.

\subsection*{Notation}

Throughout this paper, the notation $\sim$ indicates equality in distribution.
For instance, $Z\sim N(0,1)$ means that $Z$ is distributed as a standard
normal random variable. We write `$x:=y$' and `$y=:x$' to mean that $x$ is
defined by~$y$. The standard Gaussian cumulative distribution function (cdf)
is denoted by~$\Phi$; $U_{[0,1]}$ is the uniform distribution on $[0,1]$, and
$\mathbb{I}_{\{\cdot\}}$ is the indicator function. If $F$ is a cdf, $F^{-1}$
denotes the generalized inverse, i.e.\ the quantile function, $F^{-1}%
(u):=\inf\{v\in\mathbb{R}:F(v)\geq u\}$, $u\in\mathbb{R}$. Unless specified
otherwise, all limits are for $n\rightarrow\infty$. For matrices $a,b,c$ with
$n$ rows, we let $S_{ab}:=a^{\prime}b/n$ and $S_{ab.c}:=S_{ab}-S_{ac}%
S_{cc}^{-1}S_{cb}$, assuming that $S_{cc}$ has full rank.

For a (single level or first-level) bootstrap sequence, say $Y_{n}^{\ast}$, we
use $Y_{n}^{\ast}\overset{p^{\ast}}{\rightarrow}_{p}0$, or equivalently
$Y_{n}^{\ast}\overset{p^{\ast}}{\rightarrow}0$, in probability, to mean that,
for any $\epsilon>0$, $P^{\ast}(|Y_{n}^{\ast}|>\epsilon)\rightarrow_{p}0$,
where $P^{\ast}$ denotes the probability measure conditional on the original
data~$D_{n}$. An equivalent notation is $Y_{n}^{\ast}=o_{p^{\ast}}(1)$ (where
we omit the qualification \textquotedblleft in probability\textquotedblright%
\ for brevity). Similarly, for a double (or second-level) bootstrap sequence,
say $Y_{n}^{\ast\ast}$, we write $Y_{n}^{\ast\ast}=o_{p^{\ast\ast}}(1)$ to
mean that for all $\epsilon>0$, $P^{\ast\ast}(|Y_{n}^{\ast\ast}|>\epsilon
)\overset{p^{\ast}}{\rightarrow}_{p}0$, where $P^{\ast\ast}$ is the
probability measure conditional on the first-level bootstrap data $D_{n}%
^{\ast}$ and on~$D_{n}$.

We use $Y_{n}^{\ast}\overset{d^{\ast}}{\rightarrow}_{p}\xi$, or equivalently
$Y_{n}^{\ast}\overset{d^{\ast}}{\rightarrow}\xi$, in probability, to mean
that, for all continuity points $u\in\mathbb{R}$ of the cdf of $\xi$, say
$G(u):=P(\xi\leq u)$, it holds that $P^{\ast}(Y_{n}^{\ast}\leq
u)-G(u)\rightarrow_{p}0$. Similarly, for a double bootstrap sequence
$Y_{n}^{\ast\ast}$, we use $Y_{n}^{\ast\ast}\overset{d^{\ast\ast}%
}{\rightarrow}_{p^{\ast}}\xi$, in probability, to mean that $P^{\ast\ast
}(Y_{n}^{\ast\ast}\leq u)-G(u)\overset{p^{\ast}}{\rightarrow}_{p}0$ for all
continuity points $u$ of~$G$.

\section{Examples}

In this section we introduce our three leading examples. Example-specific
regularity conditions, formally stated results, and additional definitions are
given in Appendix~\ref{Sec Applications}. For each of these examples, we argue
that (\ref{eq 1}), (\ref{eq 2}), and (\ref{eq 3}) hold, such that the
bootstrap p-values $\hat{p}_{n}$ are not uniformly distributed rendering
standard bootstrap inference invalid. We then return to each example in
Section~\ref{sec:examples2}, where we discuss how to implement our proposed
method and prove its validity.

\label{sec:examples}

\subsection{Inference after model averaging}

\label{sec:MA-intro}

\noindent\textsc{Setup}. We consider inference based on a model averaging
estimator obtained as a weighted average of least squares estimates (Hansen,
2007). Assume that data are generated according to the linear model%
\begin{equation}
y=x\beta+Z\delta+\varepsilon, \label{MA model}%
\end{equation}
where $\beta$ is the (scalar) parameter of interest and $\varepsilon$ is an
$n$-vector of identically and independently distributed random variables with
mean zero and variance~$\sigma^{2}$ (henceforth i.i.d.$(0,\sigma^{2})$),
conditional on $W:=(x,Z)$.

The researcher fits a set of $M$ models, each of them based on different
exclusion restrictions on the $q$-dimensional vector~$\delta$. This setup
allows for model averaging both explicitly and implicity. The former follows,
e.g., Hansen (2007). The latter includes the common practice of robustness
checks in applied research, where the significance of a target coefficient is
evaluated through an (often informal) assessment of its significance across a
set of regressions based on different sets of controls; see Oster (2019) and
the references therein. Specifically, letting $R_{m}$ denote a $q\times q_{m}$
selection matrix, the $m^{\text{th}}$ model includes $x$ and $Z_{m}:=ZR_{m}$
as regressors, and the corresponding OLS\ estimator of $\beta$ is
$\tilde{\beta}_{m,n}=S_{xx.Z_{m}}^{-1}S_{xy.Z_{m}}$. Given a set of fixed
weights $\omega:=(\omega_{1},\dots,\omega_{M})^{\prime}$ such that $\omega
_{m}\in\lbrack0,1]$ and $\sum_{m=1}^{M}\omega_{m}=1$, the model averaging
estimator is $\tilde{\beta}_{n}:=\sum_{m=1}^{M}\omega_{m}\tilde{\beta}_{m,n}$.
Then $T_{n}:=n^{1/2}(\tilde{\beta}_{n}-\beta)$ satisfies $T_{n}-B_{n}%
\rightarrow_{d}\xi_{1}\sim N(0,v^{2})$, where $v^{2}>0$ and%
\[
B_{n}:=Q_{n}n^{1/2}\delta,\text{\quad}Q_{n}:=\sum_{m=1}^{M}\omega
_{m}S_{xx.Z_{m}}^{-1}S_{xZ.Z_{m}}.
\]
Thus, the magnitude of the asymptotic bias $B_{n}$ depends on $n^{1/2}\delta$.
If $\delta$ is local to zero in the sense that $\delta=cn^{-1/2}$ for some
vector $c\in\mathbb{R}^{q}$ (as in, e.g., Hjort and Claeskens, 2003; Liu,
2015; Hounyo and Lahiri, 2023), then $B_{n}\rightarrow_{p}B:=Qc$ with
$Q:=\operatorname*{plim}Q_{n}$, so that (\ref{eq 1}) is satisfied with nonzero
$B$ in general. Because $B$ depends on~$c$, which is not consistently
estimable, we cannot obtain valid inference from a Gaussian distribution based
on sample analogues of $B$ and~$v^{2}$.

\smallskip

\noindent\textsc{Fixed regressor bootstrap}. We generate the bootstrap sample
as $y^{\ast}=x\hat{\beta}_{n}+Z\hat{\delta}_{n}+\varepsilon^{\ast}$, where
$\varepsilon^{\ast}|D_{n}\sim N(0,\hat{\sigma}_{n}^{2}I_{n})$, $(\hat{\beta
}_{n},\hat{\delta}_{n}^{\prime},\hat{\sigma}_{n}^{2})$ is the OLS\ estimator
from the full model, and $D_{n}=\{y,W\}$. Similar results can be established
for the nonparametric bootstrap where $\varepsilon^{\ast}$ is resampled from
the full model residuals. The bootstrap model averaging estimator is given by
$\tilde{\beta}_{n}^{\ast}:=\sum_{m=1}^{M}\omega_{m}\tilde{\beta}_{m,n}^{\ast}%
$, where $\tilde{\beta}_{m,n}^{\ast}:=S_{xx.Z_{m}}^{-1}S_{xy^{\ast}.Z_{m}}$.
Letting $T_{n}^{\ast}:=n^{1/2}(\tilde{\beta}_{n}^{\ast}-\hat{\beta}_{n})$, we
can show that (\ref{eq 2}) holds with $\hat{B}_{n}:=Q_{n}n^{1/2}\hat{\delta
}_{n}$ such that, as in (\ref{eq 3}),
\[
\hat{B}_{n}-B_{n}=Q_{n}n^{1/2}(\hat{\delta}_{n}-\delta)\overset{d}{\rightarrow
}\xi_{2}\sim N(0,v_{22}),\quad v_{22}>0,
\]
given in particular the asymptotic normality of $n^{1/2}(\hat{\delta}%
_{n}-\delta)$. Because the bias term in the bootstrap world is random in the
limit, the conditional distribution of $T_{n}^{\ast}$ is also random in the
limit, and in particular does not mimic the asymptotic distribution of the
original statistic~$T_{n}$.

\smallskip

\noindent\textsc{Pairs bootstrap}. Consider now a pairs (random design)
bootstrap sample $\{y_{t}^{\ast},x_{t}^{\ast},z_{t}^{\ast};t=1,\dots,n\}$,
based on resampling with replacement from the tuples $\{y_{t},x_{t}%
,z_{t};t=1,\dots,n\}$. As is standard, it is useful to recall that the
bootstrap data have the representation%
\[
y^{\ast}=x^{\ast}\hat{\beta}_{n}+Z^{\ast}\hat{\delta}_{n}+\varepsilon^{\ast},
\]
where $\varepsilon^{\ast}=(\varepsilon_{1}^{\ast},\ldots,\varepsilon_{n}%
^{\ast})^{\prime}$ and $\varepsilon_{t}^{\ast}$ is an i.i.d.\ draw from
$\hat{\varepsilon}_{t}=y_{t}-x_{t}\hat{\beta}_{n}-z_{t}^{\prime}\hat{\delta
}_{n}$. The pairs bootstrap model averaging estimator is%
\[
\tilde{\beta}_{n}^{\ast}:=\sum_{m=1}^{M}\omega_{m}\tilde{\beta}_{m,n}^{\ast
}\text{ with }\tilde{\beta}_{m,n}^{\ast}:=S_{x^{\ast}x^{\ast}.Z_{m}^{\ast}%
}^{-1}S_{x^{\ast}y^{\ast}.Z_{m}^{\ast}}%
\]
and $Z_{m}^{\ast}=Z^{\ast}R_{m}$. The pairs bootstrap statistic is then%
\[
T_{n}^{\ast}:=n^{1/2}(\tilde{\beta}_{n}^{\ast}-\hat{\beta}_{n})=B_{n}^{\ast
}+n^{1/2}S_{x^{\ast}x^{\ast}}^{-1}S_{x^{\ast}\varepsilon^{\ast}},
\]
where%
\[
B_{n}^{\ast}:=\sum_{m=1}^{M}\omega_{m}S_{x^{\ast}x^{\ast}.Z_{m}^{\ast}}%
^{-1}S_{x^{\ast}Z^{\ast}.Z_{m}^{\ast}}n^{1/2}\hat{\delta}_{n}.
\]
Therefore, and in contrast with the fixed regressor bootstrap (FRB), the term
$B_{n}^{\ast}$ is stochastic under the bootstrap probability measure and
replaces the bias term$~\hat{B}_{n}$. This difference is not innocuous because
it implies that $T_{n}^{\ast}-\hat{B}_{n}$ no longer replicates the asymptotic
distribution of $T_{n}-B_{n}$ and (\ref{eq 2}) does not hold. However, this
does not prevent our method from working, but it will require a different set
of conditions which we will give in
Section~\ref{Sec more general high level conditions}.

\subsection{Ridge regression}

\label{sec:ridge-intro}

\noindent\textsc{Setup}. We consider estimation of a vector of regression
parameters through regularization; in particular, by using a ridge estimator.
The model is $y_{t}=\theta^{\prime}x_{t}+\varepsilon_{t}$, $t=1,\dots,n$,
where $x_{t}$ is a $p\times1$ non-stochastic vector and $\varepsilon_{t}\sim
~$i.i.d.$(0,\sigma^{2})$. Interest is on testing $\mathsf{H}_{0}:g^{\prime
}\theta=r$, based on ridge estimation of~$\theta$. Specifically, the ridge
estimator has closed form expression $\tilde{\theta}_{n}=\tilde{S}_{xx}%
^{-1}S_{xy}$, where $\tilde{S}_{xx}:=S_{xx}+n^{-1}c_{n}I_{p}$ and $c_{n}$ is a
tuning parameter that controls the degree of shrinkage towards zero. Clearly,
$c_{n}=0$ corresponds to the OLS\ estimator,~$\hat{\theta}_{n}$. We are
interested in the case where the regressors have limited explanatory power,
i.e.,\ where $\theta=\delta n^{-1/2}$ is local to zero, which can in fact be
taken as a motivation for shrinkage towards zero and hence for ridge
estimation. To test $\mathsf{H}_{0}$, we consider the test statistic
$T_{n}=n^{1/2}(g^{\prime}\tilde{\theta}_{n}-r)$. If $n^{-1}c_{n}\rightarrow
c_{0}\geq0$ (as in, e.g., Fu and Knight, 2000) then, under the null, it holds
that $T_{n}-B_{n}\rightarrow_{d}\xi_{1}\sim N(0,v^{2})$, where
\[
B_{n}:=-c_{n}n^{-1/2}g^{\prime}\tilde{S}_{xx}^{-1}\theta=-c_{n}n^{-1}%
g^{\prime}\tilde{S}_{xx}^{-1}\delta\rightarrow B:=-c_{0}g^{\prime}%
\tilde{\Sigma}_{xx}^{-1}\delta
\]
with $\tilde{\Sigma}_{xx}:=\Sigma_{xx}+c_{0}I_{p}$ and $\Sigma_{xx}:=\lim
S_{xx}$. Hence, for $c_{0}>0$, $\tilde{\theta}_{n}$ is asymptotically biased
and the bias term cannot be consistently estimated. Consequently, (\ref{eq 1})
is satisfied, and inference based on the quantiles of the $N(0,v^{2})$
distribution is invalid unless $c_{0}=0$.

\smallskip

\noindent\textsc{Bootstrap}. Consider a pairs (random design) bootstrap sample
$\{y_{t}^{\ast},x_{t}^{\ast};t=1,\dots,n\}$ built by i.i.d.\ resampling from
the tuples $\{y_{t},x_{t};t=1,\dots,n\}$. The bootstrap analogue of the ridge
estimator is $\tilde{\theta}_{n}^{\ast}:=\tilde{S}_{x^{\ast}x^{\ast}}%
^{-1}S_{x^{\ast}y^{\ast}}$, where $\tilde{S}_{x^{\ast}x^{\ast}}:=S_{x^{\ast
}x^{\ast}}+n^{-1}c_{n}I_{p}$. The bootstrap statistic is $T_{n}^{\ast
}:=n^{1/2}g^{\prime}(\tilde{\theta}_{n}^{\ast}-\hat{\theta}_{n})$, which is
centered using $\hat{\theta}_{n}$ to guarantee that $\varepsilon_{t}^{\ast}$
and $x_{t}^{\ast}$ are uncorrelated in the bootstrap world. Because we have
used a pairs bootstrap, we now have $T_{n}^{\ast}-B_{n}^{\ast}%
\overset{d}{\rightarrow}_{p^{\ast}}\xi_{1}$ for $B_{n}^{\ast}:=-c_{n}%
n^{-1/2}g^{\prime}\tilde{S}_{x^{\ast}x^{\ast}}^{-1}\hat{\theta}_{n}$. However,
$B_{n}^{\ast}-\hat{B}_{n}=o_{p^{\ast}}(1)$ with $\hat{B}_{n}:=-c_{n}%
n^{-1/2}g^{\prime}\tilde{S}_{xx}^{-1}\hat{\theta}_{n}$, such that $T_{n}%
^{\ast}-\hat{B}_{n}$ still satisfies (\ref{eq 2}). Then (\ref{eq 3}) holds
with%
\[
\hat{B}_{n}-B_{n}=-c_{n}n^{-1}g^{\prime}\tilde{S}_{xx}^{-1}n^{1/2}(\hat
{\theta}_{n}-\theta)\overset{d}{\rightarrow}\xi_{2}\sim N(0,v_{22}),\quad
v_{22}>0,
\]
so the bootstrap fails to approximate the asymptotic distribution of $T_{n}$
(see also Chatterjee and Lahiri, 2010, 2011).

\subsection{Nonparametric regression}

\label{sec:nonpar-intro}

\noindent\textsc{Setup}. Consider the model
\begin{equation}
y_{t}=\beta(x_{t})+\varepsilon_{t},\text{\quad}t=1,\dots,n,
\label{nonpar model}%
\end{equation}
where $\beta(\cdot)$ is a smooth function and $\varepsilon_{t}\sim
$~i.i.d.$(0,\sigma^{2})$. For simplicity, we consider a fixed-design model;
i.e., $x_{t}=t/n$. The goal is inference on $\beta(x)$ for a fixed $x\in
(0,1)$. We apply the standard Nadaraya-Watson (fixed-design) estimator
$\hat{\beta}_{h}(x)=(nh)^{-1}\sum_{t=1}^{n}K((x_{t}-x)/h)y_{t}$, where
$h=cn^{-1/5}$ for some $c>0$ is the MSE-optimal bandwidth and $K$ is the
kernel function. We do not consider the more general local polynomial
regression case, although we conjecture that very similar results will hold.
We leave that case for future research. The statistic $T_{n}=(nh)^{1/2}%
(\hat{\beta}_{h}(x)-\beta(x))$ satisfies $T_{n}-B_{n}\rightarrow_{d}\xi
_{1}\sim N(0,v^{2})$, where $v^{2}:=\sigma^{2}\int K(u)^{2}du>0$ and
\begin{equation}
B_{n}:=(nh)^{1/2}\left(  \frac{1}{nh}\sum_{t=1}^{n}k_{t}\beta(x_{t}%
)-\beta(x)\right)  \label{L142}%
\end{equation}
with $k_{t}:=K((x_{t}-x)/h)$. The bias $B_{n}$ satisfies%
\begin{equation}
B_{n}=(nh)^{1/2}(h^{2}\beta^{\prime\prime}(x)\kappa_{2}/2+o(h^{2}))\rightarrow
B:=c^{5/2}\beta^{\prime\prime}(x)\kappa_{2}/2, \label{np bias}%
\end{equation}
where $\kappa_{2}:=\int u^{2}K(u)du$ and $\beta^{\prime\prime}(x)\ $denotes
the second-order derivative of~$\beta(x)$. Thus, (\ref{eq 1}) is satisfied.
Estimating $B$ or $B_{n}$ is challenging because it involves estimating~$\beta
^{\prime\prime}(x)$, and although theoretically valid estimators exist, they
perform poorly in finite samples. This issue is pointed out by Calonico,
Cattaneo, and Titunik (2014) and Calonico, Cattaneo, and Farrell (2018), who
propose more accurate bias correction techniques specifically for regression
discontinuity designs and nonparametric curve estimation.

\smallskip

\noindent\textsc{Bootstrap.} The (parametric) bootstrap sample is generated as
$y_{t}^{\ast}=\hat{\beta}_{h}(x_{t})+\varepsilon_{t}^{\ast}$, $t=1,\dots
,n$,\ where $\varepsilon_{t}^{\ast}|D_{n}\sim$~i.i.d.$N(0,\hat{\sigma}_{n}%
^{2})$ with $D_{n}=\{y_{t},t=1,\dots,n\}$ and $\hat{\sigma}_{n}^{2}$ denotes a
consistent estimator of~$\sigma^{2}$; e.g.\ the residual variance. Let
$\hat{\beta}_{h}^{\ast}(x)=(nh)^{-1}\sum_{t=1}^{n}k_{t}y_{t}^{\ast}$ and
$T_{n}^{\ast}=(nh)^{1/2}(\hat{\beta}_{h}^{\ast}(x)-\hat{\beta}_{h}(x))$. Then
(\ref{eq 2}) is satisfied with%
\[
\hat{B}_{n}:=(nh)^{1/2}\left(  \frac{1}{nh}\sum_{t=1}^{n}k_{t}\hat{\beta}%
_{h}(x_{t})-\hat{\beta}_{h}(x)\right)  .
\]
Because $h=cn^{-1/5}$, (\ref{eq 3}) holds with
\[
\hat{B}_{n}-B_{n}=(nh)^{1/2}\left(  \frac{1}{nh}\sum_{t=1}^{n}k_{t}(\hat
{\beta}_{h}(x_{t})-\beta(x_{t}))-(\hat{\beta}_{h}(x)-\beta(x))\right)
\overset{d}{\rightarrow}\xi_{2}\sim N(0,v_{22}),
\]
where $v_{22}>0$, so the bootstrap is invalid. Two possible\textbf{ }solutions
to this problem are to generate the bootstrap sample as $y_{t}^{\ast}%
=\hat{\beta}_{g}(x_{t})+\varepsilon_{t}^{\ast}$, where $g$ is an oversmoothing
bandwidth satisfying $ng^{5}\rightarrow\infty$ (e.g., H\"{a}rdle and Marron,
1991) or to center the bootstrap statistic at its expected value and add a
consistent estimator of $B$ (e.g., H\"{a}rdle and Bowman, 1988; Eubank and
Speckman, 1993). Both approaches require selecting two bandwidths, which is
not straightforward. An alternative approach suggested by Hall and Horowitz
(2013) focuses on an asymptotic theory-based confidence interval and applies
the bootstrap to calibrate its coverage probability. However, this requires an
additional averaging step across a grid of $x$ (their step 6) to
asymptotically eliminate\ $\xi_{2}$, and it results in an asymptotically
conservative interval. Finally, a non-bootstrap-based solution is
undersmoothing using a bandwidth $h$ satisfying $nh^{5}\rightarrow0$, although
of course that is not MSE-optimal and may result in trivial power against
certain local alternatives; see Section~\ref{sec:nonpar-revist}.

\section{General results}

\label{Sec:general}

\subsection{Framework and invalidity of the standard bootstrap}

The general framework is as follows. We have a statistic $T_{n}$ defined as a
general function of a sample $D_{n}$, for which we would like to compute a
valid bootstrap p-value. Usually $T_{n}$ is a test statistic or a (possibly
normalized) parameter estimator; for example, $T_{n}=g(n)(\hat{\theta}%
_{n}-\theta_{0})$. Let $D_{n}^{\ast}$ denote the bootstrap sample, which
depends on the original data and on some auxiliary bootstrap variates (which
we assume defined jointly with $D_{n}$ on a possibly extended probability
space). Let $T_{n}^{\ast}$ denote the bootstrap version of $T_{n}$ computed on
$D_{n}^{\ast}$; for example, $T_{n}^{\ast}=g(n)(\hat{\theta}_{n}^{\ast}%
-\hat{\theta}_{n})$. Let $\hat{L}_{n}(u):=P^{\ast}(T_{n}^{\ast}\leq u)$,
$u\in\mathbb{R}$, denote its distribution function, conditional on the
original data. The bootstrap p-value is defined as
\[
\hat{p}_{n}:=P^{\ast}(T_{n}^{\ast}\leq T_{n})=\hat{L}_{n}(T_{n}).
\]

First-order asymptotic validity of $\hat{p}_{n}$ requires that $\hat{p}_{n}$
converges in distribution to a standard uniform distribution; i.e.,\ that
$\hat{p}_{n}\rightarrow_{d}U_{[0,1]}$. In this section we focus on a class of
statistics $T_{n}$ and $T_{n}^{\ast}$ for which this condition is not
necessarily satisfied. The main reason is the presence of an additive
`bias'\ term $B_{n}$ that contaminates the distribution of $T_{n}$ and cannot
be replicated by the bootstrap distribution of~$T_{n}^{\ast}$.

\begin{condition}
\label{Assn T}$T_{n}-B_{n}\rightarrow_{d}\xi_{1}$, where $\xi_{1}$ is centered
at zero and the cdf $G(u)=P(\xi_{1}\leq u)$ is continuous and strictly
increasing over its support.
\end{condition}

When $B_{n}$ converges to a nonzero constant~$B$, Assumption~\ref{Assn T} can
be written $T_{n}\rightarrow_{d}B+\xi_{1}$ as in (\ref{eq 1}). If $T_{n}$ is a
normalized version of a (scalar) parameter estimator, i.e.,\ $T_{n}%
=g(n)(\hat{\theta}_{n}-\theta_{0})$, then we can think of $B$ as the
asymptotic bias of\ $\hat{\theta}_{n}$ because $\xi_{1}$ is centered at zero.
Although we allow for the possibility that $B_{n}$ does not have a limit (and
it may even diverge), we will still refer to $B_{n}$ as a `bias term'. More
generally, in Assumption~\ref{Assn T} we cover any statistic $T_{n}$ that is
not necessarily Gaussian (even asymptotically) and whose limiting distribution
is $G$ only after we subtract the sequence~$B_{n}$. The limiting distribution
$G$ may depend on a parameter such that $T_{n}-B_{n}$ is not an asymptotic pivot.

Inference based on the asymptotic distribution of $T_{n}$ requires estimating
$B_{n}$ and any parameter in~$G$. Alternatively, we can use the bootstrap to
bypass parameter estimation and directly compute a bootstrap p-value that
relies on $T_{n}^{\ast}$ and $T_{n}$ alone; that is, we consider $\hat{p}%
_{n}:=P^{\ast}(T_{n}^{\ast}\leq T_{n})$. A set of high-level conditions on
$T_{n}^{\ast}$ and $T_{n}$ that allow us to derive the asymptotic properties
of this p-value are described next.

\begin{condition}
\label{Assn BS}For some $D_{n}$-measurable random variable $\hat{B}_{n}$, it
holds that: (i) $T_{n}^{\ast}-\hat{B}_{n}\overset{d^{\ast}}{\rightarrow}%
_{p}\xi_{1}$, where $\xi_{1}$ is described in Assumption~\ref{Assn T}; (ii)%
\[
\left(
\begin{array}
[c]{c}%
T_{n}-B_{n}\\
\hat{B}_{n}-B_{n}%
\end{array}
\right)  \overset{d}{\rightarrow}\left(
\begin{array}
[c]{c}%
\xi_{1}\\
\xi_{2}%
\end{array}
\right)  ,
\]
where $\xi_{2}$ is centered at zero and $F(u)=P(\xi_{1}-\xi_{2}\leq u)$ is a
continuous cdf.
\end{condition}

Assumption~\ref{Assn BS}(i) states that $T_{n}^{\ast}-\hat{B}_{n}$ converges
in distribution to a random variable $\xi_{1}$ having the same distribution
function $G$ as $T_{n}-B_{n}$.\footnote{Note that we write $T_{n}^{\ast}%
-\hat{B}_{n}\overset{d^{\ast}}{\rightarrow}_{p}\xi_{1}$ to mean that
$T_{n}^{\ast}-\hat{B}_{n}$ has (conditionally on $D_{n}$) the same asymptotic
distribution function as the random variable $\xi_{1}$. We could alternatively
write that $T_{n}^{\ast}-\hat{B}_{n}\overset{d^{\ast}}{\rightarrow}_{p}\xi
_{1}^{\ast}$ and $T_{n}-B_{n}\overset{d}{\rightarrow}\xi_{1}$ where $\xi
_{1}^{\ast}$ and $\xi_{1}$ are two independent copies of the same
distribution, i.e.\ $P(\xi_{1}\leq u)=$ $P(\xi_{1}^{\ast}\leq u)$. We do not
make this distinction because we care only about distributional results, but
it should be kept in mind.} Thus, $\hat{B}_{n}$ can be thought of as an
implicit bootstrap bias that affects the statistic~$T_{n}^{\ast}$, in the same
way that $B_{n}$ affects the original statistic~$T_{n}$.
Assumption~\ref{Assn BS}(ii) complements Assumption~\ref{Assn T} by requiring
the joint convergence of $T_{n}-B_{n}$ and $\hat{B}_{n}-B_{n}$ towards
$\xi_{1}$ and $\xi_{2}$, respectively; see also (\ref{eq 1})--(\ref{eq 3}).

Given Assumption~\ref{Assn BS}(i), we could use the bootstrap distribution of
$T_{n}^{\ast}-\hat{B}_{n}$ to approximate the distribution of $T_{n}-B_{n}$.
Since $B_{n}$ is typically unknown, this result is not very useful for
inference unless $\hat{B}_{n}$ is consistent for~$B_{n}$. In this case,
Assumption~\ref{Assn BS} together with Assumption~\ref{Assn T} imply that
$\hat{p}_{n}$ is asymptotically distributed as $U_{[0,1]}$. This follows by
noting that if $\hat{B}_{n}-B_{n}=o_{p}(1)$, then $\xi_{2}=0$ a.s., implying
that $F(u)=G(u)$. Consequently,%
\begin{align*}
\hat{p}_{n}  &  :=P^{\ast}(T_{n}^{\ast}\leq T_{n})=P^{\ast}(T_{n}^{\ast}%
-\hat{B}_{n}\leq T_{n}-\hat{B}_{n})\\
&  =G(T_{n}-\hat{B}_{n})+o_{p}(1)\text{ (by Assumption~\ref{Assn BS}(i))}\\
&  \overset{d}{\rightarrow}G(\xi_{1}-\xi_{2})\text{ (by
Assumption~\ref{Assn BS}(ii) and continuity of }G\text{)}\\
&  \sim U_{[0,1]},
\end{align*}
where the last distributional equality holds by $F=G$\ and the probability
integral transform.\textbf{ }However, this result does not hold if $\hat
{B}_{n}-B_{n}$ does not converge to zero in probability. Specifically, if
$\hat{B}_{n}-B_{n}\rightarrow_{d}\xi_{2}$ (jointly with $T_{n}-B_{n}%
\rightarrow_{d}\xi_{1}$), then
\[
T_{n}-\hat{B}_{n}=(T_{n}-B_{n})-(\hat{B}_{n}-B_{n})\overset{d}{\rightarrow}%
\xi_{1}-\xi_{2}\sim F^{-1}(U_{[0,1]})
\]
under Assumptions~\ref{Assn T} and~\ref{Assn BS}(ii). When $\xi_{2}$ is
nondegenerate, $F\neq G$, implying that $\hat{p}_{n}=G(T_{n}-\hat{B}%
_{n})+o_{p}(1)$ is not asymptotically distributed as a standard uniform random
variable. This result is summarized in the following theorem.

\begin{theorem}
\label{Theor1}Suppose Assumptions~\ref{Assn T} and~\ref{Assn BS} hold. Then
$\hat{p}_{n}\rightarrow_{d}G(F^{-1}(U_{[0,1]}))$.
\end{theorem}

\noindent\textsc{Proof.\ }First notice that $\hat{p}_{n}$ and $G(T_{n}-\hat
{B}_{n})$ have the same asymptotic distribution because
Assumption~\ref{Assn BS}(i) and continuity of $G$ imply that, by Polya's
Theorem,
\[
|\hat{p}_{n}-G(T_{n}-\hat{B}_{n})|\leq\sup_{u\in\mathbb{R}}|P^{\ast}%
(T_{n}^{\ast}-\hat{B}_{n}\leq u)-G(u)|\overset{p}{\rightarrow}0.
\]
Next, by Assumption~\ref{Assn BS}(ii), $T_{n}-\hat{B}_{n}\rightarrow_{d}%
\xi_{1}-\xi_{2}$, such that
\[
G(T_{n}-\hat{B}_{n})\overset{d}{\rightarrow}G(\xi_{1}-\xi_{2})
\]
by continuity of $G$ and the continuous mapping theorem. Since $\xi_{1}%
-\xi_{2}$ has continuous cdf $F$, it holds that $\xi_{1}-\xi_{2}\sim
F^{-1}(U_{[0,1]})$, which completes the proof.$\hfill\square$

\begin{remark}
\label{RemarkBhat=0}The value of $\hat{B}_{n}$ in Assumption~\ref{Assn BS}(i)
depends on the chosen bootstrap algorithm. It is possible that $\hat{B}%
_{n}\rightarrow_{p}0$ for some bootstrap algorithms; examples are given in
Remark~\ref{Rem InfV m out of n} and Appendix~\ref{sec:panel-details}. If this
is the case, then $\xi_{2}=-B$ a.s., which implies that
\[
F(u):=P(\xi_{1}-\xi_{2}\leq u)=P(\xi_{1}\leq u-B)=G(u-B),
\]
and hence Assumption~\ref{Assn BS}(ii)\ is not satisfied. In this case the
bootstrap p-value satisfies
\[
\hat{p}_{n}\overset{d}{\rightarrow}G(G^{-1}(U_{[0,1]})+B).
\]
Note that this distribution is uniform only if $B=0$. Hence, the p-value
depends on $B$, even in the limit.
\end{remark}

\begin{remark}
\label{Remark con CI}Under Assumptions~\ref{Assn T} and~\ref{Assn BS},
standard bootstrap (percentile) confidence sets are also in general invalid.
Consider, e.g., the case where $T_{n}=g(n)(\hat{\theta}_{n}-\theta_{0})$ and
$T_{n}^{\ast}$ is its bootstrap analogue with (conditional) distribution
function~$\hat{L}_{n}(u)$. A right-sided confidence set for $\theta_{0}$ at
nominal confidence level $1-\alpha\in(0,1)$ can be obtained as (e.g.,
Horowitz, 2001, p.~3171) $CI_{n}^{1-\alpha}:=[\hat{\theta}_{n}-g(n)^{-1}%
\hat{q}_{n}(1-\alpha),+\infty)$, where $\hat{q}_{n}(1-\alpha):=\hat{L}%
_{n}^{-1}(1-\alpha)$. Then%
\begin{align*}
P(\theta_{0}\in CI_{n}^{1-\alpha})  &  =P(\hat{\theta}_{n}-g(n)^{-1}\hat
{q}_{n}(1-\alpha)\leq\theta_{0})=P(T_{n}\leq\hat{q}_{n}(1-\alpha))\\
&  =P(\hat{L}_{n}(T_{n})\leq1-\alpha)=P(\hat{p}_{n}\leq1-\alpha)\nrightarrow
1-\alpha
\end{align*}
because by Theorem~\ref{Theor1} $\hat{p}_{n}$ is not asymptotically uniformly distributed.
\end{remark}

\begin{remark}
\label{Rem on random limits}It is worth noting that, under
Assumptions~\ref{Assn T} and~\ref{Assn BS}, the bootstrap
(conditional)\ distribution is random in the limit whenever $\xi_{2}$ is
non-degenerate. Specifically, assume for simplicity that $B_{n}\rightarrow
_{p}B$. Recall that $\hat{L}_{n}(u):=P^{\ast}(T_{n}^{\ast}\leq u)$,
$u\in\mathbb{R}$, and let $\hat{G}_{n}(u):=P^{\ast}(T_{n}^{\ast}-\hat{B}%
_{n}\leq u)$. It then holds that
\[
\hat{L}_{n}(u)=\hat{G}_{n}(u-\hat{B}_{n})=G(u-B-(\hat{B}_{n}-B))+\hat{a}%
_{n}(u),
\]
where $\hat{a}_{n}(u)\leq\sup_{u\in\mathbb{R}}|\hat{G}_{n}(u)-G(u)|=o_{p}(1)$
by Assumption~\ref{Assn BS}(i), continuity of $G$, and Polya's Theorem.
Because $\hat{B}_{n}-B\rightarrow_{d}\xi_{2}$, it follows that when $\xi_{2}$
is non-degenerate, $\hat{L}_{n}(u)\rightarrow_{w}G(u-B-\xi_{2})$, where
$\rightarrow_{w}$ denotes weak convergence of cdf's as (random) elements of a
function space (see Cavaliere and Georgiev, 2020). The presence of $\xi_{2}$
in $G(u-B-\xi_{2})$ makes this a random cdf.\footnote{The same result follows
in terms of weak convergence in distribution of $T_{n}^{\ast}|D_{n}$.
Specifically, because $T_{n}^{\ast}=(T_{n}^{\ast}-\hat{B}_{n})+(\hat{B}%
_{n}-B_{n})+B_{n}$, where $T_{n}^{\ast}-\hat{B}_{n}\overset{d^{\ast
}}{\rightarrow}_{p}\xi_{1}^{\ast}$ and (jointly) $\hat{B}_{n}-B_{n}%
\overset{d}{\rightarrow}\xi_{2}$ with $\xi_{1}^{\ast}\sim\xi_{1}$ independent
of $\xi_{2}$, we have that $T_{n}^{\ast}|D_{n}\overset{w}{\rightarrow}%
(B+\xi_{1}^{\ast}+\xi_{2})|\xi_{2}$.} Therefore, the bootstrap is unable to
mimic the asymptotic distribution of $T_{n}$, which is $G(u-B)$ by
Assumption~\ref{Assn T}.
\end{remark}

Next, we describe two possible solutions to the invalidity of the standard
bootstrap p-value~$\hat{p}_{n}$. One relies on the prepivoting approach of
Beran (1987, 1988); see Section~\ref{sec:prepivot}. The basic idea is that we
modify $\hat{p}_{n}$ by applying the mapping $\hat{p}_{n}\mapsto H(\hat{p}%
_{n})$, where $H(u)$ is the asymptotic cdf of $\hat{p}_{n}$, which makes the
modified p-value $H(\hat{p}_{n})$ asymptotically standard uniform. Contrary to
Beran (1987, 1988), who proposed prepivoting as a way of providing asymptotic
refinements for the bootstrap, here we show how to use prepivoting to solve
the invalidity of the standard bootstrap p-value~$\hat{p}_{n}$. This result is
new in the bootstrap literature. The second approach relies on computing a
standard bootstrap p-value based on the modified statistic given by
$T_{n}-\hat{B}_{n}$; see Section~\ref{sec:boottbhat}. Thus, we modify the test
statistic rather than modifying the way we compute the bootstrap p-value.

\subsection{Prepivoting}

\label{sec:prepivot}

Theorem~\ref{Theor1} implies that%
\[
P(\hat{p}_{n}\leq u)\rightarrow P(G(F^{-1}(U_{[0,1]}))\leq u)=P(U_{[0,1]}\leq
F(G^{-1}(u)))=F(G^{-1}(u))=:H(u)
\]
uniformly over $u\in\lbrack0,1]$ by Polya's Theorem, given the continuity of
$G$ and~$F$. Although $H$ is not the uniform distribution, unless $G=F$, it is
continuous because $G$ is strictly increasing. Thus, the following corollary
to Theorem~\ref{Theor1} holds by the probability integral transform.

\begin{corollary}
\label{corollary to th first level p value}Under the conditions of
Theorem~\ref{Theor1}, $H(\hat{p}_{n})\rightarrow_{d}U_{[0,1]}$.
\end{corollary}

Therefore, the mapping of $\hat{p}_{n}$ into $H(\hat{p}_{n})$ transforms
$\hat{p}_{n}$ into a new p-value, $H(\hat{p}_{n})$, whose asymptotic
distribution is the standard uniform distribution on~$[0,1]$. Inference based
on $H(\hat{p}_{n})$ is generally infeasible, because we do not observe~$H(u)$.
However, if we can replace $H(u)$ with a uniformly consistent estimator
$\hat{H}_{n}(u)$ then this approach will deliver a feasible modified p-value
$\tilde{p}_{n}:=\hat{H}_{n}(\hat{p}_{n})$. Since the limit distribution of
$\tilde{p}_{n}$ is the standard uniform distribution, $\tilde{p}_{n}$ is an
asymptotically valid p-value. The mapping of $\hat{p}_{n}$ into $\tilde{p}%
_{n}=\hat{H}_{n}(\hat{p}_{n})$ by the estimated distribution of the former
corresponds to what Beran (1987) calls `prepivoting'. In the following
sections, we describe two methods of obtaining a consistent estimator
of~$H(u)$.

\begin{remark}
\label{Remark con CI valid}The prepivoting approach can also be used to solve
the invalidity of confidence sets based on the standard bootstrap; see
Remark~\ref{Remark con CI}. In particular, replace the nominal level
$1-\alpha$ by $\hat{H}_{n}^{-1}(1-\alpha)$ and consider $\widetilde{CI}%
_{n}^{1-\alpha}:=[\hat{\theta}_{n}-g(n)^{-1}\hat{q}_{n}(\hat{H}_{n}%
^{-1}(1-\alpha)),+\infty)$. Then%
\[
P(\theta_{0}\in\widetilde{CI}_{n}^{1-\alpha})=P(\hat{p}_{n}\leq\hat{H}%
_{n}^{-1}(1-\alpha))=P(\hat{H}_{n}(\hat{p}_{n})\leq1-\alpha)\rightarrow
1-\alpha,
\]
where the last convergence is implied by
Corollary~\ref{corollary to th first level p value} and consistency
of~$\hat{H}_{n}$.
\end{remark}

\begin{remark}
Corollary~\ref{corollary to th first level p value} can also be applied to
right-tailed or two-tailed tests. The right-tailed p-value, say $\hat{p}%
_{n,r}:=P^{\ast}(T_{n}^{\ast}>T_{n})=1-\hat{L}_{n}(T_{n})=1-\hat{p}_{n}$, has
cdf $P(\hat{p}_{n,r}\leq u)=P(\hat{p}_{n}\geq1-u)=1-P(\hat{p}_{n}%
<1-u)=1-H(1-u)+o(1)$ uniformly in~$u$. Note that, because the conditional\ cdf
of $T_{n}^{\ast}$ is continuous in the limit, the p-value $\hat{p}_{n,r}$ is
asymptotically equivalent to $P^{\ast}(T_{n}^{\ast}\geq T_{n})$. Thus, by
Corollary~\ref{corollary to th first level p value}, the modified right-tailed
p-value, $\tilde{p}_{n,r}:=1-\hat{H}_{n}(\hat{p}_{n,r})$, satisfies
\[
\tilde{p}_{n,r}=1-H(1-\hat{p}_{n,r})+o_{p}(1)=1-H(\hat{p}_{n})+o_{p}%
(1)\overset{d}{\rightarrow}U_{[0,1]}.
\]
Similarly, for two-tailed tests the equal-tailed bootstrap p-value, $\tilde
{p}_{n,\text{et}}:=2\min\{\tilde{p}_{n},\tilde{p}_{n,r}\}=2\min\{\tilde{p}%
_{n},1-\tilde{p}_{n}\}$, satisfies $\tilde{p}_{n,\text{et}}\rightarrow
_{d}U_{[0,1]}$ by Corollary~\ref{corollary to th first level p value} and the
continuous mapping theorem.
\end{remark}

\subsubsection{Plug-in approach}

\label{sec:plugin}

Suppose $H(u)=H_{\gamma}(u)$ depends on a finite-dimensional
parameter,~$\gamma$. In view of Theorem~\ref{Theor1}, a simple approach to
estimating $H(u)$ is to use%
\[
\hat{H}_{n}(u)=H_{\hat{\gamma}_{n}}(u),
\]
where $\hat{\gamma}_{n}$ denotes a consistent estimator of $\gamma$. This
leads to a plug-in modified p-value defined as
\[
\tilde{p}_{n}=H_{\hat{\gamma}_{n}}(\hat{p}_{n}).
\]
By consistency of $\hat{\gamma}_{n}$ and under the assumption that $H_{\gamma
}$ is continuous in $\gamma$, it follows immediately that%
\[
\tilde{p}_{n}=H(\hat{p}_{n})+o_{p}(1)\overset{d}{\rightarrow}F(G^{-1}%
(G(F^{-1}(U_{[0,1]}))))=U_{[0,1]}.
\]
This result is summarized next.

\begin{corollary}
\label{Corollary-plug-in}Let Assumptions~\ref{Assn T} and~\ref{Assn BS} hold,
and suppose $H_{\gamma}(u)$ is continuous in $\gamma$ for every~$u$. If
$\hat{\gamma}_{n}\rightarrow_{p}\gamma$ then $\tilde{p}_{n}=H_{\hat{\gamma
}_{n}}(\hat{p}_{n})\rightarrow_{d}U_{[0,1]}$.
\end{corollary}

The plug-in approach relies on a consistent estimator of the asymptotic
distribution $H$, but does not require estimating the `bias term'~$B_{n}$.
When estimating $\gamma$ is simple, this approach is attractive since it does
not require any double resampling. Examples are given in
Section~\ref{sec:examples2}. However, computation of $\gamma$ is case-specific
and may be cumbersome in practice. An automatic approach is to use the
bootstrap to estimate $H(u)$, as we describe next.

\subsubsection{Double bootstrap}

\label{sec:dbs}

Following Beran (1987, 1988), we can estimate $H(u)$ with the bootstrap. That
is, we let%
\[
\hat{H}_{n}(u)=P^{\ast}(\hat{p}_{n}^{\ast}\leq u),
\]
where $\hat{p}_{n}^{\ast}$ is the bootstrap analogue of~$\hat{p}_{n}$. Since
$\hat{p}_{n}$ is itself a bootstrap p-value, computing $\hat{p}_{n}^{\ast}$
requires a double bootstrap. In particular, let $D_{n}^{\ast\ast}$ denote a
further bootstrap sample of size $n$ based on $D_{n}^{\ast}$ and some
additional bootstrap variates (defined jointly with $D_{n}$ and $D_{n}^{\ast}$
on a possibly extended probability space), and\textbf{ }let $T_{n}^{\ast\ast}$
denote the bootstrap version of $T_{n}^{\ast}$ computed on $D_{n}^{\ast\ast}$.
With this notation, the second-level bootstrap p-value is defined as%
\[
\hat{p}_{n}^{\ast}:=P^{\ast\ast}(T_{n}^{\ast\ast}\leq T_{n}^{\ast}),
\]
where $P^{\ast\ast}$ denotes the bootstrap probability measure conditional on
$D_{n}^{\ast}$ and $D_{n}$ (making $\hat{p}_{n}^{\ast}$ a function of
$D_{n}^{\ast}$ and $D_{n}$). This leads to a double bootstrap modified
p-value, as given by
\[
\tilde{p}_{n}:=\hat{H}_{n}(\hat{p}_{n})=P^{\ast}(\hat{p}_{n}^{\ast}\leq\hat
{p}_{n}).
\]

In order to show that $\tilde{p}_{n}=\hat{H}_{n}(\hat{p}_{n})\rightarrow
_{d}U_{[0,1]}$, we add the following assumption.

\begin{condition}
\label{Assn DBS}Let $\xi_{1}$ and $\xi_{2}$ be as defined in
Assumptions~\ref{Assn T} and~\ref{Assn BS}. For some $(D_{n}^{\ast},D_{n}%
)$-measurable random variable $\hat{B}_{n}^{\ast}$, it holds that:
(i)~$T_{n}^{\ast\ast}-\hat{B}_{n}^{\ast}\overset{d^{\ast\ast}}{\rightarrow
}_{p^{\ast}}\xi_{1}$, in probability, and (ii)~$T_{n}^{\ast}-\hat{B}_{n}%
^{\ast}\overset{d^{\ast}}{\rightarrow}_{p}\xi_{1}-\xi_{2}$.
\end{condition}

Assumption~\ref{Assn DBS} complements Assumptions~\ref{Assn T}
and~\ref{Assn BS} by imposing high-level conditions on the second-level
bootstrap statistics. Specifically, Assumption~\ref{Assn DBS}(i) assumes that
$T_{n}^{\ast\ast}$ has asymptotic distribution $G$ only after we subtract
$\hat{B}_{n}^{\ast}$. This term is the second-level bootstrap analogue
of~$\hat{B}_{n}$. It depends only on the first-level bootstrap data
$D_{n}^{\ast}$ and is not random under~$P^{\ast\ast}$. The second part of
Assumption~\ref{Assn DBS} follows from Assumption~\ref{Assn BS} in the special
case that $\hat{B}_{n}^{\ast}-\hat{B}_{n}=o_{p^{\ast}}(1)$, in probability;
i.e., when $\xi_{2}=0$ a.s., implying $F=G$. When $F\neq G$, $\hat{B}%
_{n}^{\ast}$ is not a consistent estimator of~$\hat{B}_{n}$. However, under
Assumption~\ref{Assn DBS},%
\[
T_{n}^{\ast}-\hat{B}_{n}^{\ast}\,=(T_{n}^{\ast}-\hat{B}_{n})-(\hat{B}%
_{n}^{\ast}-\hat{B}_{n})\overset{d^{\ast}}{\rightarrow}_{p}\xi_{1}-\xi
_{2}=F^{-1}(U_{[0,1]})
\]
implying that $T_{n}^{\ast}-\hat{B}_{n}^{\ast}\,$\ mimics the distribution of
$T_{n}-\hat{B}_{n}$. This suffices for proving the asymptotic validity of the
double bootstrap modified p-value, $\tilde{p}_{n}=\hat{H}_{n}(\hat{p}_{n})$,
as proved next.

\begin{theorem}
\label{TheoremDouble-p-val}Under Assumptions~\ref{Assn T}, \ref{Assn BS},
and~\ref{Assn DBS}, it holds that $\tilde{p}_{n}=\hat{H}_{n}(\hat{p}%
_{n})\rightarrow_{d}U_{[0,1]}$.
\end{theorem}

\noindent\textsc{Proof.\ }To prove this result, recall that $\hat{H}%
_{n}(u)=P^{\ast}(\hat{p}_{n}^{\ast}\leq u)$ and $P(\hat{p}_{n}\leq
u)\rightarrow H(u)=F(G^{-1}(u))$ uniformly in $u\in\mathbb{R}$, since $H$ is a
continuous distribution function by Assumptions~\ref{Assn T} and
\ref{Assn BS}. Thus, we have that%
\begin{align*}
\hat{p}_{n}^{\ast}  &  =P^{\ast\ast}(T_{n}^{\ast\ast}\leq T_{n}^{\ast
})=P^{\ast\ast}(T_{n}^{\ast\ast}-\hat{B}_{n}^{\ast}\leq T_{n}^{\ast}-\hat
{B}_{n}^{\ast})\\
&  =G(T_{n}^{\ast}-\hat{B}_{n}^{\ast})+o_{p^{\ast}}(1)\text{,\quad by
Assumption~\ref{Assn DBS}(i),}\\
&  =G(F^{-1}(U_{[0,1]}))+o_{p^{\ast}}(1)\text{,\quad by
Assumption~\ref{Assn DBS}(ii),}%
\end{align*}
where $G(F^{-1}(U_{[0,1]}))$ is a random variable whose distribution function
is~$H$. Hence,%
\[
\sup_{u\in\mathbb{R}}|\hat{H}_{n}(u)-H(u)|=o_{p}(1).
\]
Since $H(\hat{p}_{n})\rightarrow_{d}U_{[0,1]}$, we can conclude that
$\tilde{p}_{n}=\hat{H}_{n}(\hat{p}_{n})\rightarrow_{d}U_{[0,1]}$%
.$\hfill\square$\medskip

Theorem~\ref{TheoremDouble-p-val} shows that prepivoting the standard
bootstrap p-value $\hat{p}_{n}$ by applying the mapping $\hat{H}_{n}$
transforms it into an asymptotically uniformly distributed random variable.
This result holds under Assumptions~\ref{Assn T}, \ref{Assn BS},
and~\ref{Assn DBS}, independently of whether $G=F$ or not. When $G=F$ then
$\hat{p}_{n}\rightarrow_{d}U_{[0,1]}$ (as implied by Theorem~\ref{Theor1}). In
this case, the prepivoting approach is not necessary to obtain a first-order
asymptotically valid test, although it might help further reducing the size
distortion of the test. This corresponds to the setting of Beran (1987, 1988),
where prepivoting was proposed as a way of reducing the level distortions of
confidence intervals. When $G\neq F$ then $\hat{p}_{n}$ is not asymptotically
uniform and a standard bootstrap test based on $\hat{p}_{n}$ is asymptotically
invalid, as shown in Theorem~\ref{Theor1}. In this case, prepivoting
transforms an asymptotically invalid bootstrap p-value into one that is
asymptotically valid. This setting was not considered by Beran (1987, 1988)
and is new to our paper.

\subsection{Power of tests}

\label{subsec:power}

In this section we explicitly consider a testing situation. Suppose we are
interested in testing $\mathsf{H}_{0}:\theta=\bar{\theta}$ against
$\mathsf{H}_{1}:\theta<\bar{\theta}$. Specifically, defining $T_{n}%
(\theta):=g(n)(\hat{\theta}_{n}-\theta)$, we consider the test statistic
$T_{n}(\bar{\theta})$. The corresponding bootstrap p-value is $\hat{p}%
_{n}(\bar{\theta})$ with $\hat{p}_{n}(\theta):=P^{\ast}(T_{n}^{\ast}\leq
T_{n}(\theta))$. When the null hypothesis is true, i.e., when $\bar{\theta
}=\theta_{0}$ with $\theta_{0}$ denoting the true value, we find $T_{n}%
(\bar{\theta})=T_{n}(\theta_{0})=T_{n}$ and $\hat{p}_{n}(\bar{\theta})=\hat
{p}_{n}(\theta_{0})=\hat{p}_{n}$, where $T_{n}$ and $\hat{p}_{n}$ are as
defined previously. If Assumptions~\ref{Assn T} and~\ref{Assn BS} hold under
the null, Theorem~\ref{Theor1} and
Corollary~\ref{corollary to th first level p value} imply that tests based on
$H(\hat{p}_{n}(\bar{\theta}))$ have correct asymptotic size, where $H$
continues to denote the asymptotic cdf of~$\hat{p}_{n}$.

To analyze power, we consider $\theta_{0}=\bar{\theta}+a_{n}$ for some
deterministic sequence~$a_{n}$. Then $a_{n}=0$ under the null hypothesis,
whereas $a_{n}=a<0$ corresponds to a fixed alternative and $a_{n}=a/g(n)$ for
$a<0$ corresponds to a local alternative. Thus, we define $\pi_{n}%
:=g(n)(\theta_{0}-\bar{\theta})=g(n)a_{n}$ so that $T_{n}(\bar{\theta}%
)=T_{n}+\pi_{n}$.

\begin{theorem}
\label{Thm:power}Suppose Assumptions~\ref{Assn T} and~\ref{Assn BS} hold. (i)
If $\pi_{n}\rightarrow\pi$ then $H(\hat{p}_{n}(\bar{\theta}))\rightarrow
_{d}F(F^{-1}(U_{[0,1]})+\pi)$. (ii) If $\pi_{n}\rightarrow-\infty$ then
$P(H(\hat{p}_{n}(\bar{\theta}))\leq\alpha)\rightarrow1$ for any nominal
level\ $\alpha>0$.
\end{theorem}

\noindent\textsc{Proof.\ }As in the proof of Theorem~\ref{Theor1} we have, by
Assumption~\ref{Assn BS}(i),
\[
\hat{p}_{n}(\bar{\theta})=P^{\ast}(T_{n}^{\ast}\leq T_{n}(\bar{\theta
}))=P^{\ast}(T_{n}^{\ast}-\hat{B}_{n}\leq T_{n}-\hat{B}_{n}+\pi_{n}%
)=G(T_{n}-\hat{B}_{n}+\pi_{n})+o_{p}(1).
\]
If $\pi_{n}\rightarrow\pi$ then $\hat{p}_{n}(\bar{\theta})\rightarrow
_{d}G(F^{-1}(U_{[0,1]})+\pi)$ by Assumption~\ref{Assn BS}(ii), so that%
\[
H(\hat{p}_{n}(\bar{\theta}))\overset{d}{\rightarrow}H(G(F^{-1}(U_{[0,1]}%
)+\pi))=F(F^{-1}(U_{[0,1]})+\pi)
\]
by definition of $H(u)$. If $\pi_{n}\rightarrow-\infty$ then $\hat{p}_{n}%
(\bar{\theta})\rightarrow_{p}0$ because $T_{n}-\hat{B}_{n}=O_{p}(1)$ by
Assumption~\ref{Assn BS}(ii), so that $H(\hat{p}_{n}(\bar{\theta}%
))\rightarrow_{p}H(0)=0$ and $P(H(\hat{p}_{n}(\bar{\theta}))\leq
\alpha)\rightarrow1$ for any $\alpha>0$.$\hfill\square\medskip$

It follows from Theorem~\ref{Thm:power}(ii) that a left-tailed test that
rejects for small values of $H(\hat{p}_{n}(\bar{\theta}))$ is consistent.
Furthermore, it follows from Theorem~\ref{Thm:power}(i) that such a test has
non-trivial asymptotic local power against $\pi<0$. Specifically, the
asymptotic local power against $\pi$ is given by $P(H(\hat{p}_{n}(\bar{\theta
}))\leq\alpha)\rightarrow F(F^{-1}(\alpha)-\pi)$. Interestingly, this only
depends on $F$ and not on~$G$. As above, to implement the modified p-value,
$H(\hat{p}_{n}(\bar{\theta}))$, in practice, we would need a (uniformly)
consistent estimator of $H$, i.e., the asymptotic distribution of the
bootstrap p-value when the null hypothesis is true.\ This could be either the
plug-in or double bootstrap estimators, as discussed in
Sections~\ref{sec:plugin} and~\ref{sec:dbs}.

Note that Assumption~\ref{Assn BS} is still assumed to hold in
Theorem~\ref{Thm:power}. That is, the bootstrap statistic $T_{n}^{\ast}$ is
assumed to have the same asymptotic behavior under the null and under the
alternative. This is commonly the case when the bootstrap algorithm does not
impose the null hypothesis when generating the bootstrap data.

\subsection{Bootstrap p-value based on $T_{n}-\hat{B}_{n}$}

\label{sec:boottbhat}

The double bootstrap modified p-value $\tilde{p}_{n}$ depends only on the
statistic $T_{n}$ and their bootstrap analogues $T_{n}^{\ast}$ and~$T_{n}%
^{\ast\ast}$. It does not involve computing explicitly $\hat{B}_{n}$ or
$\hat{B}_{n}^{\ast}$, but in some applications it can be computationally
costly as it requires two levels of resampling. As it turns out, $\tilde
{p}_{n}$ is asymptotically equivalent to a single-level bootstrap p-value that
is based on bootstrapping the statistic $T_{n}-\hat{B}_{n}$, as we show next.

By definition, the double bootstrap modified p-value is given by $\tilde
{p}_{n}:=P^{\ast}(\hat{p}_{n}^{\ast}\leq\hat{p}_{n})$, where%
\[
\hat{p}_{n}^{\ast}:=P^{\ast\ast}(T_{n}^{\ast\ast}\leq T_{n}^{\ast}%
)=P^{\ast\ast}(T_{n}^{\ast\ast}-\hat{B}_{n}^{\ast}\leq T_{n}^{\ast}-\hat
{B}_{n}^{\ast})=G(T_{n}^{\ast}-\hat{B}_{n}^{\ast})+o_{p^{\ast}}(1),
\]
in probability, given Assumption~\ref{Assn DBS}. Similarly, under
Assumptions~\ref{Assn T} and~\ref{Assn BS},%
\[
\hat{p}_{n}:=P^{\ast}(T_{n}^{\ast}\leq T_{n})=P^{\ast}(T_{n}^{\ast}-\hat
{B}_{n}\leq T_{n}-\hat{B}_{n})=G(T_{n}-\hat{B}_{n})+o_{p}(1).
\]
It follows that%
\begin{align*}
\tilde{p}_{n}  &  :=P^{\ast}(\hat{p}_{n}^{\ast}\leq\hat{p}_{n})=P^{\ast
}(G(T_{n}^{\ast}-\hat{B}_{n}^{\ast})\leq G(T_{n}-\hat{B}_{n}))+o_{p}(1)\\
&  =P^{\ast}(T_{n}^{\ast}-\hat{B}_{n}^{\ast}\leq T_{n}-\hat{B}_{n})+o_{p}(1)
\end{align*}
because $G$ is continuous. We summarize this result in the following corollary.

\begin{corollary}
\label{Corollary pvalue-BC}Under Assumptions~\ref{Assn T}, \ref{Assn BS},
and~\ref{Assn DBS}, $\tilde{p}_{n}=P^{\ast}(T_{n}^{\ast}-\hat{B}_{n}^{\ast
}\leq T_{n}-\hat{B}_{n})+o_{p}(1)$.
\end{corollary}

Theorem~\ref{TheoremDouble-p-val} shows that $\tilde{p}_{n}\rightarrow
_{d}U_{[0,1]}$ and hence is asymptotically valid. In view of this,
Corollary~\ref{Corollary pvalue-BC} shows that removing $\hat{B}_{n}$ from
$T_{n}$ and computing a bootstrap p-value based on the new statistic,
$T_{n}-\hat{B}_{n}$, also solves the invalidity problem of the standard
bootstrap p-value, $\hat{p}_{n}=P^{\ast}(T_{n}^{\ast}\leq T_{n})$. Note that
we do not require $\xi_{2}=0$, i.e.\ $\hat{B}_{n}-B_{n}$ and $\hat{B}%
_{n}^{\ast}-\hat{B}_{n}$ do not need to converge to zero.

When $\hat{B}_{n}$ and $\hat{B}_{n}^{\ast}$ are easy to compute, e.g.,\ when
they are available analytically as functions of $D_{n}$ and $D_{n}^{\ast}$,
respectively, Corollary~\ref{Corollary pvalue-BC} is useful as it avoids
implementing a double bootstrap. When this is not the case, i.e., when
deriving $\hat{B}_{n}$ and $\hat{B}_{n}^{\ast}$ explicitly is cumbersome or
impossible, we may be able to estimate $\hat{B}_{n}$ from the bootstrap and
$\hat{B}_{n}^{\ast}$ from a double bootstrap.
Corollary~\ref{Corollary pvalue-BC} then shows that the double bootstrap
modified p-value $\tilde{p}_{n}$ is a convenient alternative since it depends
only on $T_{n}$, $T_{n}^{\ast}$, and~$T_{n}^{\ast\ast}$. It is important to
note that none of these approaches requires the consistency of $\hat{B}_{n}$
and~$\hat{B}_{n}^{\ast}$.

\subsection{A more general set of high-level conditions}

\label{Sec more general high level conditions}

We conclude this section by providing an alternative set of high-level
conditions that cover bootstrap methods for which $T_{n}^{\ast}-\hat{B}_{n}$
has a different limiting distribution than $T_{n}-B_{n}$. This may happen, for
example, for the pairs bootstrap; see Section~\ref{sec:MA-intro} and
Remark~\ref{Rem pairs general}.

\begin{condition}
\label{Assn BSpairs}Assumption~\ref{Assn BS} holds with part~(i) replaced by
(i)~$T_{n}^{\ast}-\hat{B}_{n}\overset{d^{\ast}}{\rightarrow}_{p}\zeta_{1}$,
where $\zeta_{1}$ is centered at zero and the cdf $J(u)=P(\zeta_{1}\leq u)$ is
continuous and strictly increasing over its support.
\end{condition}

Under Assumption~\ref{Assn BSpairs}, $T_{n}^{\ast}-\hat{B}_{n}$ does not
replicate the distribution of $T_{n}-B_{n}$. This is to be understood in the
sense that there does not exist a $P^{\ast}$-measurable term $\hat{B}_{n}$
such that $T_{n}^{\ast}-\hat{B}_{n}$ has the same asymptotic distribution as
$T_{n}-B_{n}$.

An important generalization provided by Assumption~\ref{Assn BSpairs} compared
with Assumption~\ref{Assn BS} is to allow for bootstrap methods where the
`centering term', say $B_{n}^{\ast}$, depends on the bootstrap data. That is,
to allow cases where there is a random (with respect to~$P^{\ast}$, i.e.,
depending on the bootstrap data) term $B_{n}^{\ast}$ such that $T_{n}^{\ast
}-B_{n}^{\ast}\overset{d^{\ast}}{\rightarrow}_{p}\xi_{1}$ and hence has the
same asymptotic distribution as $T_{n}-B_{n}$. Clearly, this violates
Assumption~\ref{Assn BS} unless $B_{n}^{\ast}-\hat{B}_{n}\overset{p^{\ast
}}{\rightarrow}_{p}0$ (as in the ridge regression in
Section~\ref{sec:ridge-intro}).\ However, letting $\zeta_{1}$ be such that
$B_{n}^{\ast}-\hat{B}_{n}\overset{d^{\ast}}{\rightarrow}_{p}\zeta_{1}-\xi_{1}%
$, then Assumption~\ref{Assn BSpairs} covers the former case.

\begin{remark}
\label{Rem pairs general}A leading example where $T_{n}^{\ast}-B_{n}^{\ast
}\overset{d^{\ast}}{\rightarrow}_{p}\xi_{1}$ and hence has the same asymptotic
distribution as $T_{n}-B_{n}$ is the pairs bootstrap as in
Section~\ref{sec:MA-intro} for the model averaging example. We study this case
in more detail in Section~\ref{sec:MA-revisit}.
\end{remark}

The asymptotic distribution of the bootstrap p-value under
Assumption~\ref{Assn BSpairs} is given in the following theorem. The proof is
identical to that of Theorem~\ref{Theor1}, with $G$ replaced by $J$, and hence omitted.

\begin{theorem}
\label{Theor1pairs}If Assumptions~\ref{Assn T} and \ref{Assn BSpairs} hold
then $\hat{p}_{n}\rightarrow_{d}J(F^{-1}(U_{[0,1]}))$.
\end{theorem}

Theorem~\ref{Theor1pairs} implies that now $P(\hat{p}_{n}\leq u)\rightarrow
P(J(F^{-1}(U_{[0,1]}))\leq u)=F(J^{-1}(u))=:H(u)$. Clearly, a plug-in approach
to estimating this $H(u)$ based on $G$ as described in
Section~\ref{sec:plugin} would be invalid because $G\neq J$ in general.
However, it follows straightforwardly by the same arguments as applied in
Section~\ref{sec:plugin} that a plug-in approach based on $J$ will deliver an
asymptotically valid plug-in modified p-value.

To implement an asymptotically valid double bootstrap modified p-value we
consider the following high-level condition.

\begin{condition}
\label{Assn DBSpairs}Assumption~\ref{Assn DBS} holds with part~(i) replaced by
(i)~$T_{n}^{\ast\ast}-\hat{B}_{n}^{\ast}\overset{d^{\ast\ast}}{\rightarrow
}_{p^{\ast}}\zeta_{1}$, in probability, where $\zeta_{1}$ is defined in
Assumption~\ref{Assn BSpairs}.
\end{condition}

Under Assumption~\ref{Assn DBSpairs}, the second-level bootstrap statistic,
$T_{n}^{\ast\ast}-\hat{B}_{n}^{\ast}$, replicates the distribution of the
first-level statistic, $T_{n}^{\ast}-\hat{B}_{n}$. Thus, the second-level
bootstrap p-value is
\begin{align*}
\hat{p}_{n}^{\ast}:=P^{\ast\ast}(T_{n}^{\ast\ast}\leq T_{n}^{\ast})  &
=P^{\ast\ast}(T_{n}^{\ast\ast}-\hat{B}_{n}^{\ast}\leq T_{n}^{\ast}-\hat{B}%
_{n}^{\ast})=J(T_{n}^{\ast}-\hat{B}_{n}^{\ast})+o_{p^{\ast}}(1)\\
&  \overset{d^{\ast}}{\rightarrow}_{p}J(\xi_{1}-\xi_{2})=J(F^{-1}(U_{[0,1]}))
\end{align*}
under Assumption~\ref{Assn DBSpairs}. Hence, the second-level bootstrap
p-value has the same asymptotic distribution as the original bootstrap
p-value. It follows that the double bootstrap modified p-value, $\tilde{p}%
_{n}:=\hat{H}_{n}(\hat{p}_{n})=P^{\ast}(\hat{p}_{n}^{\ast}\leq\hat{p}_{n})$,
is asymptotically valid, which is stated next. The proof is essentially
identical to that of Theorem~\ref{TheoremDouble-p-val} and hence omitted.

\begin{theorem}
\label{TheoremDouble-p-val pairs}Under Assumptions~\ref{Assn T},
\ref{Assn BSpairs}, and~\ref{Assn DBSpairs}, it holds that $\tilde{p}_{n}%
=\hat{H}_{n}(\hat{p}_{n})\rightarrow_{d}U_{[0,1]}$.
\end{theorem}

\begin{remark}
\label{Rem pairs double general}Consider again the case with a random
bootstrap centering term in Remark~\ref{Rem pairs general}, where $B_{n}%
^{\ast}-\hat{B}_{n}\overset{d^{\ast}}{\rightarrow}_{p}\zeta_{1}-\xi_{1}$ such
that $T_{n}^{\ast}-B_{n}^{\ast}\overset{d^{\ast}}{\rightarrow}_{p}\xi_{1}$.
Within this setup, we can consider double bootstrap methods such that, for a
random (with respect to $P^{\ast\ast}$) term $B_{n}^{\ast\ast}$ we have
$T_{n}^{\ast\ast}-B_{n}^{\ast\ast}\overset{d^{\ast\ast}}{\rightarrow}%
_{p^{\ast}}\xi_{1}$, in probability. Thus, the asymptotic distribution of the
second-level bootstrap statistic mimics that of the first-level statistic.
When $B_{n}^{\ast\ast}$ and $\zeta_{1}$ are such that $B_{n}^{\ast\ast}%
-\hat{B}_{n}^{\ast}\overset{d^{\ast\ast}}{\rightarrow}_{p^{\ast}}\zeta_{1}%
-\xi_{1}$, in probability, then Assumption~\ref{Assn DBSpairs} is satisfied.
As in Remark~\ref{Rem pairs general} this setup allows us to cover the pairs bootstrap.
\end{remark}

\section{Examples continued}

\label{sec:examples2}

In this section we revisit our three leading examples from
Section~\ref{sec:examples}, where we argued that standard boostrap inference
is invalid due to the presence of bias. In this section\ we show how to apply
our general theory in each example. Again, we refer to
Appendix~\ref{Sec Applications} for detailed derivations.

\subsection{Inference after model averaging}

\label{sec:MA-revisit}

\noindent\textsc{Fixed regressor bootstrap}. Extending the arguments in
Section~\ref{sec:MA-intro}, we obtain the following result.

\begin{lemma}
\label{Lemma MA}Under regularity conditions stated in
Appendix~\ref{sec:MA-details}, Assumptions~\ref{Assn T} and~\ref{Assn BS} are
satisfied with $(\xi_{1},\xi_{2})^{\prime}\sim N(0,V)$, where $V:=(v_{ij}%
),i,j=1,2$, is positive definite and continuous in $\omega$, $\sigma^{2}$, and
$\Sigma_{WW}:=\operatorname*{plim}S_{WW}$.
\end{lemma}

By Lemma~\ref{Lemma MA}, the conditions of Theorem~\ref{Theor1} hold with
$G(u)=\Phi(u/v_{11})$ and $F(u)=\Phi(u/v_{d})$, where $v_{d}^{2}=v_{11}%
+v_{22}-2v_{12}>0$. Then Theorem~\ref{Theor1} implies that the standard
bootstrap p-value satisfies $\hat{p}_{n}\rightarrow_{d}\Phi(m\Phi
^{-1}(U_{[0,1]}))$ with $m^{2}:=v_{d}^{2}/v^{2}$. Because $\omega$ is known
and $\sigma^{2},\Sigma_{WW}$ are easily estimated, a consistent estimator
$\hat{m}_{n}\rightarrow_{p}m$ is available, and the plug-in approach in
Corollary~\ref{Corollary-plug-in} can be implemented by considering the
modified p-value, $\tilde{p}_{n}=\Phi(\hat{m}_{n}^{-1}\Phi^{-1}(\hat{p}_{n}%
))$. Inspection of the proofs shows that our modified bootstrap approach is
asymptotically valid whether $\delta$ is fixed or local-to-zero. In the former
case, $B_{n}$ is $O_{p}(n^{1/2})$ rather than $O_{p}(1)$, implying that
$B_{n}$ diverges in probability and $\tilde{\beta}_{n}$ is not even consistent
for~$\beta$. Despite this, the modified bootstrap p-value is asymptotically valid.

Alternatively, we can implement the double bootstrap as in
Section~\ref{sec:dbs}. Specifically, let%
\[
y^{\ast\ast}=x\hat{\beta}_{n}^{\ast}+Z\hat{\delta}_{n}^{\ast}+\varepsilon
^{\ast\ast},
\]
where $\varepsilon^{\ast\ast}|\{D_{n},D_{n}^{\ast}\}\sim N(0,\hat{\sigma}%
_{n}^{\ast2}I_{n})$, $(\hat{\beta}_{n}^{\ast},\hat{\delta}_{n}^{\ast^{\prime}%
},\hat{\sigma}_{n}^{\ast2})\,$is the OLS estimator obtained from the full
model estimated on the first-level bootstrap data, and $D_{n}^{\ast}%
=\{y^{\ast},W\}$. The double bootstrap statistic is $T_{n}^{\ast\ast}%
:=n^{1/2}(\tilde{\beta}_{n}^{\ast\ast}-\hat{\beta}_{n}^{\ast})$, where
$\tilde{\beta}_{n}^{\ast\ast}:=\sum_{m=1}^{M}\omega_{m}\tilde{\beta}%
_{m,n}^{\ast\ast}$ with $\tilde{\beta}_{m,n}^{\ast\ast}:=S_{xx.Z_{m}}%
^{-1}S_{xy^{\ast\ast}.Z_{m}}$ defined as the double bootstrap OLS estimator
from the $m^{\text{th}}$ model. The double bootstrap modified p-value is then
$\tilde{p}_{n}=P^{\ast}(\hat{p}_{n}^{\ast}\leq\hat{p}_{n})$ with $\hat{p}%
_{n}^{\ast}=P^{\ast\ast}(T_{n}^{\ast\ast}\leq T_{n}^{\ast})$.

\begin{lemma}
\label{Lemma MA dbs}Under the conditions of Lemma~\ref{Lemma MA},
Assumption~\ref{Assn DBS} holds with $\hat{B}_{n}^{\ast}:=Q_{n}n^{1/2}%
\hat{\delta}_{n}^{\ast}$.
\end{lemma}

Lemma~\ref{Lemma MA dbs} shows that Assumption~\ref{Assn DBS} is verified in
this example. The asymptotic validity of the double bootstrap modified p-value
now follows from Lemmas~\ref{Lemma MA} and~\ref{Lemma MA dbs} and
Theorem~\ref{TheoremDouble-p-val}.

\smallskip

\noindent\textsc{Pairs bootstrap}. For the pairs bootstrap we verify the
high-level conditions in Section~\ref{Sec more general high level conditions}.
To simplify the discussion we consider the case with scalar $z_{t}$ in
(\ref{MA model}) and where we \textquotedblleft average\textquotedblright%
\ over only one model ($M=1$), which is the simplest model in which $z_{t}$ is
omitted from the regression. That is, we estimate $\beta$ by regression of $y$
on $x$, i.e., $\tilde{\beta}_{n}=S_{xx}^{-1}S_{xy}$. In this special case,
$T_{n}-B_{n}\rightarrow_{d}N(0,v^{2})$ with $v^{2}=\sigma^{2}\Sigma_{xx}^{-1}$
and $B_{n}=S_{xx}^{-1}S_{xz}n^{1/2}\delta$.

\begin{lemma}
\label{lemma mod av pairs bootstrap}Under regularity conditions stated in
Appendix~\ref{sec:MA-details}, it holds that $T_{n}^{\ast}-\hat{B}%
_{n}\overset{d^{\ast}}{\rightarrow}_{p}N(0,v^{2}+\kappa^{2})$, where $\hat
{B}_{n}:=S_{xx}^{-1}S_{xz}n^{1/2}\hat{\delta}_{n}$ and $\kappa^{2}%
:=d_{r}(\delta)^{\prime}\Sigma_{r}d_{r}(\delta)$ with $d_{r}(\delta
):=\delta(\Sigma_{xx}^{-1},-\Sigma_{xx}^{-2}\Sigma_{xz})^{\prime}$.
\end{lemma}

Notice that, in contrast to the FRB, the asymptotic variance of $T_{n}^{\ast}$
fails to replicate that of $T_{n}$ because of the term $\kappa^{2}>0$. This
implies that the methodology developed in Theorem~\ref{Theor1} and its
corollaries no longer applies. Instead we can apply the theory of
Section~\ref{Sec more general high level conditions}. In particular,
Lemma~\ref{lemma mod av pairs bootstrap} shows that
Assumption~\ref{Assn BSpairs}(i) holds in this case with $\zeta_{1}\sim
N(0,v^{2}+\kappa^{2})$. Lemma~\ref{lemma mod av pairs bootstrap} also shows
that $\hat{B}_{n}$ is the same for the pairs bootstrap and the FRB, such that
Lemma~\ref{Lemma MA} shows that Assumptions~\ref{Assn T} and~\ref{Assn BS}(ii)
are verified. This implies that Theorem~\ref{Theor1pairs} holds for this
example. Using similar arguments, it can be shown that
Assumption~\ref{Assn DBSpairs} also holds for this example, which would imply
that the double bootstrap $p$-values are asymptotically uniformly distributed.

Under local alternatives of the form $\beta_{0}=\bar{\beta}+an^{-1/2}$, where
$\bar{\beta}$ is the value under the null (Section~\ref{subsec:power}), the
asymptotic local power function for the modified p-value is given by
$\Phi(\Phi^{-1}(\alpha)-a/v_{d})$; see Theorem~\ref{Thm:power}. It is not
difficult to verify that this is the same power function as that obtained from
a test based directly on $\hat{\beta}_{n}$ from the full model~(\ref{MA model}).

\subsection{Ridge regression}

\label{sec:ridge-revisit}

To complete the example in Section~\ref{sec:ridge-intro}, we can proceed as in
the previous example.

\begin{lemma}
\label{Lemma ridge}Under the null hypotheses and the regularity conditions
stated in Appendix~\ref{sec:ridge-details}, Assumptions~\ref{Assn T}
and~\ref{Assn BS} are satisfied with $(\xi_{1},\xi_{2})^{\prime}\sim N(0,V)$,
where $V:=(v_{ij}),i,j=1,2$, is positive definite and continuous in $c_{0}$,
$\sigma^{2}$, and~$\Sigma_{xx}$.
\end{lemma}

As in Section~\ref{sec:MA-revisit}, Lemma~\ref{Lemma ridge} and
Theorem~\ref{Theor1} imply that the standard bootstrap p-value satisfies
$\hat{p}_{n}\rightarrow_{d}\Phi(m\Phi^{-1}(U_{[0,1]}))$, where we now have
$m^{2}=(g^{\prime}\tilde{\Sigma}_{xx}^{-1}\Sigma_{xx}\tilde{\Sigma}_{xx}%
^{-1}g)^{-1}g^{\prime}\Sigma_{xx}^{-1}g$. Note that this result holds
irrespectively of $\theta$ being fixed or local to zero. Thus, the bootstrap
is invalid unless $c_{0}=0$ which implies $m=1$. For the plug-in method, a
simple consistent estimator of $m$ is given by $\hat{m}_{n}^{2}:=(g^{\prime
}\tilde{S}_{xx}^{-1}S_{xx}\tilde{S}_{xx}^{-1}g)^{-1}g^{\prime}S_{xx}^{-1}g$,
and inference based on the plug-in modified p-value $\tilde{p}_{n}=\Phi
(\hat{m}_{n}^{-1}\Phi^{-1}(\hat{p}_{n}))$ is then asymptotically valid by
Corollary~\ref{Corollary-plug-in}.

To implement the double bootstrap method, we can draw the double bootstrap
sample $\{y_{t}^{\ast\ast},x_{t}^{\ast\ast};t=1,\dots,n\}$ as i.i.d.\ from
$\{y_{t}^{\ast},x_{t}^{\ast};t=1,\dots,n\}$. Accordingly, the second-level
bootstrap ridge estimator is $\tilde{\theta}_{n}^{\ast\ast}:=\tilde
{S}_{x^{\ast\ast}x^{\ast\ast}}^{-1}S_{x^{\ast\ast}y^{\ast\ast}}$ with
associated test statistic $T_{n}^{\ast\ast}:=n^{1/2}g^{\prime}(\tilde{\theta
}_{n}^{\ast\ast}-\hat{\theta}_{n}^{\ast})$, which is centered at the
first-level bootstrap OLS estimator, $\hat{\theta}_{n}^{\ast}$. It is
straightforward to show that, without additional conditions,
Assumption~\ref{Assn DBS} holds.

\begin{lemma}
\label{Lemma ridge dbs}Under the conditions of Lemma~\ref{Lemma ridge},
Assumption~\ref{Assn DBS} holds with $\hat{B}_{n}^{\ast}:=-c_{n}%
n^{-1/2}g^{\prime}\tilde{S}_{x^{\ast}x^{\ast}}^{-1}\hat{\theta}_{n}^{\ast}$.
\end{lemma}

Validity of the double bootstrap modified p-value $\tilde{p}_{n}=P^{\ast}%
(\hat{p}_{n}^{\ast}\leq\hat{p}_{n})$ now follows by application of
Theorem~\ref{TheoremDouble-p-val}.

\subsection{Nonparametric regression}

\label{sec:nonpar-revist}

Again, we complete the example in Section~\ref{sec:nonpar-intro} by proceeding
as in the previous examples.

\begin{lemma}
\label{Lemma NP}Under regularity conditions stated in
Appendix~\ref{sec:nonpar-details}, Assumptions~\ref{Assn T} and~\ref{Assn BS}
are satisfied with $(\xi_{1},\xi_{2})^{\prime}\sim N(0,V)$, where
$V:=(v_{ij}),i,j=1,2$, is positive definite and continuous in $\sigma^{2}$ and
the kernel function.
\end{lemma}

As before, Lemma~\ref{Lemma NP} and Theorem~\ref{Theor1} imply that the
standard bootstrap p-value satisfies $\hat{p}_{n}\rightarrow_{d}\Phi
(m\Phi^{-1}(U_{[0,1]}))$, where we now have $m^{2}:=4+(\int K^{2}%
(u)du)^{-1}(\int(\int K(s-u)K(s)ds)^{2}du-4\int K(u)\int K(u-s)K(s)dsdu)$.
Thus, in this example, $m$ need not be estimated because it is observed once
$K$ is chosen. Therefore, valid inference is feasible with the modified
p-value $\tilde{p}_{n}=H(\hat{p}_{n})=\Phi(m^{-1}\Phi^{-1}(\hat{p}_{n}))$; see
Corollary~\ref{corollary to th first level p value}.

We can also apply a double bootstrap modification. Let $y_{t}^{\ast\ast}%
=\hat{\beta}_{h}^{\ast}(x_{t})+\varepsilon_{t}^{\ast\ast}$, $t=1,\dots
,n$,\ where $\varepsilon_{t}^{\ast\ast}|\{D_{n},D_{n}^{\ast}\}\sim
~$i.i.d.$N(0,\hat{\sigma}_{n}^{\ast2})$ with $D_{n}^{\ast}:=\{y_{t}^{\ast
},t=1,\dots,n\}$ and $\hat{\sigma}_{n}^{\ast2}$ denoting the residual variance
from the first-level bootstrap data. The double bootstrap analogue of $T_{n}$
is $T_{n}^{\ast\ast}:=(nh)^{1/2}(\hat{\beta}_{h}^{\ast\ast}(x)-\hat{\beta}%
_{h}^{\ast}(x))$, where $\hat{\beta}_{h}^{\ast\ast}(x):=(nh)^{-1}\sum
_{t=1}^{n}k_{t}y_{t}^{\ast\ast}$. This can be decomposed as $T_{n}^{\ast\ast
}=\xi_{1,n}^{\ast\ast}+\hat{B}_{n}^{\ast}$, where $\hat{B}_{n}^{\ast
}:=(nh)^{1/2}(({nh})^{-1}\sum_{t=1}^{n}k_{t}\hat{\beta}_{h}^{\ast}(x_{t}%
)-\hat{\beta}_{h}^{\ast}(x))$. Unfortunately, although $\xi_{1,n}^{\ast\ast}$
satisfies Assumption~\ref{Assn DBS}(i), $\hat{B}_{n}^{\ast}$ does not satisfy
Assumption~\ref{Assn DBS}(ii). The reason is that $\hat{B}_{n}^{\ast}-\hat
{B}_{n}=\xi_{2,n}^{\ast}+\hat{B}_{2,n}-\hat{B}_{n}$, where $\xi_{2,n}^{\ast}$
satisfies Assumption~\ref{Assn DBS}(ii), but $\hat{B}_{2,n}:=(nh)^{-1}%
\sum_{t=1}^{n}k_{t}\hat{B}_{n}(x_{t})$ is a smoothed version of $\hat{B}_{n}$
(evaluated at $x_{t}$) and although $\hat{B}_{2,n}-\hat{B}_{n}$ is mean zero
it is not $o_{p}(1)$. However, $\hat{B}_{2,n}-\hat{B}_{n}$ is observed, so
this is easily corrected by defining $\bar{T}_{n}^{\ast\ast}:=T_{n}^{\ast\ast
}-(\hat{B}_{2,n}-\hat{B}_{n})$. Then we have the following result.

\begin{lemma}
\label{Lemma NP dbs}Under the conditions of Lemma~\ref{Lemma NP},
Assumption~\ref{Assn DBS} holds with $T_{n}^{\ast\ast}$ and $\hat{B}_{n}%
^{\ast}$ replaced by $\bar{T}_{n}^{\ast\ast}$ and $\bar{B}_{n}^{\ast}:=\hat
{B}_{n}^{\ast}-(\hat{B}_{2,n}-\hat{B}_{n})$, respectively.
\end{lemma}

The validity of the double bootstrap modified p-value $\tilde{p}_{n}=P^{\ast
}(\hat{p}_{n}^{\ast}\leq\hat{p}_{n})$, where $\hat{p}_{n}^{\ast}:=P^{\ast\ast
}(\bar{T}_{n}^{\ast\ast}\leq T_{n}^{\ast})$, follows from
Lemma~\ref{Lemma NP dbs} and Theorem~\ref{TheoremDouble-p-val}. This in turn
implies that confidence intervals based on the double bootstrap are
asymptotically valid; see also Remark~\ref{Remark con CI valid}. We note that
Hall and Horowitz (2013) also proposed, without theory, a version of their
calibration method based on the double bootstrap. Our double bootstrap-based
method for confidence intervals corresponds to their steps~1--5, and where we
need a correction they have instead a step~6 in which they average over a grid
of~$x$.

Finally, under local alternatives of the form $\beta_{0}(x)=\bar{\beta
}+an^{-2/5}$, where $\bar{\beta}$ is the value under the null
(Section~\ref{subsec:power}), the asymptotic local power function for the
modified p-value is given by $\Phi(\Phi^{-1}(\alpha)-a/v_{d})$; see
Theorem~\ref{Thm:power}. Alternatively, we could consider a \textquotedblleft
bias-free\textquotedblright\ test based on undersmoothing; that is using a
bandwidth $h$ satisfying $nh^{5}\rightarrow0$ such that $B_{n}\rightarrow0$
and inference can be based on quantiles of $\xi_{1}\sim N(0,v_{11}^{2})$. In
contrast to our procedure, however, such a test has only trivial power against
$\bar{\beta}+an^{-2/5}$ because $(nh)^{1/2}an^{-2/5}\rightarrow0$.

\section{Concluding remarks}

\label{Sec conc}

In this paper, we have shown that in statistical problems involving bias terms
that cannot be estimated, the bootstrap can be modified to provide
asymptotically valid inference. Intuitively, the main idea is the following:
in some important cases, the bootstrap can be used to `debias' a statistic
whose bias is non-negligible, but when doing so additional `noise' is
injected. This additional noise does not vanish because the bias cannot be
consistently estimated, but it can be handled either by a `plug-in' method or
by an additional (i.e., double) bootstrap layer. Specifically, our solution is
simple and involves (i)~focusing on the bootstrap p-value; (ii)~estimating its
asymptotic distribution; (iii)~mapping the original (invalid)\ p-value into a
new (valid)\ p-value using the prepivoting approach. These steps are easy to
implement in practice and we provide sufficient conditions for asymptotic
validity of the associated tests and confidence intervals.

Our results can be generalized in several directions. For instance, there is a
growing literature where inference on a parameter of interest is combined with
some auxiliary information in the form of a bound on the bias of the estimator
in question. These bounds appear, e.g., in Oster (2019) and Li and M\"{u}ller
(2021). It is of interest to investigate how our analysis can be extended in
order to incorporate such bounds. Other possible extensions include
non-ergodic problems, large-dimensional models, and multivariate estimators or
statistics. All these extensions are left for future research.

\section*{Acknowledgements}

We thank Federico Bandi, Matias Cattaneo, Christian Gourieroux, Philip Heiler,
Michael Jansson, Anders Kock, Damian Kozbur, Marcelo Moreira, David
Preinerstorfer, Mikkel S\o lvsten, Luke Taylor, Michael Wolf, and participants
at the AiE Conference in Honor of Joon Y.\ Park, 2022 Conference on
Econometrics and Business Analytics (CEBA), 2023 Conference on Robust
Econometric Methods in Financial Econometrics, 2022 EC$^{2}$ conference,
2$^{\text{nd}}$ `High Voltage Econometrics' workshop, 2023 IAAE Conference,
3$^{\text{rd}}$ Italian Congress of Econometrics and Empirical Economics,
3$^{\text{rd}}$\ Italian Meeting on Probability and Mathematical Statistics,
19$^{\text{th}}$ School of Time Series and Econometrics, Brazilian Statistical
Association, 2023 Soci\'{e}t\'{e} Canadienne de Sciences \'{E}conomiques, 2022
Virtual Time Series Seminars, as well as seminar participants at Aarhus
University, CREST, FGV - Rio, FGV - S\~{a}o Paulo, Ludwig Maximilian
University of Munich, Queen Mary University, Singapore Management University,
UFRGS, University of the Balearic Islands, University of Oxford, University of
Pittsburgh, University of Victoria, York University, for useful comments and
feedback. Cavaliere thanks the the Italian Ministry of University and Research
(PRIN 2017 Grant 2017TA7TYC) for financial support. Gon\c{c}alves thanks the
Natural Sciences and Engineering Research Council of Canada for financial
support (NSERC grant number RGPIN-2021-02663). Nielsen thanks the Danish
National Research Foundation for financial support (DNRF Chair grant number DNRF154).

\section*{References}

\begin{description}
\item \textsc{Beran, R. }(1987). Prepivoting to reduce level error in
confidence sets. \emph{Biometrika} 74, 457--468.

\item \textsc{Beran, R. }(1988). Prepivoting test statistics: A bootstrap view
of asymptotic refinements. \emph{Journal of the American Statistical
Association} 83, 687--97.

\item \textsc{Calonico, S., M.D.\ Cattaneo, and M.H.\ Farrell }(2018). On the
effect of bias estimation on coverage accuracy in nonparametric inference.
\emph{Journal of the American Statistical Association} 113, 767--779.

\item \textsc{Calonico, S., M.D.\ Cattaneo, and R. Titiunik }(2014). Robust
nonparametric confidence intervals for regression-discontinuity designs.
\emph{Econometrica} 82, 2295--2326.

\item \textsc{Cattaneo, M.D., and M.\ Jansson }(2018). Kernel-based
semiparametric estimators:\ small bandwidth asymptotics and bootstrap
consistency. \emph{Econometrica} 86, 955--995.

\item \textsc{Cattaneo, M.D., and M.\ Jansson }(2022). Average density
estimators: efficiency and bootstrap consistency. \emph{Econometric Theory}
38, 1140--1174.

\item \textsc{Cattaneo, M.D., M.\ Jansson, and X.\ Ma\ }(2019). Two-step
estimation and inference with possibly many included covariates. \emph{Review
of Economic Studies} 86, 1095--1122.

\item \textsc{Cavaliere, G., and I.\ Georgiev }(2020). Inference under random
limit bootstrap measures. \emph{Econometrica} 88, 2547--2574.

\item \textsc{Chatterjee, A., and S.N.\ Lahiri }(2010). Asymptotic properties
of the residual bootstrap for lasso estimators. \emph{Proceedings of the
American Mathematical Society} 138, 4497--4509.

\item \textsc{Chatterjee, A., and S.N.\ Lahiri }(2011). Bootstrapping lasso
estimators. \emph{Journal of the American Statistical Association} 106, 608--625.

\item \textsc{Efron, B. }(1983). Estimating the error rate of a prediction
rule: Improvement on cross-validation. \emph{Journal of American Statistical
Association} 78, 316--331.

\item \textsc{Eubank, R.L., and P.L.\ Speckman }(1993). Confidence bands in
nonparametric regression. \emph{Journal of the American Statistical
Association} 88, 1287--1301.

\item \textsc{Fu, W., and K.\ Knight }(2000). Asymptotics for lasso-type
estimators. \emph{Annals of Statistics} 28, 1356--1378.

\item \textsc{Hall, P. }(1986). On the bootstrap and confidence intervals.
\emph{Annals of Statistics} 14, 1431--1452.

\item \textsc{Hall, P. }(1992). \emph{The Bootstrap and Edgeworth Expansion},
Springer-Verlag: Berlin.

\item \textsc{Hall, P., and J.\ Horowitz }(2013). A simple bootstrap method
for constructing nonparametric confidence bands for functions. \emph{Annals of
Statistics} 41, 1892--1921.

\item \textsc{Hansen, B.E. }(2007). Least squares model averaging.
\emph{Econometrica} 75, 1175--1189.

\item \textsc{H\"{a}rdle, W., and A.W.\ Bowman }(1988). Bootstrapping in
nonparametric regression: local adaptive smoothing and confidence bands.
\emph{Journal of the American Statistical Association} 83, 102--110.

\item \textsc{H\"{a}rdle, W., and J.S.\ Marron }(1991). Bootstrap simultaneous
error bars for nonparametric regression. \emph{Annals of Statistics} 19, 778--796.

\item \textsc{Hjort, N., and G.\ Claeskens }(2003). Frequentist model average
estimators. \emph{Journal of the American Statistical Association} 98, 879--899.

\item \textsc{Horowitz, J.L. }(2001). The bootstrap. In \emph{Handbook of
Econometrics} (Heckman, J.J., and E.\ Leamer, eds.), vol.\ 5, chp.\ 52,
Elsevier: Amsterdam.

\item \textsc{Hounyo, U. and K.\ Lahiri }(2023). Estimating the variance of a
combined forecast: Bootstrap-based approach. \emph{Journal of Econometrics}
232, 445--468.

\item \textsc{Li, C., and U.\ Muller }(2021). Linear regression with many
controls of limited explanatory power. \emph{Quantitative Economics} 12, 405--442.

\item \textsc{Liu, C.-A. }(2015). Distribution theory of the least squares
averaging estimator. \emph{Journal of Econometrics} 186, 142--159.

\item \textsc{Oster, E. }(2019). Unobservable selection and coefficient
stability: Theory and evidence. \emph{Journal of Business \&\ Economic
Statistics} 37, 187--204.

\item \textsc{Shao, X., and D.N. Politis }(2013). Fixed $b$ subsampling and
the block bootstrap:\ Improved confidence sets based on $p$-value calibration.
\emph{Journal of the Royal Statistical Society B} 75, 161--184.
\end{description}

\appendix\setcounter{section}{0}\newpage

\begin{center}
{\LARGE \vspace*{1in}}

{\Huge Supplemental Material:}

{\LARGE \vspace*{0.5in}}

{\huge Bootstrap Inference in the Presence of Bias}

{\LARGE \vspace*{1in}}

{\Large by}

\bigskip

\bigskip

{\Large G.\ Cavaliere, S.\ Gon\c{c}alves, M.\O .\ Nielsen, and E.\ Zanelli}

{\Large \bigskip}

{\Large \bigskip}

{\Large
\today
}
\end{center}

\bigskip

{\LARGE \vspace*{1in}}

This supplemental material contains two appendices.
Appendix~\ref{Sec asympt Gaussian} describes in detail the conditions and
results of the main paper under the special case of asymptotically Gaussian
statistics. Appendix~\ref{Sec Applications} contains details and proofs for
the three examples in the main paper, as well as two additional examples.
Additional references are included at the end of the supplement.%

\setcounter{page}{1}%
\newpage

\section{Special case: $T_{n}$ is asymptotically Gaussian}

\label{Sec asympt Gaussian}

In this section, we specialize Assumptions~\ref{Assn T}, \ref{Assn BS},
and~\ref{Assn DBS} to the case where $T_{n}=\sqrt{n}(\hat{\theta}_{n}%
-\theta_{0})$ is a normalized parameter estimator whose limiting distribution
is normal. We consider the following special case of Assumption~\ref{Assn T}.%

\setcounter{condition}{0}
\renewcommand{\thecondition}{\arabic{condition}$'$}%

\begin{condition}
\label{Assn T prime}It holds that $T_{n}-B_{n}\rightarrow_{d}N(0,v^{2})$,
where $v^{2}>0$.
\end{condition}

Assumption~\ref{Assn T prime} covers statistics $T_{n}$ based on
asymptotically biased estimators: when $B_{n}\rightarrow_{p}B$, we have
$T_{n}\rightarrow_{d}N\left(  B,v^{2}\right)  $, in which case $B$ is the
asymptotic bias of $\hat{\theta}_{n}$. More generally, we can interpret
$B_{n}$ as a bias term that approximates $E(\sqrt{n}(\hat{\theta}_{n}%
-\theta_{0}))$ although $B_{n}$ does not need to have a limit. Note that
Assumption~\ref{Assn T prime} obtains from Assumption~\ref{Assn T} when we let
$\xi_{1}\sim N(0,v^{2})$ and $G(u)=\Phi(u/v)$.

Let $D_{n}^{\ast}$ denote a bootstrap sample from $D_{n}$ and let $\hat
{\theta}_{n}^{\ast}$ be a bootstrap version of $\hat{\theta}_{n}$. The
bootstrap analogue of $T_{n}$ is $T_{n}^{\ast}=\sqrt{n}(\hat{\theta}_{n}%
^{\ast}-\hat{\theta}_{n})$.

\begin{condition}
\label{Assn BS prime}It holds that (i)~$T_{n}^{\ast}-\hat{B}_{n}%
\overset{d^{\ast}}{\rightarrow}_{p}N(0,v^{2})$, and (ii)
\[
\left(
\begin{array}
[c]{c}%
T_{n}-B_{n}\\
\hat{B}_{n}-B_{n}%
\end{array}
\right)  \overset{d}{\rightarrow}N(0,V),\quad V:=(v_{ij}),\quad i,j=1,2,
\]
where $v_{d}^{2}:=v_{11}+v_{22}-2v_{12}>0$ with $v_{11}:=v^{2}>0.$
\end{condition}

Assumption~\ref{Assn BS prime}(i) requires the bootstrap statistic
$T_{n}^{\ast}-\hat{B}_{n}$ to mimic the asymptotic distribution of
$T_{n}-B_{n}$, as in Assumption~\ref{Assn BS}(i). However, and contrary to
Assumption~\ref{Assn BS}(i), here this limiting distribution is the zero mean
Gaussian distribution (i.e.\ $G(u)=\Phi(u/v)$), which means that we can
interpret $\hat{B}_{n}$ as a bootstrap bias correction term; i.e.,\ $\hat
{B}_{n}=E^{\ast}(\sqrt{n}(\hat{\theta}_{n}^{\ast}-\hat{\theta}_{n}))$.
Assumption~\ref{Assn BS prime}(ii) assumes that $\hat{B}_{n}-B_{n}$ is also
asymptotically distributed as a zero mean Gaussian random variable (jointly
with $T_{n}-B_{n}$).\footnote{In terms of Assumption~\ref{Assn BS},
Assumption~\ref{Assn BS prime} corresponds to the case where the vector
$\xi=(\xi_{1},\xi_{2})^{\prime}$ is a multivariate normal distribution with
covariance matrix $V$.} An implication of this assumption is that
\begin{equation}
T_{n}-\hat{B}_{n}=(T_{n}-B_{n})-(\hat{B}_{n}-B_{n})\overset{d}{\rightarrow
}N(0,v_{d}^{2}), \label{eq Tn-Bhatn}%
\end{equation}
where $v_{d}^{2}:=v_{11}+v_{22}-2v_{12}$. We do not require $V$ to be positive
definite; for instance, $v_{22}=0$ whenever $\hat{B}_{n}-B_{n}=o_{p}(1)$, and
in fact $V$ can be rank deficient even when $v_{22}>0$. However, we do impose
the restriction that $v_{d}^{2}>0$. This ensures that the limiting
distribution function of $T_{n}-\hat{B}_{n}$, given by $F(u)=\Phi(u/v_{d})$,
is well-defined and continuous, as assumed in Assumption~\ref{Assn BS}(ii).

Let $\hat{p}_{n}$ denote the standard bootstrap p-value as defined in
Section~\ref{Sec:general}. We then obtain the following.

\begin{corollary}
\label{Cor-first-level-Gaussian}Under Assumptions~\ref{Assn T prime}
and~\ref{Assn BS prime}, $\hat{p}_{n}\rightarrow_{d}\Phi(m\Phi^{-1}%
(U_{[0,1]}))$, where $m^{2}:=v_{d}^{2}/v^{2}$.
\end{corollary}

Corollary~\ref{Cor-first-level-Gaussian} follows immediately from
Theorem~\ref{Theor1} when we let $G(u)=\Phi(u/v)$ and $F(u)=\Phi(u/v_{d})$. It
shows that the asymptotic distribution of $\hat{p}_{n}$ is uniform only when
$m=1$, or equivalently when $v_{d}^{2}=v^{2}$. In this case, the difference
$\hat{B}_{n}-B_{n}$ is $o_{p}(1)$. When $v_{d}^{2}\neq v^{2}$, $\hat{B}%
_{n}-B_{n}$ is random even in the limit, implying that the limiting bootstrap
distribution function of $T_{n}^{\ast}$ is conditionally random. Although
random limit bootstrap measures do not necessarily invalidate bootstrap
inference, as discussed by Cavaliere and Georgiev (2020), this is not the case
here. However, we can solve the problem of bootstrap invalidity by applying
the prepivoting approach or by modifying the test statistic from $T_{n}$ to
$T_{n}-\hat{B}_{n}$.

To describe the prepivoting approach, note that the limiting distribution of
$\hat{p}_{n}$ is given by%
\[
H(u):=\lim P(\hat{p}_{n}\leq u)=\Phi(m^{-1}\Phi^{-1}(u)).
\]
Hence, in this case $\gamma=m$, and a plug-in approach amounts to estimating
$m^{2}:=v_{d}^{2}/v^{2}$, where $v^{2}$ and $v_{d}^{2}$ are defined in
Assumption~\ref{Assn BS prime}. Suppose that $\hat{v}_{n}^{2}$ and $\hat
{v}_{d,n}^{2}$ are consistent estimators of $v^{2}$ and $v_{d}^{2}$ (i.e.,
assume that $(\hat{v}_{n}^{2},\hat{v}_{d,n}^{2})\rightarrow_{p}(v^{2}%
,v_{d}^{2})$) and let $\hat{m}_{n}^{2}:=\hat{v}_{d,n}^{2}/\hat{v}_{n}^{2}$.
Then, by Corollary~\ref{Corollary-plug-in}, it immediately follows that
\[
\tilde{p}_{n}=\Phi(\hat{m}_{n}^{-1}\Phi^{-1}(\hat{p}_{n}%
))\overset{d}{\rightarrow}U_{[0,1]}%
\]
under Assumptions~\ref{Assn T prime} and~\ref{Assn BS prime}. For brevity, we
do not formalize this result here.

To describe the double bootstrap modified p-value, $\tilde{p}_{n}:=\hat{H}%
_{n}(\hat{p}_{n})=P^{\ast}(\hat{p}_{n}^{\ast}\leq\hat{p}_{n})$, when applied
to the special case where $T_{n}$ satisfies Assumption~\ref{Assn T prime}, we
now introduce Assumption~\ref{Assn DBS prime}.

\begin{condition}
\label{Assn DBS prime}Let $T_{n}^{\ast\ast}=\sqrt{n}(\hat{\theta}_{n}%
^{\ast\ast}-\hat{\theta}_{n}^{\ast})$ and suppose that (i)~$T_{n}^{\ast\ast
}-\hat{B}_{n}^{\ast}\overset{d^{\ast\ast}}{\rightarrow}_{p^{\ast}}N(0,v^{2})$,
in probability, and (ii)~$T_{n}^{\ast}-\hat{B}_{n}^{\ast}\overset{d^{\ast
}}{\rightarrow}_{p}N(0,v_{d}^{2})$, where $v_{d}^{2}$ is as defined in
Assumption~\ref{Assn BS prime}(ii).
\end{condition}

Under Assumption~\ref{Assn DBS prime}(i), the double bootstrap distribution of
$T_{n}^{\ast\ast}-\hat{B}_{n}^{\ast}$ mimics the distribution of $T_{n}^{\ast
}-\hat{B}_{n}$, where the double bootstrap bias term $\hat{B}_{n}^{\ast
}=E^{\ast\ast}(\sqrt{n}(\hat{\theta}_{n}^{\ast\ast}-\hat{\theta}_{n}^{\ast}))$
is asymptotically centered at $\hat{B}_{n}$ under
Assumption~\ref{Assn DBS prime}(ii). When $v_{d}^{2}\neq v^{2}$, the double
bootstrap bias is not a consistent estimator of $\hat{B}_{n}$, but that is not
needed for the asymptotic validity of the modified double bootstrap p-value
$\tilde{p}_{n}=\hat{H}_{n}(\hat{p}_{n})$ defined in Section~\ref{Sec:general}.

By application of Theorem~\ref{TheoremDouble-p-val}, $\tilde{p}_{n}=\hat
{H}_{n}(\hat{p}_{n})\rightarrow_{d}U_{[0,1] }$ under
Assumptions~\ref{Assn T prime}, \ref{Assn BS prime}, and~\ref{Assn DBS prime}.
We can also provide a result analogous to Corollary~\ref{Corollary pvalue-BC}
under these assumptions. In this case, if closed-form expressions for $\hat
{B}_{n}$ and $\hat{B}_{n}^{\ast}$ are not available, we can approximate these
bootstrap expectations by Monte Carlo simulations and then compute $P^{\ast
}(T_{n}^{\ast}-\hat{B}_{n}^{\ast}\leq T_{n}-\hat{B}_{n})$ as a valid bootstrap
p-value. Note, however, that this approach is computationally as intensive as
the prepivoting approach based on $\tilde{p}_{n}$ since it too requires two
layers of resampling.

\begin{remark}
In the case of asymptotically Gaussian statistics discussed in this section,
the more general Assumptions~\ref{Assn BSpairs} and~\ref{Assn DBSpairs}
simplify straightforwardly. In Assumption~\ref{Assn BS prime}(i) we assume
that $T_{n}^{\ast}-\hat{B}_{n}\overset{d^{\ast}}{\rightarrow}_{p}N(0,v_{s}%
^{2})$ and in Assumption~\ref{Assn DBS prime}(i) that $T_{n}^{\ast\ast}%
-\hat{B}_{n}^{\ast}\overset{d^{\ast\ast}}{\rightarrow}_{p^{\ast}}N(0,v_{s}%
^{2})$, in probability, for some $v_{s}^{2}>0$, while the rest of
Assumptions~\ref{Assn T prime}--\ref{Assn DBS prime} are unchanged. The
results of this section continue to apply under these more general conditions,
replacing $G(u)=\Phi(u/v)$ with $J(u)=\Phi(u/v_{s})$ and consequently defining
$m:=v_{d}^{2}/v_{s}^{2}$.
\end{remark}

\begin{remark}
Contrary to Beran (1987, 1988), in our context the first level of prepivoting,
e.g.,\ by the double bootstrap, is used to obtain an asymptotically valid
bootstrap p-value. Therefore, inference based on $\tilde{p}_{n}$ does not
necessarily provide an asymptotic refinement over inference based on an
asymptotic approach that does not require the bootstrap. Nevertheless, the
Monte Carlo results in Table~\ref{tab:MC} below seem to suggest an asymptotic
refinement for the double bootstrap, at least for the non-parametric bootstrap
scheme. In the special case where the bias term $B_{n}$ is of sufficiently
small order, the arguments in Beran (1987, 1988) apply, and an asymptotic
refinement can be obtained. We also conjecture that, in the general case, an
asymptotic refinement could be obtained by further iterating the bootstrap.
\end{remark}

%

\renewcommand{\thecondition}{\arabic{condition}}%

\section{Examples with details}

\label{Sec Applications}

\subsection{Inference after model averaging}

\label{sec:MA-details}

In this section we first provide the regularity conditions required in
Lemmas~\ref{Lemma MA} and~\ref{Lemma MA dbs}, and then we give the proofs of
the lemmas. We subsequently provide some brief Monte Carlo evidence. Finally,
at the end of the section, we provide regularity conditions for the extension
to the pairs bootstrap and a proof of the associated
Lemma~\ref{lemma mod av pairs bootstrap}.

\subsubsection{Assumptions and notation}

We impose the following conditions.%

\renewcommand{\thecondition}{MA}%

\begin{condition}
\label{Assn-MA}(i)~$\varepsilon_{t}|W\sim$~i.i.d.$(0,\sigma^{2})$, where
$W:=(x,Z)$; (ii)~$S_{WW}\rightarrow_{p}\Sigma_{WW}$ with $\operatorname*{rank}%
(\Sigma_{WW})=q+1$; (iii)~$n^{1/2}S_{W\varepsilon}\rightarrow_{d}N(0,\Omega)$
with $\Omega:=\sigma^{2}\Sigma_{WW}$.
\end{condition}

\begin{remark}
We assume that the weights $\omega$ are fixed and independent of~$n$. A
popular example in forecasting is to use equal weighting. We could allow for
stochastic weights as long as these are constant in the limit. This would be
the case, for example, when the weights are based on moments that can be
consistently estimated.
\end{remark}

To proceed, we introduce the following notation. First, partition $\Sigma
_{WW}$ according to~$W$,
\[
\Sigma_{WW}:=\left(
\begin{array}
[c]{cc}%
\Sigma_{xx} & \Sigma_{xZ}\\
\Sigma_{Zx} & \Sigma_{ZZ}%
\end{array}
\right)  .
\]
Let $\Sigma_{xZ_{m}}:=\Sigma_{xZ}R_{m}$, $\Sigma_{Z_{m}Z_{m}}:=R_{m}^{\prime
}\Sigma_{ZZ}R_{m}$, $\Sigma_{xx.Z_{m}}:=\Sigma_{xx}-\Sigma_{xZ}R_{m}%
(R_{m}^{\prime}\Sigma_{ZZ}R_{m})^{-1}R_{m}^{\prime}\Sigma_{Zx}$, and
$\Sigma_{xZ.Z_{m}}:=\Sigma_{xZ}-\Sigma_{xZ}R_{m}(R_{m}^{\prime}\Sigma
_{ZZ}R_{m})^{-1}R_{m}^{\prime}\Sigma_{ZZ}$. Also let $A_{n}:=\sum_{m=1}%
^{M}\omega_{m}S_{xx.Z_{m}}^{-1}n^{-1}x^{\prime}M_{Z_{m}}$, where $M_{Z_{m}%
}:=I_{n}-Z_{m}(Z_{m}^{\prime}Z_{m})^{-1}Z_{m}^{\prime}$, such that
$A_{n}Z=Q_{n}$. With this notation,%
\begin{align}
\tilde{\beta}_{n}  &  =A_{n}y=A_{n}x\beta+Q_{n}\delta+A_{n}\varepsilon
=\beta+Q_{n}\delta+A_{n}\varepsilon,\label{beta-tilde}\\
\tilde{\beta}_{n}^{\ast}  &  =A_{n}y^{\ast}=\hat{\beta}_{n}+Q_{n}\hat{\delta
}_{n}+A_{n}\varepsilon^{\ast}. \label{beta-tilde-ast}%
\end{align}
Finally, define
\begin{align*}
\bar{d}_{M,n}^{\prime}  &  :=\sum_{m=1}^{M}\omega_{m}S_{xx.Z_{m}}%
^{-1}(1,-S_{xZ_{m}}S_{Z_{m}Z_{m}}^{-1}R_{m}^{\prime}),\\
\bar{b}_{M,n}^{\prime}  &  :=\sum\nolimits_{m=1}^{M}\omega_{m}S_{xx.Z_{m}%
}^{-1}S_{xZ.Z_{m}}S_{ZZ.x}^{-1}(-S_{Zx}S_{xx}^{-1},I_{q}),
\end{align*}
and let $\bar{d}_{M}^{\prime}$ and $\bar{b}_{M}^{\prime}$ denote their
probability limits, which exist and are well-defined under
Assumption~\ref{Assn-MA}.

\subsubsection{Proofs of lemmas}

\noindent\textsc{Proof of Lemma~\ref{Lemma MA}. }We first verify
Assumption~\ref{Assn T} (or equivalently, Assumption~\ref{Assn T prime}).
Using (\ref{beta-tilde}) we can write $T_{n}=B_{n}+\xi_{1,n}$ with%
\[
\xi_{1,n}:=n^{1/2}A_{n}\varepsilon=n^{1/2}\sum_{m=1}^{M}\omega_{m}S_{xx.Z_{m}%
}^{-1}n^{-1}x^{\prime}M_{Z_{m}}\varepsilon=n^{1/2}\sum_{m=1}^{M}\omega
_{m}S_{xx.Z_{m}}^{-1}S_{x\varepsilon.Z_{m}}.
\]
Then%
\begin{align*}
S_{x\varepsilon.Z_{m}}  &  =n^{-1}x^{\prime}M_{Z_{m}}\varepsilon
=n^{-1}(x^{\prime}\varepsilon-x^{\prime}Z_{m}(Z_{m}^{\prime}Z_{m})^{-1}%
R_{m}^{\prime}Z^{\prime}\varepsilon)\\
&  =(1,-S_{xZ_{m}}(S_{Z_{m}Z_{m}})^{-1}R_{m}^{\prime})S_{W\varepsilon}%
=:\hat{d}_{m}^{\prime}S_{W\varepsilon},
\end{align*}
so that
\[
\xi_{1,n}=\sum_{m=1}^{M}\omega_{m}S_{xx.Z_{m}}^{-1}\hat{d}_{m}^{\prime}%
n^{1/2}S_{W\varepsilon}=\bar{d}_{M,n}^{\prime}n^{1/2}S_{W\varepsilon}.
\]
Hence, $\xi_{1,n}\rightarrow_{d}N(0,v^{2})$ with $v^{2}:=\bar{d}_{M}^{\prime
}\Omega\bar{d}_{M}$.

Next, we verify Assumption~\ref{Assn BS} (or Assumption~\ref{Assn BS prime}).
From (\ref{beta-tilde-ast}) we write $T_{n}^{\ast}=\hat{B}_{n}+\xi_{1,n}%
^{\ast}$ with $\xi_{1,n}^{\ast}:=n^{1/2}A_{n}\varepsilon^{\ast}\sim
N(0,\hat{\sigma}_{n}^{2}A_{n}A_{n}^{\prime})$, conditional on $D_{n}$.
Part~(i) now follows straightforwardly because $\hat{\sigma}_{n}%
^{2}\rightarrow_{p}\sigma^{2}$ and $A_{n}A_{n}^{\prime}=\bar{d}_{M,n}^{\prime
}S_{WW}\bar{d}_{M,n}\rightarrow_{p}\bar{d}_{M}^{\prime}\Sigma_{WW}\bar{d}_{M}%
$. To prove Part~(ii), note that
\[
n^{1/2}(\hat{\delta}_{n}-\delta)=S_{ZZ.x}^{-1}S_{Z\varepsilon.x}=S_{ZZ.x}%
^{-1}(-S_{Zx}S_{xx}^{-1},I_{q})n^{1/2}S_{W\varepsilon},
\]
from which it follows that
\[
\hat{B}_{n}-B_{n}=Q_{n}n^{1/2}(\hat{\delta}_{n}-\delta)=Q_{n}S_{ZZ.x}%
^{-1}(-S_{Zx}S_{xx}^{-1},I_{q})n^{1/2}S_{W\varepsilon}=\bar{b}_{M,n}^{\prime
}n^{1/2}S_{W\varepsilon}.
\]
Hence,
\[
\left(
\begin{array}
[c]{c}%
T_{n}-B_{n}\\
\hat{B}_{n}-B_{n}%
\end{array}
\right)  =\left(
\begin{array}
[c]{c}%
\bar{d}_{M,n}^{\prime}\\
\bar{b}_{M,n}^{\prime}%
\end{array}
\right)  n^{-1/2}W^{\prime}\varepsilon\overset{d}{\rightarrow}N(0,V),\quad
V=\left(
\begin{array}
[c]{cc}%
\bar{d}_{M}^{\prime}\Omega\bar{d}_{M} & \bar{d}_{M}^{\prime}\Omega\bar{b}%
_{M}\\
\bar{b}_{M}^{\prime}\Omega\bar{d}_{M} & \bar{b}_{M}^{\prime}\Omega\bar{b}_{M}%
\end{array}
\right)  ,
\]
which completes the proof.$\hfill\square$

\smallskip

\noindent\textsc{Proof of Lemma~\ref{Lemma MA dbs}.} First note that
$\tilde{\beta}_{n}^{\ast\ast}=A_{n}y^{\ast\ast}=A_{n}x\hat{\beta}_{n}^{\ast
}+A_{n}Z\hat{\delta}_{n}^{\ast}+A_{n}\varepsilon^{\ast\ast}$. It follows that
\[
T_{n}^{\ast\ast}:=n^{1/2}(\tilde{\beta}_{n}^{\ast\ast}-\hat{\beta}_{n}^{\ast
})=\hat{B}_{n}^{\ast}+n^{1/2}A_{n}\varepsilon^{\ast\ast},
\]
where $\hat{B}_{n}^{\ast}:=n^{1/2}Q_{n}\hat{\delta}_{n}^{\ast}$ and $\xi
_{1,n}^{\ast\ast}:=n^{1/2}A_{n}\varepsilon^{\ast\ast}\sim N(0,\hat{\sigma}%
_{n}^{\ast2}A_{n}A_{n}^{\prime})$, conditional on $(D_{n},D_{n}^{\ast})$. The
conditions in Assumption~\ref{Assn DBS}(i) or~\ref{Assn DBS prime}(i) now
follows as in Part~(i) of the previous proof because $\hat{\sigma}_{n}^{\ast
2}\overset{p^{\ast}}{\rightarrow}_{p}\sigma^{2}$. For
Assumption~\ref{Assn DBS}(ii) or~\ref{Assn DBS prime}(ii) we consider the
joint convergence of $(T_{n}^{\ast}-\hat{B}_{n},\hat{B}_{n}^{\ast}-\hat{B}%
_{n})^{\prime}$. By noticing that%
\[
n^{1/2}(\hat{\delta}_{n}^{\ast}-\hat{\delta}_{n})=S_{ZZ.x}^{-1}S_{Z\varepsilon
^{\ast}.x}=S_{ZZ.x}^{-1}(-S_{Zx}S_{xx}^{-1},I_{q})n^{1/2}S_{W\varepsilon
^{\ast}},
\]
it follows that
\[
\left(
\begin{array}
[c]{c}%
T_{n}^{\ast}-\hat{B}_{n}\\
\hat{B}_{n}^{\ast}-\hat{B}_{n}%
\end{array}
\right)  =\left(
\begin{array}
[c]{c}%
\bar{d}_{M,n}^{\prime}\\
\bar{b}_{M,n}^{\prime}%
\end{array}
\right)  n^{1/2}S_{W\varepsilon^{\ast}}\sim N(0,\hat{V}_{n}),
\]
conditional on $D_{n}$, where%
\[
\hat{V}_{n}=\hat{\sigma}_{n}^{2}\left(
\begin{array}
[c]{cc}%
\bar{d}_{M,n}^{\prime}S_{WW}\bar{d}_{M,n} & \bar{d}_{M,n}^{\prime}S_{WW}%
\bar{b}_{M,n}\\
\bar{b}_{M,n}^{\prime}S_{WW}\bar{d}_{M,n} & \bar{b}_{M,n}^{\prime}S_{WW}%
\bar{b}_{M,n}%
\end{array}
\right)  \overset{p}{\rightarrow}V.
\]
The desired result follows.$\hfill\square$

\subsubsection{A small Monte Carlo experiment}

In Table~\ref{tab:MC}\ we present the results of a small Monte Carlo
simulation experiment to illustrate the above results numerically. We generate
the data from the regression model (\ref{MA model}) with sample sizes
$n=10,20,40$. The regressors $x_{t}$ and $z_{t}$ are both scalar and
multivariate normally distributed with unit variances and correlation~$0.7$,
and the errors are either standard normal, $t_{3}$, or $\chi_{1}^{2}$
distributed. The true values are $\beta=\bar{\beta}+an^{-1/2}$ with
$\bar{\beta}=1$ and $\delta=1$ (the results are invariant to $\bar{\beta}$ and
$\delta$ because we use the unrestricted estimates to construct the bootstrap
samples). We test the null hypothesis $\mathsf{H}_{0}:\beta=\bar{\beta}$
against a left-sided alternative. Results for right-tailed and two-tailed
tests are analogous to those presented here for left-tailed tests. The case
$a=0$ corresponds to rejection frequencies under the null, and $a=-1,-2,-4$
corresponds to rejection frequencies under local alternatives. The estimator
puts weight $\omega_{1}=1/2$ on the short model that includes only $x$ (and a
constant term)\ and weight $\omega_{2}=1/2$ on the long model that includes
both regressors (and a constant term). We consider two bootstrap schemes. The
first is the parametric bootstrap scheme, where $\varepsilon_{t}^{\ast}\sim
$~i.i.d.$N(0,1)$, which is denoted as \textquotedblleft par.\textquotedblright%
\ in the table. The second is the non-parametric bootstrap scheme, where
$\varepsilon_{t}^{\ast}$ is resampled independently from the (centered)
residuals from the long regression, which is denoted as \textquotedblleft
non-par.\textquotedblright\ Results are based on 10,000 Monte Carlo
simulations and $B=999$ bootstrap replications.

First consider the case $a=0$. The simulation outcomes in Table~\ref{tab:MC}
clearly illustrate our theoretical results. The standard bootstrap p-value,
$\hat{p}_{n}$, is much larger than the nominal level of the test. The plug-in
modified p-value, $\tilde{p}_{n,p}$, is close to the nominal level for the
parametric bootstrap scheme, but is still over-sized for the non-parametric
scheme with the smaller sample sizes. Finally, the double bootstrap modified
p-value, $\tilde{p}_{n,d}$, is nearly perfectly sized throughout the table.

Table~\ref{tab:MC} for $a=-1,-2,-4$ clearly shows nontrivial power, which
increases as $a$ increases. The discrepancies in finite-sample power are due
to differences in size. For example, consider the standard parametric
bootstrap with 5\% nominal level and normal errors (top left of the table). It
has finite-sample size very close to 10\%. Comparing this with our modified
bootstrap test with nominal size 10\% (towards the right in the same panel of
the table), we see that the finite-sample powers are nearly identical.%

\begin{table}[tp]
\begin{footnotesize}
\begin{center}%
\caption{Simulated rejection frequencies (\%) of bootstrap tests}\label{tab:MC}%
\begin{tabular}
[c]{lllccccccccccccccc}\hline
&  &  & \multicolumn{7}{c}{$5\%$\ nominal level} &  &
\multicolumn{7}{c}{$10\%$ nominal level}\\\cline{4-10}\cline{12-18}
&  &  & \multicolumn{3}{c}{par.} &  & \multicolumn{3}{c}{non-par.} &  &
\multicolumn{3}{c}{par.} &  & \multicolumn{3}{c}{non-par.}\\\cline{4-6}%
\cline{8-10}\cline{12-14}\cline{16-18}%
\multicolumn{1}{c}{dist.} & \multicolumn{1}{c}{$a$} & \multicolumn{1}{c}{$n$}
& $\hat{p}_{n}$ & $\tilde{p}_{n,p}$ & $\tilde{p}_{n,d}$ &  & $\hat{p}_{n}$ &
$\tilde{p}_{n,p}$ & $\tilde{p}_{n,d}$ &  & $\hat{p}_{n}$ & $\tilde{p}_{n,p}$ &
$\tilde{p}_{n,d}$ &  & $\hat{p}_{n}$ & $\tilde{p}_{n,p}$ & $\tilde{p}_{n,d}%
$\\\hline
\multicolumn{1}{c}{$N$} & \multicolumn{1}{c}{$0$} & \multicolumn{1}{c}{$10$} &
\multicolumn{1}{r}{$10.1$} & \multicolumn{1}{r}{$5.0$} &
\multicolumn{1}{r}{$5.0$} &  & \multicolumn{1}{r}{$16.2$} &
\multicolumn{1}{r}{$11.2$} & \multicolumn{1}{r}{$6.3$} &  &
\multicolumn{1}{r}{$15.9$} & \multicolumn{1}{r}{$10.0$} &
\multicolumn{1}{r}{$10.0$} &  & \multicolumn{1}{r}{$21.5$} &
\multicolumn{1}{r}{$16.2$} & \multicolumn{1}{r}{$10.5$}\\
& \multicolumn{1}{c}{} & \multicolumn{1}{c}{$20$} & \multicolumn{1}{r}{$9.7$}
& \multicolumn{1}{r}{$5.0$} & \multicolumn{1}{r}{$5.1$} &  &
\multicolumn{1}{r}{$12.6$} & \multicolumn{1}{r}{$7.8$} &
\multicolumn{1}{r}{$5.4$} &  & \multicolumn{1}{r}{$15.1$} &
\multicolumn{1}{r}{$9.8$} & \multicolumn{1}{r}{$9.8$} &  &
\multicolumn{1}{r}{$18.2$} & \multicolumn{1}{r}{$12.9$} &
\multicolumn{1}{r}{$10.4$}\\
& \multicolumn{1}{c}{} & \multicolumn{1}{c}{$\underset{}{40}$} &
\multicolumn{1}{r}{$9.8$} & \multicolumn{1}{r}{$5.1$} &
\multicolumn{1}{r}{$5.2$} &  & \multicolumn{1}{r}{$10.5$} &
\multicolumn{1}{r}{$5.8$} & \multicolumn{1}{r}{$4.9$} &  &
\multicolumn{1}{r}{$15.4$} & \multicolumn{1}{r}{$10.1$} &
\multicolumn{1}{r}{$10.2$} &  & \multicolumn{1}{r}{$16.5$} &
\multicolumn{1}{r}{$11.0$} & \multicolumn{1}{r}{$9.8$}\\
& \multicolumn{1}{c}{$-1$} & \multicolumn{1}{c}{$10$} &
\multicolumn{1}{r}{$25.5$} & \multicolumn{1}{r}{$15.9$} &
\multicolumn{1}{r}{$16.0$} &  & \multicolumn{1}{r}{$34.0$} &
\multicolumn{1}{r}{$26.1$} & \multicolumn{1}{r}{$16.2$} &  &
\multicolumn{1}{r}{$35.7$} & \multicolumn{1}{r}{$25.9$} &
\multicolumn{1}{r}{$25.8$} &  & \multicolumn{1}{r}{$42.3$} &
\multicolumn{1}{r}{$34.3$} & \multicolumn{1}{r}{$24.8$}\\
& \multicolumn{1}{c}{} & \multicolumn{1}{c}{$20$} & \multicolumn{1}{r}{$26.0$}
& \multicolumn{1}{r}{$16.5$} & \multicolumn{1}{r}{$16.7$} &  &
\multicolumn{1}{r}{$30.1$} & \multicolumn{1}{r}{$21.0$} &
\multicolumn{1}{r}{$15.9$} &  & \multicolumn{1}{r}{$36.5$} &
\multicolumn{1}{r}{$26.9$} & \multicolumn{1}{r}{$26.6$} &  &
\multicolumn{1}{r}{$40.0$} & \multicolumn{1}{r}{$30.9$} &
\multicolumn{1}{r}{$26.1$}\\
& \multicolumn{1}{c}{} & \multicolumn{1}{c}{$\underset{}{40}$} &
\multicolumn{1}{r}{$27.4$} & \multicolumn{1}{r}{$17.7$} &
\multicolumn{1}{r}{$17.9$} &  & \multicolumn{1}{r}{$29.3$} &
\multicolumn{1}{r}{$19.4$} & \multicolumn{1}{r}{$16.9$} &  &
\multicolumn{1}{r}{$37.8$} & \multicolumn{1}{r}{$28.4$} &
\multicolumn{1}{r}{$28.4$} &  & \multicolumn{1}{r}{$38.9$} &
\multicolumn{1}{r}{$30.0$} & \multicolumn{1}{r}{$27.6$}\\
& \multicolumn{1}{c}{$-2$} & \multicolumn{1}{c}{$10$} &
\multicolumn{1}{r}{$47.7$} & \multicolumn{1}{r}{$35.6$} &
\multicolumn{1}{r}{$35.7$} &  & \multicolumn{1}{r}{$57.0$} &
\multicolumn{1}{r}{$47.5$} & \multicolumn{1}{r}{$33.2$} &  &
\multicolumn{1}{r}{$58.4$} & \multicolumn{1}{r}{$48.5$} &
\multicolumn{1}{r}{$48.1$} &  & \multicolumn{1}{r}{$64.9$} &
\multicolumn{1}{r}{$57.4$} & \multicolumn{1}{r}{$45.7$}\\
& \multicolumn{1}{c}{} & \multicolumn{1}{c}{$20$} & \multicolumn{1}{r}{$51.6$}
& \multicolumn{1}{r}{$38.3$} & \multicolumn{1}{r}{$38.3$} &  &
\multicolumn{1}{r}{$56.3$} & \multicolumn{1}{r}{$44.0$} &
\multicolumn{1}{r}{$36.2$} &  & \multicolumn{1}{r}{$62.5$} &
\multicolumn{1}{r}{$52.1$} & \multicolumn{1}{r}{$52.3$} &  &
\multicolumn{1}{r}{$65.9$} & \multicolumn{1}{r}{$56.9$} &
\multicolumn{1}{r}{$51.4$}\\
& \multicolumn{1}{c}{} & \multicolumn{1}{c}{$\underset{}{40}$} &
\multicolumn{1}{r}{$52.5$} & \multicolumn{1}{r}{$39.9$} &
\multicolumn{1}{r}{$39.8$} &  & \multicolumn{1}{r}{$54.8$} &
\multicolumn{1}{r}{$43.0$} & \multicolumn{1}{r}{$39.1$} &  &
\multicolumn{1}{r}{$63.9$} & \multicolumn{1}{r}{$53.6$} &
\multicolumn{1}{r}{$53.8$} &  & \multicolumn{1}{r}{$64.9$} &
\multicolumn{1}{r}{$55.5$} & \multicolumn{1}{r}{$52.8$}\\
& \multicolumn{1}{c}{$-4$} & \multicolumn{1}{c}{$10$} &
\multicolumn{1}{r}{$84.9$} & \multicolumn{1}{r}{$75.6$} &
\multicolumn{1}{r}{$75.6$} &  & \multicolumn{1}{r}{$88.2$} &
\multicolumn{1}{r}{$82.5$} & \multicolumn{1}{r}{$71.6$} &  &
\multicolumn{1}{r}{$90.1$} & \multicolumn{1}{r}{$84.5$} &
\multicolumn{1}{r}{$84.3$} &  & \multicolumn{1}{r}{$91.9$} &
\multicolumn{1}{r}{$87.9$} & \multicolumn{1}{r}{$81.2$}\\
& \multicolumn{1}{c}{} & \multicolumn{1}{c}{$20$} & \multicolumn{1}{r}{$90.5$}
& \multicolumn{1}{r}{$82.9$} & \multicolumn{1}{r}{$82.7$} &  &
\multicolumn{1}{r}{$91.5$} & \multicolumn{1}{r}{$85.5$} &
\multicolumn{1}{r}{$80.2$} &  & \multicolumn{1}{r}{$94.2$} &
\multicolumn{1}{r}{$90.3$} & \multicolumn{1}{r}{$90.0$} &  &
\multicolumn{1}{r}{$94.4$} & \multicolumn{1}{r}{$91.3$} &
\multicolumn{1}{r}{$88.8$}\\
& \multicolumn{1}{c}{} & \multicolumn{1}{c}{$\underset{}{40}$} &
\multicolumn{1}{r}{$91.7$} & \multicolumn{1}{r}{$85.4$} &
\multicolumn{1}{r}{$85.3$} &  & \multicolumn{1}{r}{$92.5$} &
\multicolumn{1}{r}{$87.0$} & \multicolumn{1}{r}{$84.7$} &  &
\multicolumn{1}{r}{$95.3$} & \multicolumn{1}{r}{$92.2$} &
\multicolumn{1}{r}{$92.0$} &  & \multicolumn{1}{r}{$95.8$} &
\multicolumn{1}{r}{$92.7$} & \multicolumn{1}{r}{$91.7$}\\\hline
\multicolumn{1}{c}{$t_{3}$} & \multicolumn{1}{c}{$0$} &
\multicolumn{1}{c}{$10$} & $7.3$ & $3.7$ & $3.8$ &  & $15.6$ & $10.8$ & $5.8$
&  & $12.0$ & $7.3$ & $7.2$ &  & $21.5$ & $15.8$ & $10.2$\\
& \multicolumn{1}{c}{} & \multicolumn{1}{c}{$20$} & $7.5$ & $4.1$ & $4.2$ &  &
$13.2$ & $8.1$ & $5.6$ &  & $12.7$ & $7.6$ & $7.9$ &  & $19.0$ & $13.4$ &
$10.9$\\
& \multicolumn{1}{c}{} & \multicolumn{1}{c}{$\underset{}{40}$} & $7.5$ & $3.8$
& $3.9$ &  & $10.5$ & $5.7$ & $4.9$ &  & $12.8$ & $7.8$ & $7.8$ &  & $16.6$ &
$10.8$ & $9.6$\\
& \multicolumn{1}{c}{$-1$} & \multicolumn{1}{c}{$10$} & $20.9$ & $12.0$ &
$11.9$ &  & $39.4$ & $30.6$ & $19.8$ &  & $31.7$ & $21.4$ & $21.3$ &  & $47.7$
& $39.8$ & $29.5$\\
& \multicolumn{1}{c}{} & \multicolumn{1}{c}{$20$} & $23.3$ & $13.1$ & $13.3$ &
& $35.2$ & $25.3$ & $19.3$ &  & $34.2$ & $23.9$ & $24.0$ &  & $45.0$ & $36.3$
& $31.1$\\
& \multicolumn{1}{c}{} & \multicolumn{1}{c}{$\underset{}{40}$} & $24.6$ &
$14.7$ & $14.7$ &  & $31.8$ & $21.4$ & $19.2$ &  & $35.6$ & $25.3$ & $25.3$ &
& $42.3$ & $32.9$ & $30.5$\\
& \multicolumn{1}{c}{$-2$} & \multicolumn{1}{c}{$10$} & $47.5$ & $32.2$ &
$32.2$ &  & $65.2$ & $56.6$ & $42.8$ &  & $60.3$ & $47.7$ & $47.6$ &  & $72.7$
& $65.4$ & $55.1$\\
& \multicolumn{1}{c}{} & \multicolumn{1}{c}{$20$} & $51.4$ & $36.7$ & $37.0$ &
& $63.7$ & $52.3$ & $45.1$ &  & $64.4$ & $52.4$ & $52.4$ &  & $72.5$ & $63.9$
& $59.1$\\
& \multicolumn{1}{c}{} & \multicolumn{1}{c}{$\underset{}{40}$} & $52.8$ &
$38.1$ & $38.3$ &  & $60.8$ & $47.8$ & $44.6$ &  & $65.6$ & $53.9$ & $53.7$ &
& $70.9$ & $61.7$ & $58.9$\\
& \multicolumn{1}{c}{$-4$} & \multicolumn{1}{c}{$10$} & $87.7$ & $78.1$ &
$77.9$ &  & $91.3$ & $86.9$ & $78.6$ &  & $92.1$ & $87.3$ & $87.2$ &  & $94.1$
& $91.2$ & $85.9$\\
& \multicolumn{1}{c}{} & \multicolumn{1}{c}{$20$} & $91.8$ & $85.0$ & $84.9$ &
& $92.6$ & $88.1$ & $84.2$ &  & $95.1$ & $91.6$ & $91.6$ &  & $95.3$ & $92.8$
& $91.0$\\
& \multicolumn{1}{c}{} & \multicolumn{1}{c}{$\underset{}{40}$} & $93.2$ &
$87.7$ & $87.6$ &  & $93.2$ & $88.2$ & $86.8$ &  & $96.1$ & $93.3$ & $93.3$ &
& $96.0$ & $93.3$ & $92.5$\\\hline
\multicolumn{1}{c}{$\chi_{1}^{2}$} & \multicolumn{1}{c}{$0$} &
\multicolumn{1}{c}{$10$} & $8.3$ & $4.7$ & $4.7$ &  & $16.0$ & $10.7$ & $5.8$
&  & $12.6$ & $8.0$ & $8.0$ &  & $21.5$ & $16.2$ & $9.9$\\
& \multicolumn{1}{c}{} & \multicolumn{1}{c}{$20$} & $8.5$ & $4.9$ & $4.9$ &  &
$12.2$ & $7.0$ & $5.0$ &  & $13.5$ & $8.6$ & $8.6$ &  & $18.1$ & $12.4$ &
$9.8$\\
& \multicolumn{1}{c}{} & \multicolumn{1}{c}{$\underset{}{40}$} & $9.2$ & $4.9$
& $4.8$ &  & $10.9$ & $6.1$ & $5.3$ &  & $14.8$ & $9.7$ & $9.5$ &  & $17.1$ &
$11.2$ & $10.1$\\
& \multicolumn{1}{c}{$-1$} & \multicolumn{1}{c}{$10$} & $21.1$ & $12.6$ &
$12.6$ &  & $41.9$ & $33.2$ & $22.5$ &  & $30.9$ & $21.7$ & $21.2$ &  & $50.1$
& $42.0$ & $31.9$\\
& \multicolumn{1}{c}{} & \multicolumn{1}{c}{$20$} & $23.4$ & $14.3$ & $14.3$ &
& $35.1$ & $25.1$ & $19.7$ &  & $33.6$ & $24.0$ & $24.1$ &  & $45.2$ & 35.9 &
$31.2$\\
& \multicolumn{1}{c}{} & \multicolumn{1}{c}{$\underset{}{40}$} & $25.5$ &
$15.8$ & $15.9$ &  & $31.7$ & $21.2$ & $19.1$ &  & $36.2$ & $26.6$ & $26.7$ &
& $42.2$ & $32.7$ & $30.4$\\
& \multicolumn{1}{c}{$-2$} & \multicolumn{1}{c}{$10$} & $46.9$ & $31.3$ &
$31.5$ &  & $65.2$ & $57.2$ & $45.3$ &  & $60.6$ & $47.6$ & $47.6$ &  & $72.0$
& $65.4$ & $55.8$\\
& \multicolumn{1}{c}{} & \multicolumn{1}{c}{$20$} & $51.2$ & $36.3$ & $36.4$ &
& $62.2$ & $51.4$ & $44.3$ &  & $64.3$ & $52.4$ & $52.5$ &  & $71.3$ & $62.9$
& $57.9$\\
& \multicolumn{1}{c}{} & \multicolumn{1}{c}{$\underset{}{40}$} & $53.9$ &
$39.2$ & $39.1$ &  & $59.4$ & $46.9$ & $43.9$ &  & $65.2$ & $55.1$ & $54.9$ &
& $69.9$ & $60.4$ & $57.8$\\
& \multicolumn{1}{c}{$-4$} & \multicolumn{1}{c}{$10$} & $87.2$ & $78.5$ &
$78.3$ &  & $88.8$ & $84.3$ & $76.6$ &  & $91.5$ & $86.6$ & $86.4$ &  & $91.8$
& $88.6$ & $83.2$\\
& \multicolumn{1}{c}{} & \multicolumn{1}{c}{$20$} & $91.1$ & $84.7$ & $84.7$ &
& $90.6$ & $84.6$ & $80.4$ &  & $94.2$ & $91.0$ & $90.8$ &  & $93.9$ & $90.5$
& $88.4$\\
& \multicolumn{1}{c}{} & \multicolumn{1}{c}{$\underset{}{40}$} & $92.6$ &
$86.8$ & $86.8$ &  & $91.8$ & $86.7$ & $85.2$ &  & $95.6$ & $92.7$ & $92.5$ &
& $94.8$ & $92.0$ & $91.0$\\\hline
\end{tabular}%
\end{center}%

Notes: $\hat{p}_{n}$ denotes the standard bootstrap; $\tilde{p}_{n,p}$ and
$\tilde{p}_{n,d}$ denote the modified bootstrap using the plug-in and the
double bootstrap methods, respectively. The parametric bootstrap scheme, where
$\varepsilon_{t}^{\ast}\sim$~i.i.d.$N(0,1)$, is denoted as \textquotedblleft
par.\textquotedblright\ and the non-parametric bootstrap scheme, where
$\varepsilon_{t}^{\ast}$ is re-sampled independently from the long regression
(centered) residuals, is denoted as \textquotedblleft
non-par.\textquotedblright\ The $\varepsilon_{t}$'s are i.i.d.\ draws from
(standardized) $N$, $t_{3}$, and $\chi_{1}^{2}$ distributions. The parameter
$a$ denotes the drift under the local alternative $\beta_{0}=\bar{\beta
}+an^{-1/2}$. Results are based on 10,000 Monte Carlo simulations and $B=999$
bootstrap replications for each level.%
\end{footnotesize}%
\end{table}%

\subsubsection{Extension to the pairs bootstrap}

\label{sec:MA-pairs}

In addition to Assumption~\ref{Assn-MA} we also impose the following conditions.%

\renewcommand{\thecondition}{MA$_{\rm{2}}$}%

\begin{condition}
\label{Assn-MA2}With $w_{t}:=(x_{t},z_{t})^{\prime}$ it holds that
(i)~$\sup_{t}E\left\Vert w_{t}\right\Vert ^{4}<\infty$, $E\varepsilon_{t}%
^{4}<\infty$; (ii)~$n^{-1}\sum_{t=1}^{n}x_{t}^{2}\varepsilon_{t}%
^{2}\rightarrow_{p}\sigma^{2}\Sigma_{xx}$, $n^{-1}\sum_{t=1}^{n}x_{t}^{2}%
w_{t}w_{t}^{\prime}\rightarrow_{p}\Sigma_{r}>0$, and $n^{-1}\sum_{t=1}%
^{n}x_{t}^{2}w_{t}\varepsilon_{t}\rightarrow_{p}0$.
\end{condition}

\smallskip

\noindent\textsc{Proof of Lemma \ref{lemma mod av pairs bootstrap}. }We first
prove that%
\begin{align}
&  S_{W^{\ast}W^{\ast}}-S_{WW}\overset{p^{\ast}}{\rightarrow}_{p}%
0,\label{pairs LLN}\\
S_{n}^{\ast}  &  :=\left(
\begin{array}
[c]{c}%
n^{1/2}S_{x^{\ast}\varepsilon^{\ast}}\\
n^{1/2}(S_{x^{\ast}z^{\ast}}-S_{xz})\\
n^{1/2}(S_{x^{\ast}x^{\ast}}-S_{xx})
\end{array}
\right)  \overset{d^{\ast}}{\rightarrow}_{p}N(0,\Sigma_{s}),\quad\Sigma
_{s}=\left(
\begin{array}
[c]{cc}%
\sigma^{2}\Sigma_{xx} & 0\\
0 & \Sigma_{r}%
\end{array}
\right)  . \label{pairs CLT}%
\end{align}
Here, (\ref{pairs LLN}) follows by straightforward application of Chebyshev's LLN.

To prove (\ref{pairs CLT}), we first compute the mean and variance of
$S_{n}^{\ast}$. Note that the mean of $S_{n}^{\ast}$ is zero by construction;
for example, $E^{\ast}(n^{1/2}S_{x^{\ast}\varepsilon^{\ast}})=n^{-1/2}%
\sum_{t=1}^{n}E^{\ast}(x_{t}^{\ast}\varepsilon_{t}^{\ast})=n^{1/2}%
S_{x\hat{\varepsilon}}=0$ by the OLS\ first-order condition. In addition,
\[
\operatorname{Var}^{\ast}(n^{1/2}S_{x^{\ast}\varepsilon^{\ast}})=n^{-1}%
\sum_{t=1}^{n}E^{\ast}(x_{t}^{\ast2}\varepsilon_{t}^{\ast2})=n^{-1}\sum
_{t=1}^{n}x_{t}^{2}\hat{\varepsilon}_{t}^{2}\overset{p}{\rightarrow}\sigma
^{2}\Sigma_{xx}%
\]
under Assumptions~\ref{Assn-MA} and~\ref{Assn-MA2}. Similarly, letting
\[
\left(
\begin{array}
[c]{c}%
n^{1/2}(S_{x^{\ast}z^{\ast}}-S_{xz})\\
n^{1/2}(S_{x^{\ast}x^{\ast}}-S_{xx})
\end{array}
\right)  =n^{1/2}(S_{x^{\ast}W^{\ast}}-S_{xW}),
\]
we find that%
\begin{align*}
\operatorname{Var}^{\ast}(n^{1/2}\left(  S_{x^{\ast}W^{\ast}}-S_{xW}\right)
)  &  =n^{-1}\sum_{t=1}^{n}(x_{t}w_{t}-E^{\ast}(x_{t}^{\ast}w_{t}^{\ast
}))(x_{t}w_{t}-E^{\ast}(x_{t}^{\ast}w_{t}^{\ast}))^{\prime}\\
&  =n^{-1}\sum_{t=1}^{n}x_{t}^{2}w_{t}w_{t}^{\prime}-S_{xW}S_{Wx}%
\overset{p}{\rightarrow}\Sigma_{r}-\Sigma_{xW}\Sigma_{Wx}.
\end{align*}
Note also that the covariance between $n^{1/2}S_{x^{\ast}\varepsilon^{\ast}}$
and $n^{1/2}(S_{x^{\ast}W^{\ast}}-S_{xW})$ is zero because%
\begin{align*}
E^{\ast}(nS_{x^{\ast}\varepsilon^{\ast}}S_{x^{\ast}W^{\ast}})  &
=n^{-1}E^{\ast}\left(  \sum_{t=1}^{n}x_{t}^{\ast}\varepsilon_{t}^{\ast}%
\sum_{s=1}^{n}x_{s}^{\ast}w_{s}^{\ast}\right)  =n^{-1}E^{\ast}\left(
\sum_{t=1}^{n}x_{t}^{\ast2}w_{t}^{\ast}\varepsilon_{t}^{\ast}\right) \\
&  =E^{\ast}(x_{t}^{\ast2}w_{t}^{\ast}\varepsilon_{t}^{\ast})=n^{-1}\sum
_{t=1}^{n}x_{t}^{2}w_{t}\hat{\varepsilon}_{t}\overset{p}{\rightarrow}0
\end{align*}
by Assumption~\ref{Assn-MA2}(ii). Thus, we have shown that $E^{\ast}%
(S_{n}^{\ast})=0$ and $E^{\ast}(S_{n}^{\ast}S_{n}^{\ast\prime})\rightarrow
_{p}\Sigma_{s}$. The result (\ref{pairs CLT}) now follows because the stated
moment conditions imply the Lindeberg condition by standard arguments.

Next we can write%
\[
T_{n}^{\ast}-\hat{B}_{n}=n^{1/2}S_{x^{\ast}x^{\ast}}^{-1}S_{x^{\ast
}\varepsilon^{\ast}}+B_{n}^{\ast}-\hat{B}_{n},
\]
where
\[
B_{n}^{\ast}-\hat{B}_{n}=(S_{x^{\ast}x^{\ast}}^{-1}S_{x^{\ast}z^{\ast}}%
-S_{xx}^{-1}S_{xz})n^{1/2}\hat{\delta}_{n}.
\]
Adding and subtracting appropriately, we can write this difference as%
\[
B_{n}^{\ast}-\hat{B}_{n}=n^{1/2}(S_{x^{\ast}x^{\ast}}^{-1}S_{x^{\ast}z^{\ast}%
}-S_{xx}^{-1}S_{xz})\delta+(S_{x^{\ast}x^{\ast}}^{-1}S_{x^{\ast}z^{\ast}%
}-S_{xx}^{-1}S_{xz})n^{1/2}(\hat{\delta}_{n}-\delta),
\]
where $n^{1/2}(\hat{\delta}_{n}-\delta)$ is $O_{p}(1)$ by a central limit
theorem and $S_{x^{\ast}x^{\ast}}^{-1}S_{x^{\ast}z^{\ast}}-S_{xx}^{-1}%
S_{xz}=o_{p^{\ast}}(1)$, in probability, by (\ref{pairs LLN}). The first term
in $B_{n}^{\ast}-\hat{B}_{n}$ can be written as%
\begin{align*}
&  S_{x^{\ast}x^{\ast}}^{-1}n^{1/2}(S_{x^{\ast}z^{\ast}}-S_{xz})\delta
-S_{x^{\ast}x^{\ast}}^{-1}S_{xx}^{-1}(S_{x^{\ast}x^{\ast}}-S_{xx}%
)n^{1/2}S_{xz}\delta\\
&  =\delta(\Sigma_{xx}^{-1},-\Sigma_{xx}^{-2}\Sigma_{xz})\left(
\begin{array}
[c]{c}%
n^{1/2}(S_{x^{\ast}z^{\ast}}-S_{xz})\\
n^{1/2}(S_{x^{\ast}x^{\ast}}-S_{xx})
\end{array}
\right)  +o_{p^{\ast}}(1),
\end{align*}
in probability, by application of (\ref{pairs LLN}) and
Assumption~\ref{Assn-MA}(ii).\ It follows that%
\begin{align*}
T_{n}^{\ast}-\hat{B}_{n}  &  =S_{x^{\ast}x^{\ast}}^{-1}n^{1/2}S_{x\varepsilon
}^{\ast}+\delta(\Sigma_{xx}^{-1},-\Sigma_{xx}^{-2}\Sigma_{xz})\left(
\begin{array}
[c]{c}%
n^{1/2}(S_{x^{\ast}z^{\ast}}-S_{xz})\\
n^{1/2}(S_{x^{\ast}x^{\ast}}-S_{xx})
\end{array}
\right)  +o_{p^{\ast}}(1)\\
&  =(\Sigma_{xx}^{-1},\Sigma_{xx}^{-1}\delta,-\Sigma_{xx}^{-2}\Sigma
_{xz}\delta)S_{n}^{\ast}+o_{p^{\ast}}(1),
\end{align*}
in probability. The required result now follows from (\ref{pairs CLT}) because%
\begin{align*}
&  (\Sigma_{xx}^{-1},\Sigma_{xx}^{-1}\delta,-\Sigma_{xx}^{-2}\Sigma_{xz}%
\delta)\left(
\begin{array}
[c]{cc}%
\Sigma_{s} & 0\\
0 & \Sigma_{r}%
\end{array}
\right)  (\Sigma_{xx}^{-1},\Sigma_{xx}^{-1}\delta,-\Sigma_{xx}^{-2}\Sigma
_{xz}\delta)^{\prime}\\
&  =\Sigma_{xx}^{-1}\Sigma_{s}\Sigma_{xx}^{-1}+d_{r}(\delta)^{\prime}%
\Sigma_{r}d_{r}(\delta)=v^{2}+\kappa^{2},
\end{align*}
which completes the proof upon noting that $\Sigma_{s}=\sigma^{2}\Sigma_{xx}$
implies $v^{2}=\sigma^{2}\Sigma_{xx}^{-1}$.\hfill$\square$

\subsection{Ridge regression}

\label{sec:ridge-details}

\subsubsection{Assumptions and notation}

As in Fu and Knight (2000) we assume the following.%

\renewcommand{\thecondition}{RE}%

\begin{condition}
\label{Assn Ridge}(i)~$\varepsilon_{t}\sim$~i.i.d.$(0,\sigma^{2})$;
(ii)~$\max_{t=1,\dots,n}x_{t}^{\prime}x_{t}=o(n)$; (iii)~$S_{xx}$ is
nonsingular for any $n$ and converges to a positive definite matrix,
$\Sigma_{xx}$; (iv)~$\theta=\delta n^{-1/2}$; and (v)~$n^{-1}c_{n}\rightarrow
c_{0}\geq0$.
\end{condition}

For the bootstrap we will also need the following.%

\renewcommand{\thecondition}{RE$_2$}%

\begin{condition}
\label{Assn Ridge BS}Assumption~\ref{Assn Ridge} holds with (ii) replaced by
(ii')~$\max_{t=1,\dots,n}x_{t}^{\prime}x_{t}=o(n^{1/2})$ and with the
additional condition (vi)~$E\varepsilon_{t}^{4}<\infty$.
\end{condition}

Finally, we define
\[
V=\sigma^{2}\left(
\begin{array}
[c]{cc}%
g^{\prime}\tilde{\Sigma}_{xx}^{-1}\Sigma_{xx}\tilde{\Sigma}_{xx}^{-1}g &
-c_{0}g^{\prime}\tilde{\Sigma}_{xx}^{-1}\tilde{\Sigma}_{xx}^{-1}g\\
-c_{0}g^{\prime}\tilde{\Sigma}_{xx}^{-1}\tilde{\Sigma}_{xx}^{-1}g & c_{0}%
^{2}g^{\prime}\tilde{\Sigma}_{xx}^{-1}\Sigma_{xx}^{-1}\tilde{\Sigma}_{xx}%
^{-1}g
\end{array}
\right)
\]
where $v^{2}:=v_{11}$, and it holds that%
\begin{equation}
m^{2}:=\frac{v_{11}+v_{22}-2v_{12}}{v_{11}}=\frac{g^{\prime}\Sigma_{xx}^{-1}%
g}{g^{\prime}\tilde{\Sigma}_{xx}^{-1}\Sigma_{xx}\tilde{\Sigma}_{xx}^{-1}g},
\label{eq ridge app m}%
\end{equation}
where the last equality is derived in the proof of Lemma~\ref{Lemma ridge}.

\subsubsection{Proofs of lemmas}

\noindent\textsc{Proof of Lemma~\ref{Lemma ridge} and derivation of
(\ref{eq ridge app m}).} The result follows by showing that%
\begin{equation}
\left(
\begin{array}
[c]{c}%
T_{n}-B_{n}\\
\hat{B}_{n}-B_{n}%
\end{array}
\right)  =\left(
\begin{array}
[c]{c}%
g^{\prime}\tilde{S}_{xx}^{-1}\\
-c_{n}n^{-1}g^{\prime}\tilde{S}_{xx}^{-1}S_{xx}^{-1}%
\end{array}
\right)  n^{1/2}S_{x\varepsilon}\overset{d}{\rightarrow}\left(
\begin{array}
[c]{c}%
\xi_{1}\\
\xi_{2}%
\end{array}
\right)  \sim N(0,V),\text{\quad}V=(v_{ij}),
\label{eq joint convergence ridge}%
\end{equation}
and that%
\begin{equation}
T_{n}^{\ast}-B_{n}^{\ast}=T_{n}^{\ast}-\hat{B}_{n}+o_{p^{\ast}}%
(1)\overset{d^{\ast}}{\rightarrow}_{p}N(0,v^{2}). \label{ridge bs}%
\end{equation}

To prove (\ref{eq joint convergence ridge}) we first notice that, since
$c_{n}n^{-1}\rightarrow c_{0}$, under Assumption \ref{Assn Ridge}\ we have
that $n^{1/2}S_{x\varepsilon}\rightarrow_{d}N(0,\sigma^{2}\Sigma_{xx})$ and
hence
\begin{align}
\left(
\begin{array}
[c]{c}%
T_{n}-B_{n}\\
\hat{B}_{n}-B_{n}%
\end{array}
\right)   &  =(I_{2}\otimes g^{\prime}\tilde{S}_{xx}^{-1})\left(
\begin{array}
[c]{c}%
I_{p}\\
-n^{-1}c_{n}S_{xx}^{-1}%
\end{array}
\right)  n^{1/2}S_{x\varepsilon}\nonumber\\
&  \overset{d}{\rightarrow}(I_{2}\otimes g^{\prime}\tilde{\Sigma}_{xx}%
^{-1})N\left(  0,\sigma^{2}\left(
\begin{array}
[c]{cc}%
\Sigma_{xx} & -c_{0}I_{p}\\
-c_{0}I_{p} & c_{0}^{2}\Sigma_{xx}^{-1}%
\end{array}
\right)  \right)  \sim N(0,V),\label{eq joint Z1n and Z2n}\\
V  &  =\sigma^{2}\left(
\begin{array}
[c]{cc}%
g^{\prime}\tilde{\Sigma}_{xx}^{-1}\Sigma_{xx}\tilde{\Sigma}_{xx}^{-1}g &
-c_{0}g^{\prime}\tilde{\Sigma}_{xx}^{-1}\tilde{\Sigma}_{xx}^{-1}g\\
-c_{0}g^{\prime}\tilde{\Sigma}_{xx}^{-1}\tilde{\Sigma}_{xx}^{-1}g & c_{0}%
^{2}g^{\prime}\tilde{\Sigma}_{xx}^{-1}\Sigma_{xx}^{-1}\tilde{\Sigma}_{xx}%
^{-1}g
\end{array}
\right)  .\nonumber
\end{align}
This immediately implies that $m^{2}$ in (\ref{eq ridge app m}) is given by%
\begin{equation}
m^{2}=\frac{g^{\prime}\tilde{\Sigma}_{xx}^{-1}\Sigma_{xx}\tilde{\Sigma}%
_{xx}^{-1}g+2c_{0}g^{\prime}\tilde{\Sigma}_{xx}^{-1}\tilde{\Sigma}_{xx}%
^{-1}g+c_{0}^{2}g^{\prime}\tilde{\Sigma}_{xx}^{-1}\Sigma_{xx}^{-1}%
\tilde{\Sigma}_{xx}^{-1}g}{g^{\prime}\tilde{\Sigma}_{xx}^{-1}\Sigma_{xx}%
\tilde{\Sigma}_{xx}^{-1}g}. \label{eq ridge m2}%
\end{equation}
The numerator of $m^{2}$ in (\ref{eq ridge m2}) can be written as%
\[
g^{\prime}\tilde{\Sigma}_{xx}^{-1}(\Sigma_{xx}+2c_{0}I_{p}+c_{0}^{2}%
\Sigma_{xx}^{-1})\tilde{\Sigma}_{xx}^{-1}g=g^{\prime}\tilde{\Sigma}_{xx}%
^{-1}(\tilde{\Sigma}_{xx}\Sigma_{xx}^{-1}\tilde{\Sigma}_{xx})\tilde{\Sigma
}_{xx}^{-1}g=g^{\prime}\Sigma_{xx}^{-1}g,
\]
and hence (\ref{eq ridge app m}) follows.

To prove (\ref{ridge bs}) we note that $T_{n}^{\ast}-\hat{B}_{n}=\xi
_{1,n}^{\ast}+B_{n}^{\ast}-\hat{B}_{n}$, where $\xi_{1,n}^{\ast}%
:=n^{1/2}g^{\prime}\tilde{S}_{x^{\ast}x^{\ast}}^{-1}S_{x^{\ast}\varepsilon
^{\ast}}$ and%
\begin{align*}
B_{n}^{\ast}-\hat{B}_{n}  &  =-c_{n}n^{-1/2}g^{\prime}\tilde{S}_{x^{\ast
}x^{\ast}}^{-1}\hat{\theta}_{n}+c_{n}n^{-1/2}g^{\prime}\tilde{S}_{xx}^{-1}%
\hat{\theta}_{n}\\
&  =-c_{n}n^{-1}g^{\prime}(\tilde{S}_{x^{\ast}x^{\ast}}^{-1}-\tilde{S}%
_{xx}^{-1})n^{1/2}(\hat{\theta}_{n}-\theta)-c_{n}n^{-1}g^{\prime}(\tilde
{S}_{x^{\ast}x^{\ast}}^{-1}-\tilde{S}_{xx}^{-1})\delta,
\end{align*}
such that $B_{n}^{\ast}-\hat{B}_{n}\overset{p^{\ast}}{\rightarrow}_{p}0$ if
$\tilde{S}_{x^{\ast}x^{\ast}}^{-1}-\tilde{S}_{xx}^{-1}\overset{p^{\ast
}}{\rightarrow}_{p}0$. Because $||\tilde{S}_{xx}^{-1}||=O(1)$ under the stated
assumptions, it follows that $||\tilde{S}_{x^{\ast}x^{\ast}}^{-1}-\tilde
{S}_{xx}^{-1}||$ has the same rate as $||\tilde{S}_{x^{\ast}x^{\ast}}%
-\tilde{S}_{xx}||$. Thus, $\tilde{S}_{x^{\ast}x^{\ast}}-\tilde{S}%
_{xx}=S_{x^{\ast}x^{\ast}}-S_{xx}=n^{-1}\sum_{t=1}^{n}x_{t}^{\ast}x_{t}%
^{\ast\prime}-E^{\ast}(x_{t}^{\ast}x_{t}^{\ast\prime})\overset{p^{\ast
}}{\rightarrow}_{p}0$ by a straightforward application of Chebyshev's LLN
using that $\max_{t}x_{t}^{\prime}x_{t}=o(n^{1/2})$ by
Assumption~\ref{Assn Ridge BS}(ii').

The proof is completed by showing that $\xi_{1,n}^{\ast}$ satisfies the
bootstrap central limit theorem. By the above results it holds that $\xi
_{1,n}^{\ast}=n^{1/2}g^{\prime}\tilde{\Sigma}_{xx}^{-1}S_{x^{\ast}%
\varepsilon^{\ast}}+o_{p^{\ast}}(1)$, so it is only required to analyze the
term $n^{1/2}g^{\prime}\tilde{\Sigma}_{xx}^{-1}S_{x^{\ast}\varepsilon^{\ast}%
}=n^{1/2}S_{\tilde{x}^{\ast}\varepsilon^{\ast}}$, where $\tilde{x}_{t}^{\ast
}:=g^{\prime}\tilde{\Sigma}_{xx}^{-1}x_{t}^{\ast}$. First, we have $E^{\ast
}(n^{1/2}S_{\tilde{x}^{\ast}\varepsilon^{\ast}})=g^{\prime}\tilde{\Sigma}%
_{xx}^{-1}E^{\ast}(n^{1/2}S_{x^{\ast}\varepsilon^{\ast}})=n^{1/2}g^{\prime
}\tilde{\Sigma}_{xx}^{-1}S_{x\hat{\varepsilon}}=0$. Second, with $\tilde
{x}_{t}:=g^{\prime}\tilde{\Sigma}_{xx}^{-1}x_{t}$,
\begin{align*}
\operatorname{Var}^{\ast}(n^{1/2}S_{\tilde{x}^{\ast}\varepsilon^{\ast}})  &
=n^{-1}\sum_{t=1}^{n}\tilde{x}_{t}^{2}\hat{\varepsilon}_{t}^{2}=n^{-1}%
\sum_{t=1}^{n}\tilde{x}_{t}^{2}(\hat{\varepsilon}_{t}^{2}-\sigma^{2}%
+\sigma^{2})\\
&  =\sigma^{2}g^{\prime}\tilde{\Sigma}_{xx}^{-1}\Sigma_{xx}\tilde{\Sigma}%
_{xx}^{-1}g+n^{-1}\sum_{t=1}^{n}\tilde{x}_{t}^{2}(\varepsilon_{t}^{2}%
-\sigma^{2})+o_{p}(1).
\end{align*}
Because $\varepsilon_{t}$ is i.i.d.\ and $\tilde{x}_{t}^{2}$ is
non-stochastic, a sufficient condition for $n^{-1}\sum_{t=1}^{n}\tilde{x}%
_{t}^{2}(\varepsilon_{t}^{2}-\sigma^{2})\rightarrow_{p}0$ is that
$\lambda_{\min}(\sum_{t=1}^{n}\tilde{x}_{t}^{2})\rightarrow\infty$, where
$\lambda_{\min}(\cdot)$ denotes the minimum eigenvalue of the argument, and
this is implied by $n^{-1}\sum_{t=1}^{n}\tilde{x}_{t}^{2}\rightarrow
g^{\prime}\tilde{\Sigma}_{xx}^{-1}\Sigma_{xx}\tilde{\Sigma}_{xx}^{-1}g>0$.

Third, we check Lindeberg's condition, where we set $s_{n}^{2}:=nS_{\tilde
{x}\tilde{x}}$. For $\epsilon>0$ it holds that
\begin{align*}
\frac{1}{s_{n}^{2}}\sum_{t=1}^{n}E^{\ast}(\tilde{x}_{t}^{\ast2}\varepsilon
_{t}^{\ast2}\mathbb{I}_{\{|\tilde{x}_{t}^{\ast}\varepsilon_{t}^{\ast
}|>\epsilon s_{n}\}})  &  =\frac{1}{S_{\tilde{x}\tilde{x}}}E^{\ast}(\tilde
{x}_{t}^{\ast2}\varepsilon_{t}^{\ast2}\mathbb{I}_{\{(\tilde{x}_{t}^{\ast
}\varepsilon_{t}^{\ast})^{2}>\epsilon^{2}nS_{\tilde{x}\tilde{x}}\}})\\
&  \leq\frac{1}{\epsilon^{2}nS_{\tilde{x}\tilde{x}}^{2}}E^{\ast}(\tilde{x}%
_{t}^{\ast4}\varepsilon_{t}^{\ast4})\\
&  =\frac{1}{\epsilon^{2}n^{2}S_{\tilde{x}\tilde{x}}^{2}}\sum_{t=1}^{n}%
\tilde{x}_{t}^{4}\hat{\varepsilon}_{t}^{4}\leq\frac{n^{-1}\max_{t}\tilde
{x}_{t}^{4}}{\epsilon^{2}S_{\tilde{x}\tilde{x}}^{2}}\frac{1}{n}\sum_{t=1}%
^{n}\hat{\varepsilon}_{t}^{4}\overset{p}{\rightarrow}0
\end{align*}
because $n^{-1}\max_{t}\tilde{x}_{t}^{4}=o(1)$ and $\varepsilon_{t}$ has
bounded fourth-order moment.\hfill$\square$

\smallskip

\noindent\textsc{Proof of Lemma \ref{Lemma ridge dbs}. }The proof follows
closely the proofs of Lemma~\ref{Lemma ridge} and is omitted for
brevity.\hfill$\square$

\subsection{Nonparametric regression}

\label{sec:nonpar-details}

\subsubsection{Assumptions and notation}

We impose the following conditions.%

\renewcommand{\thecondition}{NP}%

\begin{condition}
\label{Assn NP}(i)~$\varepsilon_{t}\sim~$i.i.d.$(0,\sigma^{2})$;
(ii)~$E|\varepsilon_{t}|^{2+\delta}<\infty$; (iii)~$\beta:[0,1]\rightarrow
\mathbb{R}$ is three times continuously differentiable with bounded
derivatives; (iv) $K:\mathbb{R}\rightarrow\lbrack0,\infty)$ is symmetric and
satisfies $K(u)=0$ for all $u\not \in (-1,1)$, $\int K(u)du=1$, $\kappa
^{2}:=\int u^{2}K(u)du\neq0$, and $R_{K}:=\int K(u)^{2}du\in(0,\infty)$.
\end{condition}

Note that Assumption~\ref{Assn NP} allows for the most popular choices of
symmetric and truncated kernels.

To simplify notation, we define $k_{t}:=K((x_{t}-x)/h)$ and $k_{tj}%
:=K((x_{t}-x_{j})/h)$. We also define the variance matrix
\[
V:=\left(
\begin{matrix}
v^{2} & \omega_{12}-v^{2}\\
\omega_{12}-v^{2} & v^{2}+\omega_{22}-2\omega_{12}%
\end{matrix}
\right)  ,
\]
where $v^{2}:=\sigma^{2}R_{K}$, $\omega_{12}:=\sigma^{2}\int K(u)\int
K(s-u)K(s)dsdu$, and $\omega_{22}:=\sigma^{2}\int(\int K(s-u)K(s)ds)^{2}du$.

\subsubsection{Proofs of (\ref{np bias}) and lemmas}

Although it is well known (e.g., Li and Racine, 2007) that (\ref{np bias}) and
Assumption~\ref{Assn T} hold in this example, we give short proofs for completeness.

\noindent\textsc{Proof of (\ref{np bias}).} Under Assumption~\ref{Assn NP} we
obtain by Taylor expansion the following well-known result,%
\begin{align}
E\hat{\beta}_{h}(x)  &  =\frac{1}{nh}\sum_{t=1}^{n}k_{t}\beta(x_{t})=\int
K(u)\beta(x+uh)du+o((nh)^{-1})\nonumber\\
&  =\int K(u)\left(  \beta(x)+\beta^{\prime}(x)uh+\beta^{\prime\prime}%
(x)u^{2}h^{2}/2+o(h^{2})\right)  du+o((nh)^{-1})\nonumber\\
&  =\beta(x)+h^{2}\beta^{\prime\prime}(x)\kappa_{2}/2+o(h^{2})+o((nh)^{-1}),
\label{E1}%
\end{align}
where the last equality follows by $\int K(u)du=1$ and $\int uK(u)du=0$.
Setting the bandwidth as $h=cn^{-1/5}$ thus implies (\ref{np bias}). Note that
the limits of integration are $u\in((n^{-1}-x)/h,(1-x)/h)$, but for $n$
sufficiently large this is the same as $u\in(-1,1)$ because $K(u)=0$ for all
$u\not \in (-1,1)$. We use this property throughout the remaining proofs.

\smallskip

\noindent\textsc{Proof of Lemma~\ref{Lemma NP}.} First, we verify
Assumption~\ref{Assn T} by showing that $\xi_{1,n}:=T_{n}-B_{n}=(nh)^{-1/2}%
\sum_{t=1}^{n}k_{t}\varepsilon_{t}$ satisfies the central limit theorem.
Because $k_{t}\varepsilon_{t},t=1,\dots,n$, is a sequence of independent
random variables with mean zero and $\operatorname*{Var}(k_{t}\varepsilon
_{t})=k_{t}^{2}\sigma^{2}$, we have%
\begin{align*}
\operatorname*{Var}(\xi_{1,n})  &  =\frac{1}{nh}\sum_{t=1}^{n}k_{t}^{2}%
\sigma^{2}=\frac{\sigma^{2}}{h}\int K\left(  \frac{s-x}{h}\right)
^{2}ds+o((nh)^{-1})\\
&  =\frac{\sigma^{2}}{h}\int K(u)^{2}d(x+uh)+o((nh)^{-1})\rightarrow\sigma
^{2}R_{K}=v^{2}.
\end{align*}
Moreover, Lyapunov's condition holds because%
\begin{align}
(nh)^{-(1+\delta)}\sum_{t=1}^{n}E(k_{t}^{2+\delta}|\varepsilon_{t}|^{2+\delta
})  &  \leq c(nh)^{-(1+\delta)}\sum_{t=1}^{n}k_{t}^{2+\delta}\nonumber\\
&  \leq c(nh)^{-(1+\delta)}\sum_{t:|x_{t}-x|\leq h}k_{t}^{2+\delta}\leq
c(nh)^{-(1+\delta)}hn\rightarrow0. \label{lyapunov}%
\end{align}

Next, we verify Assumption~\ref{Assn BS}(i). Note that $T_{n}^{\ast}-\hat
{B}_{n}=(nh)^{-1/2}\sum_{t=1}^{n}k_{t}\varepsilon_{t}^{\ast}=:\xi_{1,n}^{\ast
}$, where, conditional on $D_{n}$, $\xi_{1,n}^{\ast}\sim N(0,\hat{\sigma}%
_{n}^{2}(nh)^{-1}\sum_{t=1}^{n}k_{t}^{2})$. Hence, the result follows from
$\hat{\sigma}_{n}^{2}\rightarrow_{p}\sigma^{2}$ and $(nh)^{-1}\sum_{t=1}%
^{n}k_{t}^{2}\rightarrow R_{K}$.

Finally, we verify Assumption~\ref{Assn BS}(ii). We first show that we can
write%
\begin{equation}
\hat{B}_{n}-B_{n}=\xi_{2,n}+o(1),\quad\xi_{2,n}:=\frac{1}{\sqrt{nh}}\sum
_{t=1}^{n}(\frac{1}{nh}\sum_{j=1}^{n}k_{j}k_{tj}-k_{t})\varepsilon_{t},
\label{joint np}%
\end{equation}
and then we show that%
\begin{equation}
\xi_{n}:=(\xi_{1,n},\xi_{2,n})^{\prime}\overset{d}{\rightarrow}N(0,V).
\label{np normal}%
\end{equation}

To prove (\ref{joint np}) we write%
\[
\hat{B}_{n}-B_{n}=(nh)^{1/2}\left(  \frac{1}{nh}\sum_{t=1}^{n}k_{t}(\hat
{\beta}_{h}(x_{t})-\beta(x_{t}))-(\hat{\beta}_{h}(x)-\beta(x))\right)  ,
\]
where%
\begin{align*}
\hat{\beta}_{h}(x)  &  =(nh)^{-1}\sum_{t=1}^{n}k_{t}\beta(x_{t})+(nh)^{-1}%
\sum_{t=1}^{n}k_{t}\varepsilon_{t},\\
k_{t}\hat{\beta}_{h}(x_{t})  &  =(nh)^{-1}\sum_{j=1}^{n}k_{t}k_{tj}\beta
(x_{j})+(nh)^{-1}\sum_{j=1}^{n}k_{t}k_{tj}\varepsilon_{j}.
\end{align*}
By reversing the summations and exploiting symmetry of $k_{tj}$, it
immediately follows that $\hat{B}_{n}-B_{n}=\xi_{2,n}+B_{2,n}-B_{1,n}$ with%
\begin{align*}
B_{1,n}(x)  &  :=B_{n}=(nh)^{1/2}\left(  (nh)^{-1}\sum_{t=1}^{n}k_{t}%
\beta(x_{t})-\beta(x)\right)  ,\\
B_{2,n}(x)  &  :=(nh)^{-1}\sum_{t=1}^{n}k_{t}B_{1,n}(x_{t})=(nh)^{-1/2}%
\sum_{t=1}^{n}k_{t}\left(  (nh)^{-1}\sum_{j=1}^{n}k_{tj}\beta(x_{j}%
)-\beta(x_{t})\right)  .
\end{align*}
By (\ref{E1}) we find%
\begin{align*}
B_{2,n}(x)-B_{1,n}(x)  &  =(nh)^{-1/2}\sum_{t=1}^{n}k_{t}\left(  h^{2}%
\beta^{\prime\prime}(x_{t})\kappa_{2}/2+o(h^{2})+o((nh)^{-1})\right) \\
&  \quad-(nh)^{1/2}h^{2}\beta^{\prime\prime}(x)\kappa_{2}/2+o(h^{2}%
)+o((nh)^{-1})\\
&  =(nh)^{1/2}\frac{\kappa_{2}}{2}h^{2}\left(  \frac{1}{nh}\sum_{t=1}^{n}%
k_{t}\beta^{\prime\prime}(x_{t})-\beta^{\prime\prime}(x)\right)
+o((nh)^{1/2}h^{2})+o((nh)^{-1/2}),
\end{align*}
where
\[
\frac{1}{nh}\sum_{t=1}^{n}k_{t}\beta^{\prime\prime}(x_{t})=\int\beta
^{\prime\prime}(x+uh)K(u)du+o((nh)^{-1})=\beta^{\prime\prime}%
(x)+O(h)+o((nh)^{-1})
\]
by first-order Taylor expansion, similar to (\ref{E1}), together with the
assumption of continuous and bounded~$\beta^{\prime\prime\prime}$. The result
now follows because $h=cn^{-1/5}$.

Finally, to prove (\ref{np normal}) we show that%
\begin{equation}
J_{n}=\frac{1}{\sqrt{nh}}\sum_{t=1}^{n}\left(
\begin{array}
[c]{c}%
k_{t}\\
(nh)^{-1}\sum_{j=1}^{n}k_{j}k_{tj}%
\end{array}
\right)  \varepsilon_{t}\overset{d}{\rightarrow}N(0,\Omega),\quad
\Omega:=(\omega_{ij})_{i,j=1,2}, \label{J normal}%
\end{equation}
from which the result follows by noting that $v^{2}=\omega_{11}$ and%
\[
\xi_{n}=\left[
\begin{array}
[c]{rr}%
1 & 0\\
-1 & 1
\end{array}
\right]  J_{n}.
\]
It is clear that $J_{n}$ has mean zero and independent increments.
Approximating summations by integrals, it can be straightforwardly shown that%
\[
\operatorname{Var}(J_{n})=\sigma^{2}\left[
\begin{matrix}
(nh)^{-1}\sum_{t=1}^{n}k_{t}^{2} & (nh)^{-2}\sum_{t,j=1}^{n}k_{t}k_{j}k_{tj}\\
(nh)^{-2}\sum_{t,j=1}^{n}k_{t}k_{j}k_{tj} & (nh)^{-1}\sum_{t=1}^{n}%
((nh)^{-1}\sum_{j=1}^{n}k_{j}k_{tj})^{2}%
\end{matrix}
\right]  \rightarrow\Omega.
\]
By the same proof as in (\ref{lyapunov}), we can show that the Lyapunov
condition is satisfied, and result (\ref{J normal}) follows.\hfill$\square$

\smallskip

\noindent\textsc{Proof of Lemma~\ref{Lemma NP dbs}.} We first verify
Assumption~\ref{Assn DBS}(i). We notice that $\bar{T}_{n}^{\ast\ast}-\bar
{B}_{n}^{\ast}=T_{n}^{\ast\ast}-\hat{B}_{n}^{\ast}=(nh)^{-1/2}\sum_{t=1}%
^{n}k_{t}\varepsilon_{i}^{\ast\ast}=:\xi_{1,n}^{\ast\ast}$, where, conditional
on $(D_{n},D_{n}^{\ast})$, $\xi_{1,n}^{\ast\ast}\sim N(0,\hat{\sigma}%
_{n}^{\ast2}(nh)^{-1}\sum_{t=1}^{n}k_{t}^{2})$. Hence, the result follows from
$\hat{\sigma}_{n}^{\ast2}\overset{p^{\ast}}{\rightarrow}_{p}\sigma^{2}$ and
$(nh)^{-1}\sum_{t=1}^{n}k_{t}^{2}\rightarrow R_{K}$.

Next, we verify Assumption~\ref{Assn DBS}(ii). We first write $T_{n}^{\ast
}-\bar{B}_{n}^{\ast}=T_{n}^{\ast}-\hat{B}_{n}-(\bar{B}_{n}^{\ast}-\hat{B}%
_{n})=\xi_{1,n}^{\ast}-(\bar{B}_{n}^{\ast}-\hat{B}_{n})$, where $\bar{B}%
_{n}^{\ast}-\hat{B}_{n}=\hat{B}_{n}^{\ast}-\hat{B}_{2,n}$. Recall $\hat
{B}_{2,n}:=(nh)^{-1}\sum_{t=1}^{n}k_{t}\hat{B}_{n}(x_{t})=(nh)^{-1/2}%
\sum_{t=1}^{n}k_{t}((nh)^{-1}\sum_{j=1}^{n}k_{tj}\hat{\beta}_{h}(x_{j}%
)-\hat{\beta}_{h}(x_{t}))$ and $\hat{B}_{n}^{\ast}:=(nh)^{1/2}((nh)^{-1}%
\sum_{t=1}^{n}k_{t}\hat{\beta}_{h}^{\ast}(x_{t})-\hat{\beta}_{h}^{\ast}(x))$,
where%
\begin{align*}
\hat{\beta}_{h}^{\ast}(x)  &  =(nh)^{-1}\sum_{t=1}^{n}k_{t}\hat{\beta}%
_{h}(x_{t})+(nh)^{-1}\sum_{t=1}^{n}k_{t}\varepsilon_{t}^{\ast},\\
k_{t}\hat{\beta}_{h}^{\ast}(x_{t})  &  =(nh)^{-1}\sum_{j=1}^{n}k_{t}k_{tj}%
\hat{\beta}_{h}(x_{j})+(nh)^{-1}\sum_{j=1}^{n}k_{t}k_{tj}\varepsilon_{j}%
^{\ast},
\end{align*}
so it follows that
\[
\bar{B}_{n}^{\ast}-\hat{B}_{n}=\hat{B}_{n}^{\ast}-\hat{B}_{2,n}=\xi
_{2,n}^{\ast},\quad\xi_{2,n}^{\ast}:=\frac{1}{\sqrt{nh}}\sum_{t=1}^{n}%
(\frac{1}{nh}\sum_{j=1}^{n}k_{j}k_{tj}-k_{t})\varepsilon_{t}^{\ast}.
\]
Thus, the proof is completed by showing that%
\begin{equation}
\xi_{n}^{\ast}:=(\xi_{1,n}^{\ast},\xi_{2,n}^{\ast})^{\prime}\overset{d^{\ast
}}{\rightarrow}_{p}N(0,V). \label{np normal dbs}%
\end{equation}
Conditional on $D_{n}$, it holds that $\xi_{n}^{\ast}\sim N(0,\hat{V}_{n})$,
where
\[
\hat{V}_{n}=\hat{\sigma}_{n}^{2}\frac{1}{nh}\sum_{t=1}^{n}\left[
\begin{array}
[c]{cc}%
k_{t}^{2} & k_{t}(\frac{1}{nh}\sum_{j=1}^{n}k_{j}k_{tj}-k_{t})\\
k_{t}(\frac{1}{nh}\sum_{j=1}^{n}k_{j}k_{tj}-k_{t}) & (\frac{1}{nh}\sum
_{j=1}^{n}k_{j}k_{tj}-k_{t})^{2}%
\end{array}
\right]  \overset{p}{\rightarrow}V
\]
by approximating the summations by integrals and using $\hat{\sigma}_{n}%
^{2}\rightarrow_{p}\sigma^{2}$. This proves (\ref{np normal dbs}) and hence
completes the proof of Lemma~\ref{Lemma NP dbs}.\hfill$\square$

\subsection{Inference under heavy tails}

\label{sec:heavy-details}

\noindent\textsc{Setup}. We consider a simple location model with heavy-tailed
data, thus demonstrating that our analysis applies to a non-Gaussian
asymptotic framework. Specifically, consider a sample of $n$ i.i.d.\ random
variables $\{y_{t}\}$. Interest is in inference on $\theta$ in the location
model%
\[
y_{t}=\theta+\varepsilon_{t},\quad E(\varepsilon_{t})=0,
\]
when the $\varepsilon_{t}$'s follow a symmetric, stable random variable
$S(\alpha)$ with tail index $\alpha\in(1,2)$ and the location parameter is
local to zero; i.e.,\ $\theta=n^{1/\alpha-1}c$.\footnote{The results in this
section can easily be generalized to the case where the $\varepsilon_{t}$'s
are not necessarily symmetric and/or are in the domain of attraction of a
stable law with index $\alpha\in(0,1)$, as in Cornea-Madeira and Davidson
(2015). Moreover, the results apply to the case of non-local $\theta$ as well;
i.e., $\theta\neq0$ fixed.} Under these assumptions, $E(|\varepsilon
_{t}|^{\alpha+\delta})=+\infty$ for any $\delta\geq0$; in particular,
$\varepsilon_{t}$ has infinite variance. Notice that $\theta$ is local of
order $n^{1/\alpha-1}$ rather than the usual $n^{-1/2}$ because of the slower
convergence rate of the OLS-type estimator when the variance of $\varepsilon
_{t}$ is infinite. We consider the biased estimator%
\[
\hat{\theta}_{n}:=\omega\bar{y}_{n},\quad\bar{y}_{n}:=n^{-1}\sum_{t=1}%
^{n}y_{t},
\]
where $\omega\in(0,1)$. In the finite variance case, this estimator improves
upon $\bar{y}_{n}$ in terms of MSE when $\theta$ is local to zero. It holds
that%
\begin{equation}
T_{n}:=n^{1-1/\alpha}(\hat{\theta}_{n}-\theta)=(\omega-1)c+\omega
n^{1-1/\alpha}\bar{\varepsilon}_{n}\sim B+\omega S(\alpha)
\label{eq InfV distribution of Tn}%
\end{equation}
with $B:=(\omega-1)c$; equivalently, $T_{n}-B\sim\xi_{1}:=\omega S(\alpha)$.
Hence, Assumption~\ref{Assn T} is satisfied with $G(u)=P(\omega S(\alpha)\leq
u)=\Psi_{\alpha}(\omega^{-1}u)$, where $\Psi_{\alpha}(u):=P(S(\alpha)\leq u)$
is continuous. Inference based on quantiles of $\xi_{1}$ is invalid because it
misses the term~$B$.

\bigskip

\noindent\textsc{Bootstrap}. It is well known that the standard bootstrap
fails to be valid under infinite variance (Knight, 1989). The `$m$ out of $n$'
bootstrap (see Politis et al., 1999, and the references therein) is an
attractive option, but it fails to mimic the non-centrality parameter~$B$; see
Remark~\ref{Rem InfV m out of n} below. Instead, we consider the parametric
bootstrap of Cornea-Madeira and Davidson (2015), which only requires a
consistent estimator $\hat{\alpha}_{n}$ of the tail index $\alpha$, assumed to
lie in a compact set. The bootstrap sample is generated as
\[
y_{t}^{\ast}=\bar{y}_{n}+\varepsilon_{t}^{\ast},\quad\varepsilon_{t}^{\ast
}\sim~\text{i.i.d.}S(\hat{\alpha}_{n}),
\]
and the bootstrap estimator is $\hat{\theta}_{n}^{\ast}:=\omega\bar{y}%
_{n}^{\ast}=\omega(\bar{y}_{n}+\bar{\varepsilon}_{n}^{\ast})$ with
$\bar{\varepsilon}_{n}^{\ast}:=n^{-1}\sum_{t=1}^{n}\varepsilon_{t}^{\ast}$.
The bootstrap analogue of $T_{n}$ then satisfies
\[
T_{n}^{\ast}:=n^{1-1/\alpha}(\hat{\theta}_{n}^{\ast}-\bar{y}_{n})=\omega
n^{1-1/\alpha}\bar{\varepsilon}_{n}^{\ast}+\hat{B}_{n}\text{ with }\hat{B}%
_{n}:=(\omega-1)n^{1-1/\alpha}\bar{y}_{n}.
\]
Now, $n^{1-1/\alpha}\bar{\varepsilon}_{n}^{\ast}\overset{d^{\ast}%
}{\rightarrow}_{p}S(\alpha)$ by Proposition~1 in Cornea-Madeira and Davidson
(2015) and, therefore,
\[
T_{n}^{\ast}-\hat{B}_{n}\overset{d^{\ast}}{\rightarrow}_{p}\xi_{1}:=\omega
S(\alpha).
\]
This shows that Assumption~\ref{Assn BS}(i) is satisfied in this example.
Notice that the bias term in the bootstrap world satisfies, jointly with
(\ref{eq InfV distribution of Tn}),
\[
\hat{B}_{n}-B=(\omega-1)n^{1-1/\alpha}\bar{\varepsilon}_{n}\sim(\omega
-1)S(\alpha)=:\xi_{2}.
\]
Specifically, because both $T_{n}$ and $\hat{B}_{n}$ depend on the data
through $\bar{\varepsilon}_{n}$ only, we have that $(\xi_{1},\xi_{2}%
)\sim(\omega,\omega-1)S(\alpha)$, implying that $\xi_{1}-\xi_{2}\sim
S(\alpha)$. Hence, Assumption~\ref{Assn BS}(ii) is satisfied with
$F(u)=P(S(\alpha)\leq u)=\Psi_{\alpha}(u)$. Since the cdf of $\xi_{1}%
\sim\omega S(\alpha)$ can be written as $G(u)=\Psi_{\alpha}(\omega^{-1}u)$, it
follows by Theorem~\ref{Theor1} that $\hat{p}_{n}\rightarrow_{d}%
G(F^{-1}(U_{[0,1]}))=\Psi_{\alpha}(\omega^{-1}\Psi_{\alpha}^{-1}(U_{[0,1]}))$
and, therefore,
\[
P(\hat{p}_{n}\leq u)\rightarrow H(u):=P(\Psi_{\alpha}(\omega^{-1}\Psi_{\alpha
}^{-1}(U_{[0,1]}))\leq u)=\Psi_{\alpha}(\omega\Psi_{\alpha}^{-1}(u)),
\]
which differs from $u$ unless $\omega=1$.

Because $\omega$ is known and we can estimate $\alpha$ consistently with
$\hat{\alpha}_{n}$, we can estimate $H(u)$ consistently with $\hat{H}%
_{n}(u):=\Psi_{\hat{\alpha}_{n}}(\omega\Psi_{\hat{\alpha}_{n}}^{-1}(u))$ and
obtain a valid plug-in modified p-value,
\[
\tilde{p}_{n}=\hat{H}_{n}(\hat{p}_{n})=\Psi_{\hat{\alpha}_{n}}(\omega
\Psi_{\hat{\alpha}_{n}}^{-1}(\hat{p}_{n})),
\]
by application of Corollary~\ref{Corollary-plug-in}.

Alternatively, we can estimate $H(u)$ using the double bootstrap estimator
$\hat{H}_{n}(u):=P^{\ast}(\hat{p}_{n}^{\ast\ast}\leq u)$, where $\hat{p}%
_{n}^{\ast\ast}:=P^{\ast\ast}(T_{n}^{\ast\ast}\leq T_{n}^{\ast})$.
Specifically, let the double bootstrap sample $\{y_{t}^{\ast\ast}\}$ be
generated as%
\[
y_{t}^{\ast\ast}=\bar{y}_{n}^{\ast}+\varepsilon_{t}^{\ast\ast},\quad
\varepsilon_{t}^{\ast\ast}\sim~\text{i.i.d.}S(\hat{\alpha}_{n}),
\]
and set $\hat{\theta}_{n}^{\ast\ast}:=\omega\bar{y}_{n}^{\ast\ast}=\omega
\bar{y}_{n}^{\ast}+\omega\bar{\varepsilon}_{n}^{\ast\ast}$, where
$\bar{\varepsilon}_{n}^{\ast\ast}:=n^{-1}\sum_{t=1}^{n}\varepsilon_{t}%
^{\ast\ast}$. The (second-level) bootstrap analogue of $T_{n}^{\ast}$ then
satisfies
\[
T_{n}^{\ast\ast}:=n^{1-1/\alpha}(\hat{\theta}_{n}^{\ast\ast}-\bar{y}_{n}%
^{\ast})=\omega n^{1-1/\alpha}\bar{\varepsilon}_{n}^{\ast\ast}+\hat{B}%
_{n}^{\ast}\text{ with }\hat{B}_{n}^{\ast}:=(\omega-1)n^{1-1/\alpha}\bar
{y}_{n}^{\ast}.
\]
Since $\varepsilon_{t}^{\ast\ast}$ is generated from $S(\hat{\alpha}_{n})$,
where $\hat{\alpha}_{n}$ depends only on $D_{n}$, the distribution of
$\varepsilon_{t}^{\ast\ast}$, conditionally on $D_{n}^{\ast}$ and $D_{n}$, is
the same as the distribution of $\varepsilon_{t}^{\ast}$, conditionally on
$D_{n}$. This implies that%
\[
n^{1-1/\alpha}\bar{\varepsilon}_{n}^{\ast\ast}\overset{d^{\ast\ast
}}{\rightarrow}_{p\ast}S(\alpha),
\]
in probability, by Proposition~1 of Cornea-Madeira and Davidson (2015).
Therefore,
\[
T_{n}^{\ast\ast}-\hat{B}_{n}^{\ast}\overset{d^{\ast\ast}}{\rightarrow}_{p\ast
}\xi_{1}=\omega S(\alpha),
\]
in probability, showing that Assumption~\ref{Assn DBS}(i) is satisfied. Since%
\[
\hat{B}_{n}^{\ast}-\hat{B}_{n}=(\omega-1)n^{1-1/\alpha}(\bar{y}_{n}^{\ast
}-\bar{y}_{n})=(\omega-1)n^{1-1/\alpha}\bar{\varepsilon}_{n}^{\ast}%
\]
and $T_{n}^{\ast}-\hat{B}_{n}=\omega n^{1-1/\alpha}\bar{\varepsilon}_{n}%
^{\ast}$, Assumption~\ref{Assn DBS}(ii) is also satisfied in this example.
Thus, $\tilde{p}_{n}=\hat{H}_{n}(\hat{p}_{n})\rightarrow_{d}U_{[0,1]}$ by
Theorem~\ref{TheoremDouble-p-val}.

\begin{remark}
\label{Rem InfV m out of n}Consider the `$m$ out of $n$' bootstrap data
generating process,%
\[
y_{t}^{\ast}=\bar{y}_{n}+\varepsilon_{t}^{\ast},\quad t=1,\ldots,m,
\]
where $\varepsilon_{t}^{\ast}$ is an i.i.d.\ sample from the residuals
$\hat{\varepsilon}_{t}=y_{t}-\bar{y}_{n}$, $t=1,\ldots,n$. Then, with
$\hat{\theta}_{m}^{\ast}:=\omega\bar{y}_{m}^{\ast}$, $\bar{y}_{m}^{\ast
}:=m^{-1}\sum_{t=1}^{m}y_{t}^{\ast}$, the `$m$ out of $n$' bootstrap statistic
is%
\[
T_{m}^{\ast}:=m^{1-1/\alpha}(\hat{\theta}_{m}^{\ast}-\bar{y}_{n})=\omega
m^{1-1/\alpha}\bar{\varepsilon}_{m}^{\ast}+(\omega-1)m^{1-1/\alpha}\bar{y}%
_{n},
\]
where $m^{1-1/\alpha}\bar{\varepsilon}_{m}^{\ast}\overset{d^{\ast
}}{\rightarrow}_{p}S(\alpha)$ as $m\rightarrow\infty$; see Arcones and
Gin\'{e} (1989). Moreover, if $m=o(n)$,
\[
\hat{B}_{m}:=(\omega-1)m^{1-1/\alpha}\bar{y}_{n}=(\omega-1)m^{1-1/\alpha
}n^{1/\alpha-1}(n^{1-1/\alpha}\bar{y}_{n})=O_{p}((m/n)^{1/\alpha-1}%
)=o_{p}(1),
\]
which shows that $T_{m}^{\ast}\overset{d^{\ast}}{\rightarrow}_{p}\omega
S(\alpha)$. Hence, Assumption~\ref{Assn BS}(i) is satisfied with $\xi
_{1}:=\omega S(\alpha)$ and $\hat{B}_{n}=0$. Since $B:=(\omega-1)c\neq0$, we
have $\xi_{2}:=-B$ a.s., so that Assumption~\ref{Assn BS}(ii) does not hold.
As in Remark~\ref{RemarkBhat=0}, it then follows that%
\[
\hat{p}_{m}:=P^{\ast}(T_{m}^{\ast}\leq T_{n})\overset{d}{\rightarrow}%
G(G^{-1}(U_{[0,1]})+B)=\Psi_{\alpha}(\Psi_{\alpha}^{-1}(U_{[0,1]})+B).
\]
This shows that the limiting distribution of $\hat{p}_{m}$ depends on~$B$.
Since $B$ cannot be consistently estimated, the `$m$ out of $n$' bootstrap
cannot be used to solve the problem.
\end{remark}

\subsection{Nonlinear dynamic panel data models with incidental parameter
bias}

\label{sec:panel-details}

Another example that fits our framework is inference based on panel data
estimators subject to incidental parameter bias. We consider the properties of
the cross-sectional pairs bootstrap considered by Kaffo (2014), Dhaene and
Jochmans (2015), and Gon\c{c}alves and Kaffo (2015) in the context of a
general nonlinear panel data model. Although this bootstrap cannot replicate
the bias, we show that our prepivoting approach based on a plug-in estimator
of the bias is valid. Recently, Higgins and Jochmans (2022) proposed a
(double) bootstrap procedure that retains asymptotic validity without an
explicit plug-in estimator of the bias, but their procedure relies heavily on
the parametric distribution assumption.

\bigskip

\noindent\textsc{Setup}. Let $z_{it}$ denote a vector of random variables for
a set of $n$ individuals, $i=1,\ldots,n$, over $T$ time periods,
$t=1,\ldots,T$. Given a model for the density function $f_{it}(\theta
,\alpha_{i}):=f(z_{it},\theta,\alpha_{i})$, the parameter of interest is
$\theta\in\Theta$, which is common to all the individuals, while $\alpha
_{i}\in\mathcal{A}$ denote the individual fixed effects. The fixed effects
estimator of $\theta$ is the maximum likelihood estimator defined as%
\begin{equation}
\hat{\theta}_{n}=\arg\max_{\theta\in\Theta}\sum_{i=1}^{n}\sum_{t=1}^{T}\log
f_{it}(\theta,\hat{\alpha}_{i}(\theta)),\text{\ where\ }\hat{\alpha}%
_{i}(\theta)=\arg\max_{\alpha_{i}\in\mathcal{A}}\sum_{t=1}^{T}\log
f_{it}(\theta,\alpha_{i}). \label{MLE}%
\end{equation}

Under certain regularity conditions (see, e.g., Hahn and Kuersteiner, 2011),
including letting $n,T\rightarrow\infty$ jointly such that $n/T\rightarrow
\rho<\infty,$%
\begin{equation}
T_{n}:=\sqrt{nT}(\hat{\theta}_{n}-\theta)\overset{d}{\rightarrow}N(B,v^{2}),
\label{Asy-dist-panel}%
\end{equation}
where $B$ denotes the incidental parameter bias and $v^{2}$ is the asymptotic
variance of~$\hat{\theta}_{n}$. Hence, Assumption~\ref{Assn T} is satisfied
with $\xi_{1}\sim N(0,v^{2})$ (equivalently, Assumption~\ref{Assn T prime} is satisfied).

The exact forms of $B$ and $v^{2}$ may be quite involved and depend on the
type of heterogeneity and dependence assumptions imposed on~$z_{it}$. A
standard assumption is that $z_{it}$ is independent across $i$ while allowing
for time series dependence of unknown form; see Hahn and Kuersteiner~(2011).

\bigskip

\noindent\textsc{Bootstrap}. Given the cross-sectional independence
assumption, a natural bootstrap method in this context is the cross-sectional
pairs\ bootstrap. The idea is to resample $z_{i}=(z_{i1},\ldots,z_{iT}%
)^{\prime}$ in an i.i.d.\ fashion in the cross-sectional dimension. If
$z_{it}=(y_{it},x_{it})^{\prime}$ and $f(z_{it},\theta,\alpha_{i}%
)=f(y_{it}|x_{it},\theta,\alpha_{i})$ is the conditional density of $y_{it}$
given $x_{it}$, this is equivalent to a cross-sectional pairs bootstrap. As
the results of Kaffo (2014, Theorem~3.1) show, this bootstrap fails to capture
the bias term~$B$. In particular, letting $\hat{\theta}_{n}^{\ast}$ denote the
bootstrap analogue of $\hat{\theta}_{n}$, we have that
\[
T_{n}^{\ast}:=\sqrt{nT}(\hat{\theta}_{n}^{\ast}-\hat{\theta}_{n}%
)\overset{d^{\ast}}{\rightarrow}_{p}N(0,v^{2}),
\]
which implies that, as in Remarks~\ref{RemarkBhat=0}
and~\ref{Rem InfV m out of n},
\[
\hat{p}_{n}:=P^{\ast}(T_{n}^{\ast}\leq T_{n})=\Phi(v^{-1}T_{n})+o_{p}%
(1)\overset{d}{\rightarrow}\Phi(v^{-1}B+\Phi^{-1}(U_{[0,1]})).
\]
Thus,%
\[
P(\hat{p}_{n}\leq u)\rightarrow H(u):=P(\Phi(\Phi^{-1}(U_{[0,1]})+v^{-1}B)\leq
u)=\Phi(\Phi^{-1}(u)-v^{-1}B),
\]
which shows that the bootstrap test based on $\hat{p}_{n}$ is asymptotically
invalid since its limiting distribution is not uniform.

\begin{remark}
Note that, in this example, $\hat{L}_{n}(u):=P^{\ast}(T_{n}^{\ast}\leq
u)\rightarrow_{p}\Phi(u/v)$, showing that the bootstrap conditional
distribution of $T_{n}^{\ast}$ is not random in the limit. The invalidity of
$\hat{p}_{n}$ is due to the fact that the cross-sectional pairs bootstrap
induces $\hat{B}_{n}=0$, whereas $B\neq0$. This implies that $\hat{B}%
_{n}-B=-B:=\xi_{2}$ is not random. The fact that $\xi_{2}$ is not zero is the
cause of the bootstrap invalidity. See Remark~\ref{RemarkBhat=0}, which
contains this example as a special case.
\end{remark}

Contrary to previous examples (e.g.,\ Remark~\ref{Rem InfV m out of n}), $B$
and $v$ can both be consistently estimated. Hence, in this example we can
restore bootstrap validity by modifying the bootstrap p-value using a plug-in
approach. More specifically, let $\tilde{B}_{n}$ and $\hat{v}_{n}$ denote
consistent estimators of $B$ and $v$, respectively.\footnote{Since we reserve
the notation~$\hat{B}_{n}$ for the bootstrap-induced bias estimator (which is
zero for the cross sectional pairs bootstrap), we use the notation $\tilde
{B}_{n}$ to denote any consistent estimator of $B$ in this setup. For
instance, $\tilde{B}_{n}$ could be the plug-in estimator proposed by Hahn and
Kuersteiner (2011), which is based on a closed-form expression of $B_{1}$.
Another option is the half-split panel jackknife estimator of Dhaene and
Jochmans~(2015).} By Corollary~\ref{Corollary-plug-in},
\[
\tilde{p}_{n}=\hat{H}_{n}(\hat{p}_{n})=\Phi(\Phi^{-1}(\hat{p}_{n})-\hat{v}%
_{n}^{-1}\tilde{B}_{n})\overset{d}{\rightarrow}U_{[0,1]}%
\]
because $\hat{H}_{n}(u):=\Phi(\Phi^{-1}(u)-\hat{v}_{n}^{-1}\tilde{B}_{n})$ is
a consistent estimator of~$H(u)$.

\begin{remark}
A double bootstrap modified p-value version of $\tilde{p}_{n}$ is not valid in
this setting. The reason is that the double bootstrap mimics the behavior of
the first-level bootstrap, i.e.%
\[
T_{n}^{\ast\ast}:=\sqrt{nT}(\hat{\theta}_{n}^{\ast\ast}-\hat{\theta}_{n}%
^{\ast})\overset{d^{\ast\ast}}{\rightarrow}_{p}N(0,v^{2}),
\]
so that $\hat{B}_{n}^{\ast}$\ in Assumption~\ref{Assn DBS}(i) is zero. Since
$\hat{B}_{n}=0$, Assumption~\ref{Assn DBS}(ii) holds with $\hat{B}_{n}^{\ast
}-\hat{B}_{n}=0$, whereas Assumption~\ref{Assn BS}(ii) has $\hat{B}_{n}%
-B_{n}=-B$ a.s. Then,%
\[
\hat{p}_{n}^{\ast}=P^{\ast\ast}(v^{-1}T_{n}^{\ast\ast}\leq v^{-1}T_{n}^{\ast
})=\Phi(v^{-1}T_{n}^{\ast})\overset{d^{\ast}}{\rightarrow}_{p}\Phi(\Phi
^{-1}(U_{[0,1]}))=U_{[0,1]},
\]
whereas
\[
\hat{p}_{n}\overset{d}{\rightarrow}\Phi(\Phi^{-1}(U_{[0,1]})+v^{-1}B).
\]
Thus, $\hat{H}_{n}(u):=P^{\ast}(\hat{p}_{n}^{\ast}\leq u)$\ is not a
consistent estimator of $H(u)$, invalidating $\tilde{p}_{n}=\hat{H}_{n}%
(\hat{p}_{n})$.
\end{remark}

\begin{remark}
A special case of the previous setup is a linear panel dynamic model, where
$z_{it}=(y_{it},x_{it}^{\prime})^{\prime}$ and $x_{it}$ is a vector containing
lags of $y_{it}$ (Hahn and Kuersteiner, 2002). In this case, the plug-in
modified p-value, $\tilde{p}_{n}$, based on the cross-sectional pairs
bootstrap can be implemented using any consistent estimator of $B$, as
described above. However, we can also use a recursive bootstrap that exploits
the linearity of the model to obtain an asymptotically valid standard
bootstrap p-value, $\hat{p}_{n}$. The validity of $\hat{p}_{n}$ follows from
the fact that the recursive bootstrap estimates $B$ consistently, contrary to
the pairs bootstrap (Gon\c{c}alves and Kaffo, 2015). In light of this,
prepivoting $\hat{p}_{n}$ by computing a double bootstrap modified p-value
$\tilde{p}_{n}=\hat{H}_{n}(\hat{p}_{n})$ is not needed in this example, but it
is still a valid alternative.
\end{remark}

\section*{References}

\begin{description}
\item \textsc{Arcones, M.A., and E.\ Gine }(1989). The bootstrap of the mean
with arbitrary bootstrap sample size. \emph{Annales de l'Institut Henri
Poincar\'{e}, Probabilit\'{e}s et Statistiques} 25, 457--481.

\item \textsc{Beran, R. }(1987). Prepivoting to reduce level error in
confidence sets. \emph{Biometrika} 74, 457--468.

\item \textsc{Beran, R. }(1988). Prepivoting test statistics: A bootstrap view
of asymptotic refinements. \emph{Journal of the American Statistical
Association} 83, 687--97.

\item \textsc{Cavaliere, G., and I.\ Georgiev }(2020). Inference under random
limit bootstrap measures. \emph{Econometrica} 88, 2547--2574.

\item \textsc{Cornea-Madeira, A., and R.\ Davidson }(2015). A parametric
bootstrap for heavy-tailed distributions. \emph{Econometric Theory} 31, 449--470.

\item \textsc{Dhaene, G., and K. Jochmans }(2015). Split-panel jackknife
estimation of fixed-effect models. \emph{Review of Economic Studies} 82, 991--1030.

\item \textsc{Fu, W., and K.\ Knight }(2000). Asymptotics for lasso-type
estimators. \emph{Annals of Statistics} 28, 1356--1378.

\item \textsc{Goncalves, S., and M.\ Kaffo }(2015). Bootstrap inference for
linear dynamic panel data models with individual fixed effects. \emph{Journal
of Econometrics }186, 407--426.

\item \textsc{Hahn, J., and G.\ Kuersteiner }(2002). Asymptotically unbiased
inference for a dynamic panel model with fixed effects when both $n$ and $T$
are large. \emph{Econometrica} 70, 1639--1657.

\item \textsc{Hahn, J., and G.\ Kuersteiner }(2011). Bias reduction for
dynamic nonlinear panel models with fixed effects. \emph{Econometric Theory}
27, 1152--1191.

\item \textsc{Higgins, A., and K.\ Jochmans }(2022). Bootstrap inference for
fixed-effect models. Working paper, Cambridge University.

\item \textsc{Kaffo, M. }(2014). Bootstrap inference for nonlinear dynamic
panel data models with individual fixed effects. Manuscript.

\item \textsc{Knight, K. }(1989). On the bootstrap of the sample mean in the
infinite variance case. \emph{Annals of Statistics} 17, 1168--1175.

\item \textsc{Li, Q., and J.S. Racine\ }(2007). \emph{Nonparametric
Econometrics}. Princeton, NJ: Princeton University Press.

\item \textsc{Politis, D.N., J.P.\ Romano, and M.\ Wolf }(1999).
\emph{Subsampling}. New York: Springer.
\end{description}

\end{document}